\documentclass[prl,twocolumn,preprintnumbers,superscriptaddress,amsmath,amssymb]{revtex4-1}
\usepackage{graphicx}
\usepackage{subfigure}
\usepackage{mathrsfs}
\usepackage{amsfonts}
\usepackage{times}
\usepackage{amsmath}
\usepackage{leftidx}
\usepackage{tikz}
\usepackage{tikz-network}
\usepackage{color}
\usepackage[colorlinks,linkcolor=blue,citecolor=blue]{hyperref}

\newcommand{\ind}{\operatorname{ind}}
\newcommand{\Tr}{\operatorname{Tr}}
\newtheorem{definition}{Definition}
\newtheorem{theorem}{Theorem}
\newtheorem{lemma}{Lemma}
\newtheorem{proposition}{Proposition}
\usepackage{bbold}
\usepackage{braket}
\usepackage{mathtools}

\begin{document}
\title{Classification of Matrix-Product Unitaries with Symmetries}
\author{Zongping Gong}
\affiliation{Department of Physics, University of Tokyo, 7-3-1 Hongo, Bunkyo-ku, Tokyo 113-0033, Japan}
\affiliation{Max-Planck-Institut f\"ur Quantenoptik, Hans-Kopfermann-Stra{\ss}e 1, D-85748 Garching, Germany}
\author{Christoph S\"underhauf}
\affiliation{Max-Planck-Institut f\"ur Quantenoptik, Hans-Kopfermann-Stra{\ss}e 1, D-85748 Garching, Germany}
\author{Norbert Schuch}
\affiliation{Max-Planck-Institut f\"ur Quantenoptik, Hans-Kopfermann-Stra{\ss}e 1, D-85748 Garching, Germany}
\author{J. Ignacio Cirac}
\affiliation{Max-Planck-Institut f\"ur Quantenoptik, Hans-Kopfermann-Stra{\ss}e 1, D-85748 Garching, Germany}
\date{\today}

\begin{abstract}
We prove that matrix-product unitaries (MPUs) with on-site unitary symmetries are \emph{completely} classified by the (chiral) index and the cohomology class of the symmetry group $G$, provided that we can add trivial and symmetric ancillas with arbitrary on-site representations of $G$. If the representations in both system and ancillas are fixed to be the same, we can define \emph{symmetry-protected indices} (SPIs) which quantify the imbalance in the transport associated to each group element and greatly refines the classification. These SPIs are stable against disorder and measurable in interferometric experiments. Our results lead to a systematic construction of two-dimensional Floquet symmetry-protected topological (SPT) phases beyond the standard classification, and thus shed new light on understanding nonequilibrium phases of quantum matter.
\end{abstract}
\maketitle

\emph{Introduction.---} Classification of topological phases of matter is a central issue in modern condensed matter physics \cite{Ryu2016}. Particular recent interest is attracted by the classification of topological systems far from thermal equilibrium \cite{Kitagawa2010,Diehl2013,Else2016,Curt2016,Potter2016,Harper2017,SY2017,Gong2018,Gong2018b,Higashikawa2019,Cooper2018b,Cooper2018c,Max2019,Perez2019}. This tendency is largely driven by the remarkable experimental developments in atomic, molecular and optical physics, which have opened up unprecedented flexibility for controlling and probing quantum many-body dynamics \cite{Blatt2011,Blatt2013,Choi2017,Bernien2017,Zhang2017,Zhang2017b}. Moreover, understanding nonequilibrium phases of matter per se is of fundamental theoretical importance in extending the conventional paradigm of statistical mechanics to the largely unexplored nonequilibrium regime \cite{Eisert2015,Nandkishore2015,Moessner2017}.

For equilibrium interacting systems, the arguably most well-understood classification is that of one-dimensional (1D) bosonic symmetry-protected topological (SPT) phases \cite{Haldane1983,Tasaki1987,Gu2009,Pollmann2010} as ground states of gapped local Hamiltonians with symmetries. These 1D SPT phases are well described by the matrix-product states (MPSs) \cite{Werner1992,Verstraete2006,Perez2007,Verstraete2008}, and are completely classified by the second cohomology group \cite{Perez2008,Schuch2011,Chen2011,Chen2011b}, provided that the symmetries are not spontaneously broken. An analogous minimal setting in the nonequilibrium context is the classification of \emph{matrix-product unitaries} (MPUs) \cite{Chen2011b,Po2016,Cirac2017,Chen2018}, which have been shown to be equivalent to quantum cellular automata \cite{Cirac2017}. They efficiently approximate finite-time 1D dynamics generated by local Hamiltonians \cite{Osborne2006}. While an MPU can be regarded as an MPS with an enlarged local Hilbert space, the classification of MPUs can be very different from that of MPSs due to the unitarity requirement. Indeed, without symmetry protection, MPSs can always be continuously deformed into product states, while MPUs are classified by the (chiral) index quantized as the logarithm of a rational number \cite{Gross2012,Po2016,Cirac2017,Chen2018}. Efforts have also been made to classify 1D SPT many-body-localized (MBL) phases, and the result turns out to be the same as that of ground-state SPT MPSs \cite{Wahl2018}.  

In stark contrast to the case of MPSs, the problem of classifying MPUs commuting with a local symmetry operation stays unsolved. In this Letter, we address this problem for general on-site unitary symmetries forming a finite group $G$. First, we allow adding arbitrary symmetric ancillas (identities) with arbitrary on-site representations of $G$. We prove that the combination of the index and the second cohomology class \emph{completely} classifies all the MPUs with given symmetries. This actually proves a conjecture raised by Hastings \cite{Hastings2013} for quantum cellular automata. Second, we allow ancillas only with the same symmetry representation as the original system. Here, we unveil a series of quantized \emph{symmetry-protected indices} (SPIs). Nonzero SPIs 
quantify an imbalance of the left and right transport of each group element in the Heisenberg picture. We identify an observable signature of SPIs as the asymmetries in the two edges of symmetry-string operators evolved by the MPU, and propose an interferometry experiment for probing the SPIs relative to the index.

Our results have direct implications in the classification of Floquet SPT phases \cite{Harper2019}. Given a 2D Floquet system with boundary in the MBL regime, its edge dynamics is well described by an MPU \cite{Po2016}. Here, we construct a class of 2D Floquet systems with edge MPUs characterized by nontrivial SPIs, and provide a unified picture for understanding the edge dynamics of 2D intrinsic Floquet SPT phases as symmetry-charge pumps.

\emph{Matrix-product unitaries (MPUs).---}
We are interested in 1D quantum dynamics that keep locality, i.e., the unitaries $U$'s that map any operator $O$ supported on a finite region $R$ into $U^\dag OU$ supported on another finite region $R'\supseteq R$. In particular, we wish to classify all possible $U$ that can continuously be deformed into each other by keeping locality, which defines different dynamical phases. According to Ref.~\cite{Cirac2017}, this allows us to use MPUs, which we define in the following.

An MPU of length $L$ is a unitary operator $U^{(L)}: (\mathbb{C}^d)^{\otimes L} \to (\mathbb{C}^d)^{\otimes L}$ generated by a rank-four tensor $\mathcal{U}$:
\begin{equation}
 U^{(L)} = \sum_{\mathclap{i_1,\ldots,i_L, j_1,\ldots,j_L}}\Tr( \mathcal{U}_{i_1j_1}\ldots\mathcal{U}_{i_Lj_L}) \ket{i_1,\ldots,i_L}\!\bra{j_1,\ldots,j_L}.
\end{equation}
The dimension $D$ of the matrices $\mathcal{U}_{ij}$ is called bond dimension.
After blocking $k$, at most $D^4$ times (combining multiple physical indices into one index), $\mathcal{U}\to\mathcal{U}_k$ is termed \emph{simple} and $U^{(L)}$ acquires the standard form
\begin{equation}
\label{standard form}
\begin{tikzpicture}[scale=0.8]
\draw[ultra thick] (0.2,0.5) -- (0.2,0.8) (1.4,0.5) -- (1.4,0.8) (2.6,0.5) -- (2.6,0.8);
\draw[ultra thick,dotted] (0.8,0.5) -- (0.8,0.8) (2,0.5) -- (2,0.8) ;
\draw[thick] (0.2,-0.25) -- (0.2,0) (0.2,1.3) -- (0.2,1.55) (0.8,-0.25) -- (0.8,0) (0.8,1.3) -- (0.8,1.55) (1.4,-0.25) -- (1.4,0) (1.4,1.3) -- (1.4,1.55) (2,-0.25) -- (2,0) (2,1.3) -- (2,1.55) (2.6,-0.25) -- (2.6,0) (2.6,1.3) -- (2.6,1.55)   ;
\draw[thick] (0,0) rectangle (1,0.5);
\draw[thick] (1.2,0) rectangle (2.2,0.5);
\draw[thick] (2.9, 0.5) -- (2.4, 0.5) -- (2.4, 0) -- (2.9, 0);
\draw[thick] (-0.1,0.8) -- (0.4,0.8) -- (0.4,1.3) -- (-0.1,1.3);
\draw[thick] (0.6,0.8) rectangle (1.6,1.3);
\draw[thick] (1.8,0.8) rectangle (2.8,1.3);
\Text[x=-0.7,y=0.62]{$=$}
\Text[x=0.5,y=0.25]{$u$}
\Text[x=1.7,y=0.25]{$u$}
\Text[x=1.1,y=1.05]{$v$}
\Text[x=2.3,y=1.05]{$v$}

\draw[thick] (-2, 0.4) rectangle (-1.5, 0.9);
\draw[thick] (-2.8, 0.4) rectangle (-2.3, 0.9);
\draw[thick] (-3.6, 0.4) rectangle (-3.1, 0.9);
\draw[thick] (-4.4, 0.4) rectangle (-3.9, 0.9);
\draw[thick] (-5.2, 0.4) rectangle (-4.7, 0.9);

\Text[x=-1.72, y=0.65,fontsize=\footnotesize]{$\mathcal{U}_k$}
\Text[x=-2.52, y=0.65,fontsize=\footnotesize]{$\mathcal{U}_k$}
\Text[x=-3.32, y=0.65,fontsize=\footnotesize]{$\mathcal{U}_k$}
\Text[x=-4.12, y=0.65,fontsize=\footnotesize]{$\mathcal{U}_k$}
\Text[x=-4.92, y=0.65,fontsize=\footnotesize]{$\mathcal{U}_k$}

\draw[thick] (-5.4, 0.65) -- (-5.2, 0.65) (-4.7, 0.65) -- (-4.4, 0.65) (-3.9, 0.65) -- (-3.6, 0.65) (-2.8, 0.65) -- (-3.1, 0.65) (-2.3, 0.65) -- (-2, 0.65) (-1.5, 0.65) -- (-1.3, 0.65);
\draw[thick] (-1.75, 0.9) -- (-1.75, 1.4) (-2.55, 0.9) -- (-2.55, 1.4) (-3.35, 0.9) -- (-3.35, 1.4) (-4.15, 0.9) -- (-4.15, 1.4) (-4.95, 0.9) -- (-4.95, 1.4);
\draw[thick] (-1.75, 0.4) -- (-1.75, -0.1) (-2.55, 0.4) -- (-2.55, -0.1) (-3.35, 0.4) -- (-3.35, -0.1) (-4.15, 0.4) -- (-4.15, -0.1) (-4.95, 0.4) -- (-4.95, -0.1);

\end{tikzpicture}
\end{equation}
in terms of unitaries $u:(\mathbb{C}^{d^k})^{\otimes2}\to\mathbb{C}^l\otimes\mathbb{C}^r$ and $v:\mathbb{C}^r\otimes\mathbb{C}^l\to(\mathbb{C}^{d^k})^{\otimes2}$. We apply operators in the graphical notation from bottom to top. The unitaries are unique up to gauge transformations $u\to  (X^\dag \otimes Y^\dag)u, v \to v(Y \otimes X)$, where $d$ is the local Hilbert-space dimension before blocking, $X\in{\rm U}(l)$ and $Y\in{\rm U}(r)$. Conversely, two arbitrary unitaries $u$ and $v$ generate an MPU, possibly with the unit cell doubled. 

We will focus our attention on $G$-symmetric MPUs which commute with a unitary representation $\rho_g$ of the finite group $G$, $[\rho_g^{\otimes L}, U^{(L)}] = 0$. Henceforth, we omit the length $L$ due to the translation-invariance of MPUs and assume the standard form. Although we assume translation invariance throughout the letter, all the topological indicators we unveil can be shown to be stable against disorder \cite{SM}. The essential physics behind the stability is the locality-preserving constraint, which is obviously satisfied by the standard form (\ref{standard form}) even if $u$ and $v$ are position-dependent.

\emph{Equivalence and complete classification.---}
We classify the MPUs according to
\begin{definition}[Equivalence]
\label{def:equivalence}
 Two $G$-symmetric MPUs $U_0$ and $U_1$ are equivalent if we allow for blocking (i.e., treat multiple sites as a single site), and the addition of local ancillas with the identity operator, such that the MPUs can then be continuously connected within the manifold of symmetric MPUs. 
\end{definition}
Here, by adding local ancillas to an MPU $U$, we mean that we take the enlarged MPU $U'=U\otimes\mathbb{1}_{\rm a}^{\otimes L}$ on $(\mathbb{C}^d\otimes\mathbb{C}^{d_{\rm a}})^{\otimes L}$, and consider the representation $\rho_g'=\rho_g\otimes\sigma_g$, where $\sigma$ can be an \emph{arbitrary} representation of $G$ on $\mathbb{C}^{d_{\rm a}}$. 

MPUs can be considered as MPSs by bunching the two physical indices of each tensor into one. Thus, $G$-symmetric MPUs can be associated to a cohomology class in $H^2(G, {\rm U}(1))$ \cite{3GC}. However, the fact that they are unitary gives extra restrictions. In order to analyze these restrictions, we employ the standard form (\ref{standard form}) and note that, due to the gauge redundancy, the action of the symmetry on the building blocks consists of two projective representations $x_g$ and $y_g$:
\begin{equation}
\begin{tikzpicture}[scale=0.8]
\draw[thick] (-1.8,-0.35) circle (0.23);
\draw[thick] (-1.2,-0.35) circle (0.23);
\draw[thick] (-2,0) rectangle (-1,0.5);
\draw[thick] (-1.8,-0.12) -- (-1.8,0) (-1.8,-0.75) -- (-1.8,-0.57) (-1.2,-0.12) -- (-1.2,0) (-1.2,-0.75) -- (-1.2,-0.57);
\draw[ultra thick] (-1.8,0.5) -- (-1.8,0.75);
\draw[ultra thick,dotted] (-1.2,0.5) -- (-1.2,0.75);
\Text[x=-1.5,y=0.25]{$u$}
\Text[x=-1.76,y=-0.35,fontsize=\scriptsize]{$\rho_g^\dag$}
\Text[x=-1.16,y=-0.35,fontsize=\scriptsize]{$\rho_g^\dag$}
\Text[x=-0.5,y=0.25]{$=$}
\draw[thick] (0,0) rectangle (1,0.5);
\draw[thick] (0.2,0.85) circle (0.23);
\draw[thick] (0.8,0.85) circle (0.23);
\draw[thick] (0.2,-0.25) -- (0.2,0) (0.8,-0.25) -- (0.8,0);
\draw[ultra thick] (0.2,0.5) -- (0.2,0.63) (0.2,1.07) -- (0.2,1.22);
\draw[ultra thick,dotted] (0.8,0.5) -- (0.8,0.63) (0.8,1.07) -- (0.8,1.22);
\Text[x=0.21,y=0.85,fontsize=\scriptsize]{$x^\dag_g$}
\Text[x=0.81,y=0.85,fontsize=\scriptsize]{$y^\dag_g$}
\Text[x=0.5,y=0.25]{$u$}
\draw[thick] (3.2,0.85) circle (0.23);
\draw[thick] (3.8,0.85) circle (0.23);
\draw[thick] (3,0) rectangle (4,0.5);
\draw[thick] (3.8,0.5) -- (3.8,0.63) (3.8,1.07) -- (3.8,1.22) (3.2,0.5) -- (3.2,0.63) (3.2,1.07) -- (3.2,1.22);
\draw[ultra thick] (3.8,-0.25) -- (3.8,0);
\draw[ultra thick,dotted] (3.2,-0.25) -- (3.2,0);
\Text[x=3.5,y=0.25]{$v$}
\Text[x=3.2,y=0.85,fontsize=\scriptsize]{$\rho_g$}
\Text[x=3.8,y=0.85,fontsize=\scriptsize]{$\rho_g$}
\Text[x=4.5,y=0.25]{$=$}
\draw[thick] (5,0) rectangle (6,0.5);
\draw[thick] (5.2,-0.35) circle (0.23);
\draw[thick] (5.8,-0.35) circle (0.23);
\draw[thick] (5.8,0.5) -- (5.8,0.75) (5.2,0.5) -- (5.2,0.75);
\draw[ultra thick] (5.8,0) -- (5.8,-0.13) (5.8,-0.57) -- (5.8,-0.72);
\draw[ultra thick,dotted] (5.2,0) -- (5.2,-0.13) (5.2,-0.57) -- (5.2,-0.72);
\Text[x=5.2,y=-0.35,fontsize=\scriptsize]{$y_g$}
\Text[x=5.8,y=-0.35,fontsize=\scriptsize]{$x_g$}
\Text[x=5.5,y=0.25]{$v$}
\Text[x=2,y=0.25]{and}
\Text[x=6.25,y=0]{.}
\end{tikzpicture}
\label{proj MPU}
\end{equation}
Both $x_g$ and $y_g^*$ belong to the same cohomology class as the associated MPS \cite{SM}.

The index \cite{Gross2012,Cirac2017,Chen2018} of the MPU is defined as
\begin{equation}
 \label{index}
 \ind \equiv\frac{1}{2}\log\frac{r}{l}=\frac{1}{2} \log \frac{\Tr y_e}{\Tr x_e}
\end{equation}
for the identity $e\in G$; it captures the imbalance of right and left-moving information. Both index and cohomology class are stable under blocking and additive under tensoring as well as composition of MPUs \cite{SM}.

As Hastings conjectured \cite{Hastings2013}, equivalent phases are indeed completely classified by index and cohomology:
\begin{theorem}[Equivalence]
\label{thm:equivalence}
 Two symmetric MPUs $U_0$ and $U_1$ with the same or different symmetry representations are equivalent if and only if they share the same indices and same cohomology classes.
\end{theorem}
Note that the necessity of same indices was shown by Cirac \emph{et al}. \cite{Cirac2017}; that of same cohomology classes follows from \cite{Schuch2011}, just as for MPS. We then only have to construct an explicit path that continuously connects $U_0$ with $U_1$. This turns out to be always possible after a symmetrization of the on-site symmetry representations of $U_0$ and $U_1$ and a regularization through attaching ancillas with regular representations \cite{SM}.

\begin{figure}
\begin{center}
       \includegraphics[width=8.5cm, clip]{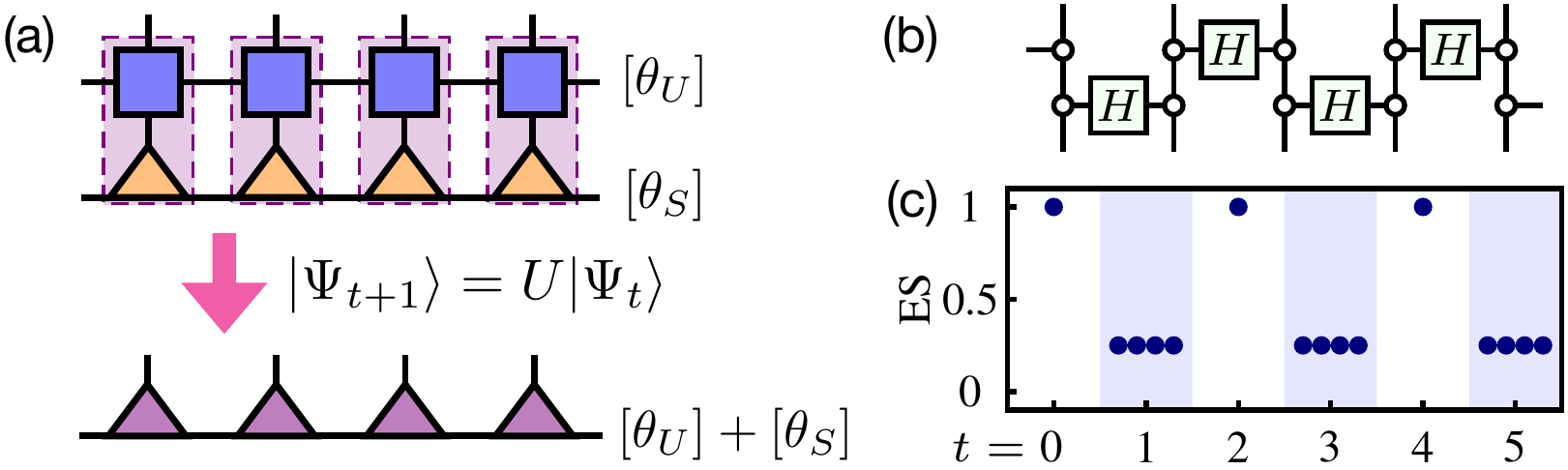}
       \end{center}
   \caption{(color online). (a) $G$-symmetric MPS evolved by a $G$-symmetric MPU with the cohomology classes summing up. (b) $\mathbb{Z}_2\times\mathbb{Z}_2$-symmetric MPU in the nontrivial cohomology class. Here 
   $\circ$ is the delta tensor and $H$ is the Hadamard matrix. (c) Stroboscopic dynamics of the entanglement spectrum (ES) governed by (b) starting from a symmetric product state.}
   \label{fig1}
\end{figure}

Examples of MPUs with nontrivial cohomology classes are already found in Refs.~\cite{Potter2017,Harper2017c} as the edge dynamics of 2D intrinsic Floquet SPT phases \footnote{By intrinsic, we mean that they have no equilibrium counterparts and correspond to $H^2(G,{\rm U}(1))$ part in the cohomology classification $H^3(\mathbb{Z}\times G,{\rm U}(1))=H^3(G,{\rm U}(1))\times H^2(G,{\rm U}(1))$ \cite{Else2016}.}. Initialized as a symmetric state, a nontrivial 1D edge evolves from one SPT phase into another after each Floquet period, reminiscent of the discrete time crystals which toggle between different symmetry-broken phases \cite{Khemani2016,Else2016b,Curt2016b,Yao2017}. In the tensor-network picture, we can understand this ``topological discrete time-crystalline oscillation" from the virtual level --- when a symmetric MPS is evolved by a symmetric MPU, their cohomology classes simply sum up (see Fig.~\ref{fig1}(a)). To diagnose this phenomenon, we may trace the stroboscopic evolution of the entanglement spectrum \cite{Gong2018b,Cooper2018b,Cooper2018c}, which is experimentally accessible by many-body-state tomography \cite{Blatt2017} or interferometric measurement \cite{Pichler2016}. For the $G=\mathbb{Z}_n\times\mathbb{Z}_n$ SPT MPU corresponding to the generator of $H^2(G,U(1))=\mathbb{Z}_n$ \cite{Potter2017}, starting from a symmetric trivial state, we will obtain (at least) $(n/{\rm gcd}(n,t))^2$-fold degeneracy in the entanglement spectrum after $t$ time steps \cite{SM}. See Figs.~\ref{fig1}(b) and (c) for the simplest case $n=2$. More general examples with nontrivial cohomology classes are available in the Supplemental Materials \cite{SM}.

\emph{Strong equivalence and symmetry-protected indices.---}
In real physical systems with symmetries, 
the 
representation is usually determined by the microscopic details and cannot be changed freely. This motivates us to ask how the classification will be modified if the representation is fixed. Forbidding arbitrary representations for the ancillas in Def.~\ref{def:equivalence} leads to
\begin{definition}[Strong Equivalence]
 Two $G$-symmetric MPUs $U_0$ and $U_1$ are strongly equivalent if (i) their on-site representations are (generally different) powers of a single fixed representation $\rho$ of $G$ and (ii) they can be continuously connected within the manifold of symmetric MPUs upon blocking and/or adding identities as ancillas with representation $\rho$.
\end{definition}
If $\rho$ is regular, we will return to Thm.~\ref{def:equivalence}. Otherwise, there is at least one $g\neq e$ with character $\chi_g\equiv\Tr\rho_g\neq0$. In this case, the notion of strong equivalence refines the phase structure beyond Thm.~\ref{def:equivalence}, as revealed by the SPIs which are natural generalization of the index (\ref{index}) to other group elements:
\begin{definition}[Symmetry Protected Index] Given a $G$-symmetric MPU $U$ for which we can determine $x_g$ and $y_g$ from a standard form, the SPI with respect to $g\in G$ with $\chi_g\neq0$ is defined as
\begin{equation}
{\rm ind}_g\equiv\frac{1}{2}\log\left|\frac{\Tr y_g}{\Tr x_g}\right|.
\label{indg}
\end{equation}
\end{definition}
Given a blocking number $k$, the SPI is well defined since the absolute value removes the phase ambiguity and the trace is gauge-invariant. Moreover, we can show that, just like $\ind=\ind_e$ \cite{Cirac2017}, $\ind_g$ is invariant under blocking and additive under tensoring and composition \cite{SM}.

We further claim that the SPI is a topological invariant for strong equivalence. Recall that ${\rm ind}_g$ does not rely on blocking and is obviously invariant if we add identities with the fixed representation. Moreover, it can be shown that ${\rm ind}_g$ is continuous and stays discretized during a continuous deformation \cite{SM}. Therefore, the SPI is a quantizied topological invariant, implying
\begin{theorem}
 Two symmetric, strongly equivalent MPUs share the same SPI for all group elements with $\chi_g\neq0$. 
\label{thm2}
\end{theorem}

\begin{figure}
\begin{center}
       \includegraphics[width=8.5cm, clip]{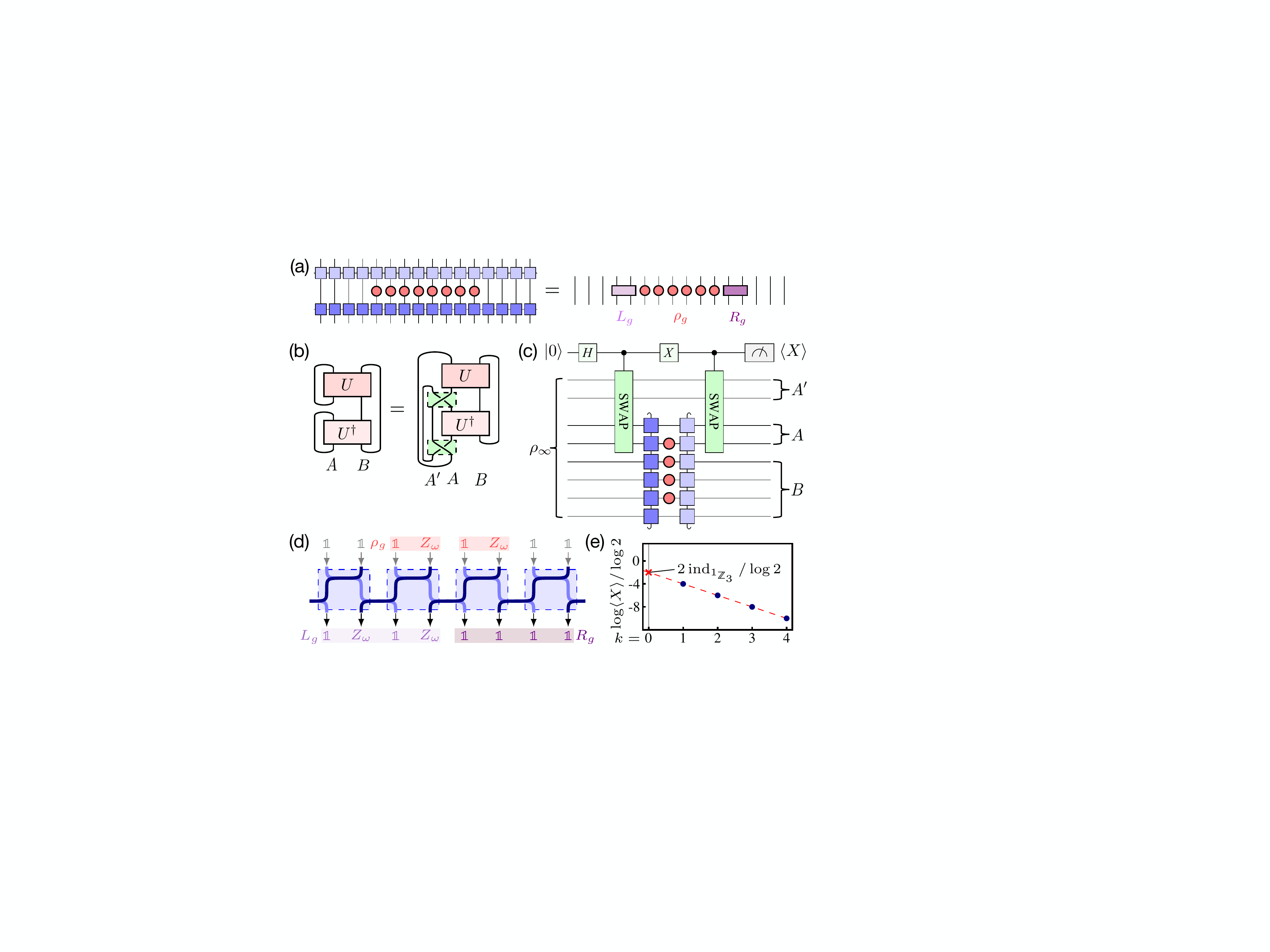}
       \end{center}
   \caption{(color online). (a) Symmetry string operator evolved by a symmetric MPU. Only the left and right edges are modified into $L_g$ and $R_g$, respectively. (b) Tensor-network representation of $\Tr_B[\Tr_AU\Tr_AU^\dag]=\Tr[U\mathbb{S}U^\dag\mathbb{S}]$, where $\mathbb{S}$ (green rectangles) is the SWAP operator between subsystem $A$ and its copy $A'$. (c) Interferometric approach to probing the relative SPI (\ref{LRind}). Initially, the qubit is set to be $|0\rangle$ while the remaining part is prepared as the infinite-temperature state $\rho_\infty$. Here $H$ is the Hadamard gate and the controlled-SWAP gates read $U_{\rm CS}=|0\rangle\langle 0|\otimes\mathbb{1}_{A'A}+|1\rangle\langle 1|\otimes\mathbb{S}$. The final expectation value $\langle X\rangle$ of the qubit is related to $|\Tr L_g|$ and thus the relative SPI. (d) Bilayer SWAP circuit subject to $\mathbb{Z}_n$ symmetry. (e) SPI of (d) with respect to $1_{\mathbb{Z}_{n=3}}$ determined by linear fitting (\ref{indgXk}).}
   \label{fig2}
\end{figure}

The contraposition of Thm.~\ref{thm2} allows us to use SPIs to distinguish topologically different MPUs. For cyclic groups $G=\mathbb{Z}_n$ with $n\ge3$, the minimal nontrivial example is the bilayer SWAP circuit of qubits \cite{Cirac2017}, where a single site contains two qubits and $\rho_{1_{\mathbb{Z}_n}}=\mathbb{1}\otimes Z_{\omega_n}$ ($1_{\mathbb{Z}_n}$: generator of $\mathbb{Z}_n$)
and $Z_{\omega_n}\equiv|0\rangle\langle0|+e^{2\pi i/n}|1\rangle\langle1|$ (see Fig.~\ref{fig2}(d)). We can check that $x_{1_{\mathbb{Z}_n}}=\mathbb{1}^{\otimes2}$ and $y_{1_{\mathbb{Z}_n}}=Z_{\omega_n}^{\otimes2}$, leading to $\ind_{1_{\mathbb{Z}_n}}=\log|\cos\frac{\pi}{n}|\neq0$, which is sufficient to rule out the \emph{strong} equivalence between the bilayer SWAP circuit and the identity. However, having $\ind=0$ and trivial cohomology, it is still equivalent to the identity. While the SPI therefore allows for an enriched classification for strong equivalence, the classification provided by Thm.~\ref{thm2} is \emph{not} complete \cite{SM}.

\emph{Physical implication and experimental probing of SPIs.---} Similar to the cohomology character, the SPI (\ref{indg}) is defined on the virtual level, so its physical meaning is not clear at first glance. Having in mind that SPT phases with nontrivial cohomology classes usually exhibit exotic edge physics \cite{Else2014}, we are naturally led to think about a similar situation for SPIs, which depend on $g$. In fact, we can consider a sufficiently long string operator $\rho^{\otimes N}_g$ evolved by the MPU and show that the $g$-string operator will almost stay unchanged, except that near the left and right edges two $2k$-site unitaries $L_g$ and $R_g$ emerge (see Fig.~\ref{fig2}(a)). These two unitaries on the physical level are related to $x_g$ and $y_g$ on the virtual level via $L_g=u^\dag(x_e\otimes y_g)u$ and $R_g=u^\dag(x_g\otimes y_e)u$, leading to
\begin{equation}
{\rm ind}_g-{\rm ind}=\frac{1}{2}\log\left|\frac{\Tr L_g}{\Tr R_g}\right|.
\label{LRind}
\end{equation}
It is now clear from Eq.~(\ref{LRind}) that $\ind_g$ gives a measure of the edge imbalance in the $g$-string operator evolved by the MPU.

Equation (\ref{LRind}) also opens up the possibility for practically measuring the SPI relative to the index. Note that $\Tr L_g\Tr R_g=d^{2k}\chi^{2k}_g$, it is sufficient to measure either $|\Tr L_g|$ or $|\Tr R_g|$. This problem can be simplified into how to measure $|\Tr U_A|$ for a subsystem unitary $U_A$ embedded in $U=U_A\otimes U_B$, where the Hilbert-space dimension $d_B$ of subsystem $B$ can be much larger than $d_A$, that of subsystem $A$. Combining $|\Tr U_A|^2=d^{-1}_B{\rm Tr}_B[\Tr_AU\Tr_AU^\dag]$ with the identity in Fig.~\ref{fig2}(b), we obtain $|\Tr U_A|^2=d^2_A\Tr[U\mathbb{S}\rho_\infty U^\dag\mathbb{S}]$, where $\mathbb{S}$ is the SWAP operator acting on $A$ and a copy $A'$ and $\rho_\infty\equiv d^{-2}_Ad^{-1}_B\mathbb{1}_{A'AB}$ is the infinite-temperature state of the entire system including $A'$. Since eventually we rewrite $|\Tr U_A|^2$ into the form of a Loschmidt echo, we can measure it by means of the standard interferometric approach \cite{Vedral2013,Swingle2016,Zhu2016,Yao2016}.

We sketch out the experimental scheme in Fig.~\ref{fig2}(c), where an auxiliary qubit is introduced and either of two controlled-SWAP gates consists of $2k$ two-site ones acting on a region $A$ near the left domain wall and its copy $A'$. By measuring the final expectation value $\langle X \rangle$ for the Pauli $X$ of the auxiliary qubit, we can determine the relative SPI from
\begin{equation}
{\rm ind}_g-{\rm ind}=\frac{1}{2}\log\langle X\rangle+k\log\frac{d}{|\chi_g|}.
\label{indgXk}
\end{equation}
If $d$ and $\chi_g$ are unknown, we are still able to measure $\langle X \rangle$ with increasing length $2k$ of $A$ and then extract ${\rm ind}_g$ from a linear fitting. See Fig.~\ref{fig2}(e) for the example of the bilayer SWAP circuit subject to $\mathbb{Z}_3$ symmetry.

\emph{General parent Floquet systems as symmetry-charge pumps.---} Recalling the relation between MPUs and Floquet systems, the (strong) equivalence between MPUs are \emph{necessary} for the (strong) equivalence between the corresponding $G$-symmetric 2D MBL Floquet systems -- they are continuously connected without crossing a delocalization point \cite{Po2016,Potter2017,Harper2017c}. This is because MBL implies a spatial factorization of the bulk Floquet unitary and its separation from the boundary unitary, which is 1D, locality-preserving and thus well described by an MPU \cite{Po2016}. A continuous deformation of the Floquet system thus gives rise to that of the edge MPU. Conversely, two inequivalent MPUs \emph{sufficiently} distinguish their parent Floquet systems.

\begin{figure}
\begin{center}
       \includegraphics[width=7.5cm, clip]{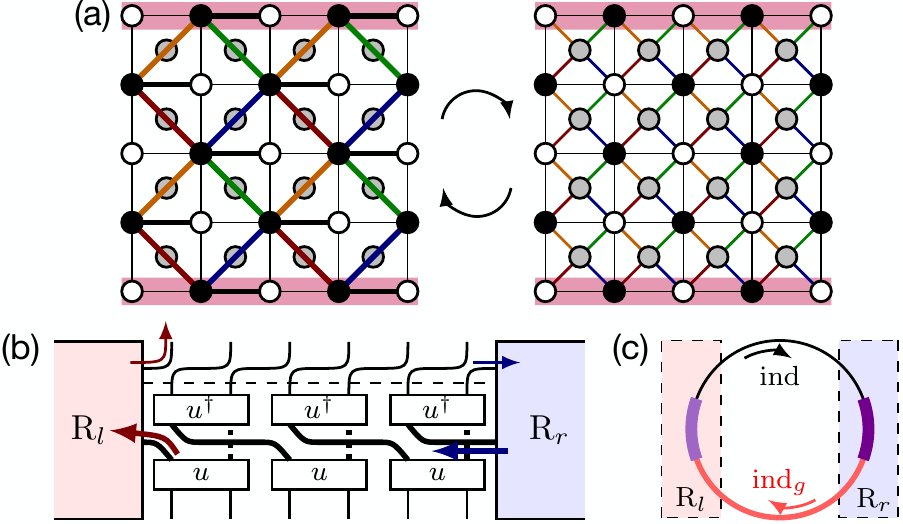}
       \end{center}
   \caption{(color online). (a) 2D Floquet system with a trivial bulk and a nontrivial edge dynamics (shaded in magenta) governed by an MPU. The open (periodic) boundary condition is imposed to the vertical (horizontal) direction. In the first (left panel)/second (right panel) half period, we apply $u$-conjugated (thick black bonds) SWAP gates (thick color bonds)/physical-level SWAP gates (color bonds) sequentially as \textcolor{red!50!black}{red}$\to$\textcolor{blue!50!black}{blue}$\to$\textcolor{green!50!black}{green}$\to$\textcolor{orange!75!black}{orange}. (b) MPU segment as a symmetry-charge pump that transfers $q_\varrho$ from ${\rm R}_l$ to ${\rm R}_r$ and $q_x$ from ${\rm R}_r$ to ${\rm R}_l$. The circuits above/below the dashed line are generated by the left/right panel in (a). (c) Edge imbalance (light/dark purple) in an evovled $g$-string operator (pink) from current imbalance (\ref{LRind}).}
   \label{fig3}
\end{figure}

It is thus natural to ask whether an MPU with nontrivial SPIs can be embedded into a parent Floquet system, just like those with nontrivial indices \cite{Po2016} and cohomology classes \cite{Potter2017,Harper2017c}. Since topologically different MPUs distinguish different MBL parent Floquet systems, the embeddability would imply a new class of 2D SPT Floquet phases characterized by SPIs. We answer in the affirmative by giving a general construction shown and explained in Fig.~\ref{fig3}(a), whose bulk is trivial and thus many-body localizable \cite{Potter2015}, while the edge dynamics is governed by an MPU generated by $u$ and $v=u^\dag\mathbb{S}_{\rm v}$, where $\mathbb{S}_{\rm v}$ exchanges the virtual Hilbert spaces $\mathbb{C}^l$ and $\mathbb{C}^r$. This construction is inspired by the standard form (\ref{standard form}) and the four-step SWAP model \cite{Rudner2013,Po2016,Harper2017b} --- we compose two four-step SWAP processes, one on the left virtual Hilbert spaces and pulled back by $u$, and the other on the physical level.

The above general construction of parent Floquet systems in turn gives a simple \emph{symmetry-charge-pump} picture for topological MPUs. Here a $G$-symmetry charge refers to a Hilbert space on which $G$ acts as a linear (integer charge) or projective representation (fractional charge). These charges can fuse or split following the fusion rules set by the group structure. With the physical and the left virtual charge in the standard form denoted as $q_{\varrho}\equiv k q_\rho$ 
and $q_x$ \cite{QX}, an MPU segment coupled to two symmetry reservoirs ${\rm R}_{l,r}$ right-translates $q_\varrho$ and left-translates $q_x$ (see Fig.~\ref{fig3}(b)). In fact, the cohomology class and the SPIs (including the index) are all characters of the net symmetry-charge current $q_\varrho-q_x$. This picture unifies all the related previous works as special situations, such as $G=\{e\}$ \cite{Po2016,Harper2017b} and $\Tr x_g=\Tr \varrho_g=\delta_{ge}\dim\varrho$ \cite{Potter2017,Harper2017c}. Remarkably, this picture gives an intuition into Eq.~(\ref{LRind}): We regard two equally long segments centered at the edges of a $g$-string operator as ${\rm R}_{l,r}$, which are connected by two pumps with inputs $g$ and $e$ (see Fig.~\ref{fig3}(c)). We can then interpret Eq.~(\ref{LRind}) as an \emph{equation of continuity}, with the left- and right-hand sides being the current and the change of charge, respectively. There is a factor $\frac{1}{2}$ since a net flow of charge $q$ causes $2q$ charge imbalance.

\emph{Summary and outlook.---} We have focused on the classification problem of symmetric MPUs, where the symmetry representation can be arbitrary or fixed. In the former case, we achieve a complete classification based on the index and the cohomology class. In the latter case, we unveil a set of experimentally accessible SPIs that enrich the classification and lead to the discovery of a new class of 2D Floquet SPT phases. However, the complete classification in the latter case stays an open problem, which we leave for future work. Other directions for future studies include the generalization to anti-unitary \cite{Cirac2017} and continuous symmetries, fermionic systems \cite{Fidkowski2019} and higher dimensions \cite{Hastings2018}. Since both SPIs and cohomology classes apply to inhomogeneous unitaries, it would also be interesting to study the impact of topology on information scrambling in random circuits \cite{Curt2018,Tibor2018,Nahum2018,Khemani2018,Christoph2018,Chan2018,Chan2018b}.

We acknowledge M. Ueda, K. Shiozaki, M. Furuta, M. Sato, D. T. Stephen, H. Tasaki, and H. Katsura for valuable discussions. This project has received funding from the European Research Council (ERC) under the European Union's Horizon 2020 research and innovation programme through the ERC Starting Grant WASCOSYS (No. 636201) and the ERC Advanced Grant QENOCOBA (No. 742102). Z. G. acknowledges support from the University of Tokyo through the Graduate Research Abroad in Science Program (GRASP) and the Max-Planck-Institut f\"ur Quantenoptik for hospitality, where this work was completed.

\bibliography{GZP_references}

\clearpage
\begin{center}
\textbf{\large Supplemental Materials}
\end{center}
\setcounter{equation}{0}
\setcounter{figure}{0}
\setcounter{table}{0}
\makeatletter
\renewcommand{\theequation}{S\arabic{equation}}
\renewcommand{\thefigure}{S\arabic{figure}}
\renewcommand{\bibnumfmt}[1]{[S#1]}

We provide the details on the proof of Theorem \ref{thm:equivalence}, some analytical and numerical results on the $\mathbb{Z}_d\times\mathbb{Z}_d$ SPT MPU, a rigorous formalism of the (refined) SPIs for MPUs and general locality-preserving unitaries, and a general construction of parent Floquet systems.

\section{Detailed proof of Theorem \ref{thm:equivalence}}
In this section, we give details about equivalence and complete classification of symmetric MPUs.
First, we will show how to define an MPU's cohomology class and demonstrate some basic facts thereof mentioned in the main text. Then, we will provide detailed proofs of a Lemma and the equivalence classification Theorem~\ref{thm:equivalence}.

\subsection{Cohomology class of a symmetric MPU}
\label{cohoSMPU}
Let us assume that the tensor $\mathcal{U}$ generating the $L$-site MPU $U^{(L)}$ is simple such that it has a standard form (Eq.~(2) in the main text) unique up to gauge invariance. This can always be achieved by blocking $\mathcal{U}$. Since 
\begin{equation}
U^{(L)} = (\rho_g^{\otimes L}) U^{(L)}(\rho_g^{\otimes L})^\dagger 
\end{equation}
we know that $u$ and $v$ from the standard form of the LHS generate the same MPU as $u(\rho_g^\dag\otimes\rho_g^\dagger)$ and $(\rho_g \otimes \rho_g)v$ from the standard form of the RHS. The gauge freedom of the standard form then immediately results in the existence of unitary $x_g, y_g$ as in Eq.~(3) in the main text, relating the standard forms of LHS and RHS. Applying the the group elements $g$ and $h$ separately or jointly,
\begin{equation}
 v (y_g y_h \otimes x_g x_h)  = (\rho_{gh}\otimes \rho_{gh}) v = v (y_{gh} \otimes x_{gh}),
\end{equation}
such that $y_g \otimes x_g$ forms a linear representation. Hence $y_g$ and $x_g$ form projective representations that have opposite phase terms $\omega(g,h) \in \mathrm{U}(1)$, $x_g x_h = \omega(g,h)x_{gh}, y_g y_h = \omega^{-1}(g,h) y_{gh}$, and thereby opposite cohomology classes.

According to Ref.~\cite{Cirac2017}, an MPU $U$ can be considered as a \emph{normal} MPS. We can thus identify a projective representation $z_g$ on the virtual level of $\mathcal{U}$ \cite{Perez2008}:
\begin{equation}
\begin{tikzpicture}[scale=0.8]
\draw[thick] (-0.5,0) -- (0.5,0) (0,-1) -- (0,1);
\Vertex[shape =circle,size=0.4,x=0,y=0.62,label=$\rho_g$,fontsize=\scriptsize,color=white]{a}
\Vertex[shape =circle,size=0.4,x=0,y=-0.62,label=$\rho_g^{\dag}$,fontsize=\scriptsize,color=white]{b}
\Vertex[shape = rectangle,size=0.4,x=0,label=$\mathcal{U}$,fontsize=\small,color=white]{A}
\Text[x=1.25]{$=$}
\draw[thick] (3,-0.5) -- (3,0.5) (2,0) -- (4,0);
\Vertex[shape =circle,size=0.4,x=2.38,label=$z^\dag_g$,fontsize=\scriptsize,color=white]{c}
\Vertex[shape =circle,size=0.4,x=3.61,label=$z_g$,fontsize=\scriptsize,color=white]{d}
\Vertex[shape = rectangle,size=0.4,x=3,label=$\mathcal{U}$,fontsize=\small,color=white]{B}
\Text[x=4.3,y=-0.25]{$.$}
\end{tikzpicture}
\label{proj MPS}
\end{equation}
To relate the cohomology class of $x_g$ (or $y_g$) to that of $z_g$ arising from the viewpoint of $U$ as an MPS, consider the building blocks of a singular value decomposition of $v$
\begin{equation}
 \begin{tikzpicture}[scale=0.8]
\draw[thick] (8.3,0) -- (8.6,0) (8,0) -- (8,0.3) (8.9,0) -- (8.9,0.3);
\filldraw[thick,fill=white] (8,0) -- (8.3,0) -- (8,-0.3) -- cycle;
\filldraw[thick,fill=lightgray] (8.9,0) -- (8.6,0) -- (8.9,-0.3) -- cycle;
\draw[ultra thick,dotted] (8,-0.3) -- (8,-0.6);
\draw[ultra thick] (8.9,-0.3) -- (8.9,-0.6);
\Text[x=9.25,y=-0.15]{$=$}
\draw[thick] (9.5,-0.4) rectangle (10.5,0.1);
\draw[thick] (9.7,0.1) -- (9.7,0.3) (10.3,0.1) -- (10.3,0.3);
\draw[ultra thick,dotted] (9.7,-0.6) -- (9.7,-0.4);
\draw[ultra thick] (10.3,-0.6) -- (10.3,-0.4);
\Text[x=10,y=-0.15]{$v$}
\Text[x=10.75,y=-0.35]{,}
\end{tikzpicture}
\end{equation}
that transform as follows:
\begin{equation}
\begin{tikzpicture}[scale=0.8]
\draw[thick] (0,0.34) circle (0.22);
\Text[y=0.34,fontsize=\footnotesize]{$\rho_g$}
\filldraw[thick,fill=lightgray] (0,0) -- (-0.3,0) -- (0,-0.3) -- cycle;
\draw[thick] (-0.6,0) -- (-0.3,0) (0,0) -- (0,0.12) (0,0.56) -- (0,0.7);
\draw[ultra thick] (0,-0.6) -- (0,-0.3);
\Text[x=0.58]{$=$}
\draw[thick] (1.34,0) circle (0.22);
\draw[thick] (2,-0.64) circle (0.22);
\Text[x=1.34,fontsize=\footnotesize]{$z^\dag_g$}
\Text[x=2,y=-0.64,fontsize=\footnotesize]{$x_g$}
\draw[thick] (1.58,0) -- (1.7,0) (0.98,0) -- (1.12,0) (2,0) -- (2,0.3);
\draw[ultra thick] (2,-0.3) -- (2,-0.42) (2,-0.86) -- (2,-1);
\filldraw[thick,fill=lightgray] (2,0) -- (1.7,0) -- (2,-0.3) -- cycle;
\Text[x=3]{and}
\draw[thick] (4,0.34) circle (0.22);
\Text[x=4,y=0.34,fontsize=\footnotesize]{$\rho_g$}
\filldraw[thick,fill=white] (4,0) -- (4.3,0) -- (4,-0.3) -- cycle;
\draw[thick] (4.3,0) -- (4.6,0) (4,0) -- (4,0.12) (4,0.56) -- (4,0.7);
\draw[ultra thick,dotted] (4,-0.6) -- (4,-0.3);
\Text[x=5]{$=$}
\draw[thick] (6.14,0) circle (0.22);
\draw[thick] (5.5,-0.64) circle (0.22);
\Text[x=6.14,fontsize=\footnotesize]{$z_g$}
\Text[x=5.5,y=-0.64,fontsize=\footnotesize]{$y_g$}
\draw[thick] (5.8,0) -- (5.92,0) (6.36,0) -- (6.5,0) (5.5,0) -- (5.5,0.3);
\draw[ultra thick,dotted](5.5,-0.3) -- (5.5,-0.42) (5.5,-0.86) -- (5.5,-1);
\filldraw[thick,fill=white] (5.5,0) -- (5.8,0) -- (5.5,-0.3) -- cycle;
\Text[x=7.25,y=0]{.}
\end{tikzpicture}
\label{rhoxyz}
\end{equation}
Similarly to above, $z_g$ and $x_g$ are projective representations of the same cohomology class. It is apparent that $z_g$ does not change when blocking. When composing or tensoring two symmetric MPUs with representations $z'_g$ and $z''_g$, the projective representation on the MPS level becomes the tensor product $z_g = z'_g \otimes z''_g$, summing the cohomology classes. Note that $z_g$ may act on a redundant virtual Hilbert space, where the unique (right) fixed point $\rho\ge0$ determined by $\mathcal{U}$ contracted from $\mathcal{U}'$ and $\mathcal{U}''$ is not full-ranked. Nevertheless, denoting $V^P$ as the maximal truncated virtual Hilbert space within which $\rho>0$ and $V^Q$ as its complement, we can show from $[z_g,\rho]=0$ (due to the uniqueness of the fixed point) that $z_g=z^P_g\oplus z^Q_g$, where $z^P_g$ and $z^Q_g$ act only on $V^P$ and $V^Q$, respectively. Moreover, both $z^P_g$ and $z^Q_g$ form a projective representation with the same factor set, and thus the same cohomology class as $z_g$. Therefore, with the virtual Hilbert space truncated to make $\mathcal{U}$ normal, its cohomology class is still the same as the sum of those of $\mathcal{U}'$ and $\mathcal{U}''$.

Finally, let us remark that there actually exist MPUs with zero index and nontrivial cohomology class. The $\mathbb{Z}_d\times\mathbb{Z}_d$-symmetric MPUs in Sec.~\ref{ZdZd} can serve as such an example.

\subsection{Proof of a Lemma}
The following lemma is crucial for proving Theorem~\ref{thm:equivalence}:
\begin{lemma}
 \label{lemma:equivalence}
 Each symmetric MPU $U$ of zero index and trivial cohomology is equivalent to the identity.
\end{lemma}
\emph{Proof:} We may assume that $U$ is simple, since blocking is allowed in the definition 
of equivalence.
Since $U$ has trivial cohomology, $\alpha_g x_g$ and $\beta_g y_g$ are linear representations for suitable $\alpha_g,\beta_g \in \mathrm{U}(1)$. We now want to get rid of the phases such that $x_g$ and $y_g$ directly are linear representations. To this end, perform the transformation
\begin{equation}
x'_g = \alpha_g x_g\ \text{and}\ y'_g = (\alpha_g\beta_g)^{-1} \beta_g y_g
\end{equation}
which does not alter Eq.~(3) in the main text 
and leaves the linear representation
\begin{equation}
x_g\otimes y_g = x'_g\otimes y'_g = (\alpha_g\beta_g)^{-1} \otimes (\alpha_g x_g) \otimes (\beta_g y_g)
\end{equation}
invariant. On the RHS, the leftmost factor must be a (one-dimensional) linear representation because the other tensor product factors already are linear representations. Therefore, both $x_g'$ and $y_g'$ are linear representations. To unclutter notation, we call them $x_g$ and $y_g$ in the following.

Now we regularise by adding ancillas (identities) to $U$. We are free to choose the regular representation $\rho^\text{reg}_g$ as the action of the symmetry $G$ on the ancillas. This yields the tensor $\mathcal{U}_k\otimes\mathbb{1}_{\dim \rho^\text{reg}}$, and the standard form is affected in the following way:
\begin{equation}
 \begin{tikzpicture}[scale=0.8]
\draw[ultra thick] (0.2,0.5) -- (0.2,0.8) (1.4,0.5) -- (1.4,0.8) (2.6,0.5) -- (2.6,0.8);
\draw[ultra thick,dotted] (0.8,0.5) -- (0.8,0.8) (2,0.5) -- (2,0.8) ;
\draw[thick] (0.2,-0.25) -- (0.2,0) (0.2,1.3) -- (0.2,1.55) (0.8,-0.25) -- (0.8,0) (0.8,1.3) -- (0.8,1.55) (1.4,-0.25) -- (1.4,0) (1.4,1.3) -- (1.4,1.55) (2,-0.25) -- (2,0) (2,1.3) -- (2,1.55) (2.6,-0.25) -- (2.6,0) (2.6,1.3) -- (2.6,1.55)   ;
\draw[thick] (0.3, -0.25) -- (0.3, 1.55) (0.9, -0.25) -- (0.9, 1.55) (1.5, -0.25) -- (1.5, 1.55) (2.1, -0.25) -- (2.1, 1.55) (2.7, -0.25) -- (2.7, 1.55);

\draw[thick] (0,0) rectangle (1,0.5);
\draw[thick] (1.2,0) rectangle (2.2,0.5);
\draw[thick] (2.9, 0.5) -- (2.4, 0.5) -- (2.4, 0) -- (2.9, 0);
\draw[thick] (-0.1,0.8) -- (0.4,0.8) -- (0.4,1.3) -- (-0.1,1.3);
\draw[thick] (0.6,0.8) rectangle (1.6,1.3);
\draw[thick] (1.8,0.8) rectangle (2.8,1.3);
\Text[x=-0.7,y=0.62]{$=$}
\Text[x=0.5,y=0.25]{$u$}
\Text[x=1.7,y=0.25]{$u$}
\Text[x=1.1,y=1.05]{$v$}
\Text[x=2.3,y=1.05]{$v$}
\Text[x=3.2,y=-0.25]{$.$}

\draw[thick] (-2, 0.4) rectangle (-1.5, 0.9);
\draw[thick] (-2.8, 0.4) rectangle (-2.3, 0.9);
\draw[thick] (-3.6, 0.4) rectangle (-3.1, 0.9);
\draw[thick] (-4.4, 0.4) rectangle (-3.9, 0.9);
\draw[thick] (-5.2, 0.4) rectangle (-4.7, 0.9);

\Text[x=-1.7, y=0.65,fontsize=\footnotesize]{$\mathcal{U}_k$}
\Text[x=-2.5, y=0.65,fontsize=\footnotesize]{$\mathcal{U}_k$}
\Text[x=-3.3, y=0.65,fontsize=\footnotesize]{$\mathcal{U}_k$}
\Text[x=-4.1, y=0.65,fontsize=\footnotesize]{$\mathcal{U}_k$}
\Text[x=-4.9, y=0.65,fontsize=\footnotesize]{$\mathcal{U}_k$}

\draw[thick] (-5.4, 0.65) -- (-5.2, 0.65) (-4.7, 0.65) -- (-4.4, 0.65) (-3.9, 0.65) -- (-3.6, 0.65) (-2.8, 0.65) -- (-3.1, 0.65) (-2.3, 0.65) -- (-2, 0.65) (-1.5, 0.65) -- (-1.3, 0.65);
\draw[thick] (-1.75, 0.9) -- (-1.75, 1.4) (-2.55, 0.9) -- (-2.55, 1.4) (-3.35, 0.9) -- (-3.35, 1.4) (-4.15, 0.9) -- (-4.15, 1.4) (-4.95, 0.9) -- (-4.95, 1.4);
\draw[thick] (-1.75, 0.4) -- (-1.75, -0.1) (-2.55, 0.4) -- (-2.55, -0.1) (-3.35, 0.4) -- (-3.35, -0.1) (-4.15, 0.4) -- (-4.15, -0.1) (-4.95, 0.4) -- (-4.95, -0.1);
\draw[thick] (-1.65, -0.1) -- (-1.65, 1.4) (-2.45, -0.1) -- (-2.45, 1.4) (-3.25, -0.1) -- (-3.25, 1.4) (-4.05, -0.1) -- (-4.05, 1.4) (-4.85, -0.1) -- (-4.85, 1.4) ;
\end{tikzpicture}
\end{equation}
Blocking the original physical legs together with the ancillas, the blocked local symmetry is $\tilde{\rho}_g = \rho_g \otimes \rho^\text{reg}_g$. It acts on the unitaries of the blocked standard form $\tilde{v} = v\otimes\mathbb{1}_{(\dim \rho^\text{reg})^2}$ as
\begin{equation}
( \tilde{\rho}_g \otimes \tilde{\rho}_g ) \tilde v = \tilde v \left( (y_g\otimes \rho^\text{reg}_g) \otimes (x_g\otimes \rho^\text{reg}_g)\right) \equiv \tilde v (\tilde{y}_g \otimes \tilde{x}_g).
\end{equation}

In fact, $\tilde{\rho}_g, \tilde{x}_g,$ and $\tilde{y}_g$ are equivalent linear representations. To see this, let us demonstrate that
$\tau\otimes \rho^\text{reg}$ is equivalent to a $\dim \tau$-fold copy of $\rho^\text{reg}$ for any linear representation $\tau$. Within the character theory of finite groups \cite{Serre1977}, we can calculate
\begin{equation}
\Tr( \tau_g\otimes \rho^\text{reg}_g ) = \Tr(\tau_g) (\dim \rho^\text{reg}_g) \delta_{ge} = (\dim\tau_e) \Tr(\rho^\text{reg}_g),
\end{equation}
with the identity $e\in G$. Since the MPU is index zero, $\rho_g, x_g,$ and $y_g$ have the same dimensions and $\tilde\rho_g,\tilde x_g,$ and $\tilde y_g$ are therefore equivalent representations.

The equivalence of representations means that there exist two unitaries $X$ and $Y$ such that $\tilde x_g=X\tilde\rho_gX^\dag$ and $\tilde y_g=Y\tilde\rho_gY^\dag$, implying
\begin{equation}
[\tilde\rho_g\otimes\tilde\rho_g,\tilde v']=[\tilde\rho_g\otimes\tilde\rho_g,\tilde u']=0,\ \forall g\in G,
\label{uvsym}
\end{equation}
where $\tilde v'=\tilde v(Y\otimes X)$ and $\tilde u'=(X^\dag\otimes Y^\dag)\tilde u$ are related to $\tilde v$ and $\tilde u$ by a gauge transformation, so they generate the same MPU. Thanks to Eq.~(\ref{uvsym}), we can find two Hermitian operators $h_u$ and $h_v$ such that $\tilde u'=e^{-ih_u}$ and $\tilde v'=e^{-ih_v}$ and
\begin{equation}
[\tilde\rho_g\otimes\tilde\rho_g,h_u]=[\tilde\rho_g\otimes\tilde\rho_g,h_v]=0,\ \forall g\in G.
\end{equation}
Defining $\tilde u'(\lambda)\equiv e^{-i\lambda h_u}$ and $\tilde v'(\lambda)\equiv e^{-i\lambda h_v}$, we immediately know that the generated MPU $U(\lambda)$ gives a continuous path which respects all the symmetries and interpolates $U$ and the (global) identity $\mathbb{I}\equiv\mathbb{1}^{\otimes L}$. \hfill$\Box$

\subsection{Proof of Theorem~\ref{thm:equivalence}}
As mentioned in the main text, it was already proven in \cite{Cirac2017,Schuch2011} that identical indices and identical cohomology classes are necessary for two MPUs to be equivalent. For the reverse direction, we will now construct an explicit path connecting the MPUs.

If $U_0$ and $U_1$ have different representations $\rho_0$ and $\rho_1$ of the symmetry, we may add  ancillas (identities) with $\rho_1$ and $\rho_0$ to $U_0$ and $U_1$, respectively. The composition $U_1^\dag U_0$ is then symmetric with the representation $\rho_0\otimes\rho_1$.
Since $U_0$ and $U_1$ have the same cohomology class and index, their additivity under composition leads to $U_1^\dag U_0$ having trivial cohomology class and zero index. Therefore, we can apply Lemma~\ref{lemma:equivalence} to $U^\dag_1U_0$ to find a one-parameter class of symmetric MPUs $V_\lambda$ such that $V_0=\mathbb{I}$ and $U_1^\dag U_0=V_1$.  This gives a continuous path $U_{\lambda}\equiv U_0V^\dag_\lambda$ which is symmetric under $G$ for $\forall \lambda\in[0,1]$ and connects $U_0$ and $U_1$.

\section{$\mathbb{Z}_d\times\mathbb{Z}_d$ SPT MPUs}
\label{ZdZd}
In this section, we represent the $\mathbb{Z}_n\times\mathbb{Z}_n$ SPT edge unitary in Ref.~\cite{Potter2017} into the MPU form and show that its cohomology class is nontrivial. We also 
discuss the general feature in the time evolution of the entanglement spectra 
and provide both analytical and numerical demonstrations.
Since the $\mathbb{Z}_n\times\mathbb{Z}_n$ SPT MPUs act on qudit systems with the local Hilbert-space dimension being $d=n$, we would like to use $\mathbb{Z}_d\times\mathbb{Z}_d$ symmetry throughout this section.

\subsection{Building block and the standard form}
Specifying the representation of $\mathbb{Z}_d\times\mathbb{Z}_d$ symmetry group as $\rho_{(m,n)\in\mathbb{Z}_d\times\mathbb{Z}_d}=Z^m\otimes Z^n$, where $Z_{ij}\equiv\delta_{ij}\omega^j_d$, we can write down the generator of nontrivial SPT MPUs (cohomology group) as \cite{Potter2017}
\begin{equation}
U_{\rm SPT}=e^{-i\frac{2\pi}{d}\sum_j(-)^jN_{j,X}N_{j+1,X}},
\label{USPT}
\end{equation}
where $N_X\equiv\sum^{d-1}_{j=1}\frac{X^j-\mathbb{1}}{\omega^{-j}_d-1}$ with $X_{ij}=\delta_{i+1,j}$ \footnote{The variables in the Kronecker delta are elements in $\mathbb{Z}_d$, thus $\delta_{d,0}=1$.}, which satisfies $XZ=\omega_dZX$. With the delta tensor and the generalized Hadamard matrix defined diagramatically as
\begin{equation}
\begin{tikzpicture}
\draw[thick] (0,-0.5) -- (0,0.5) (0,0) -- (0.5,0);
\draw[thick,fill=white] (0,0) circle (0.05);
\Text[y=0.7,fontsize=\footnotesize]{$i$}
\Text[x=0.7,fontsize=\footnotesize]{$j$}
\Text[y=-0.7,fontsize=\footnotesize]{$k$}
\Text[x=1.6]{$\equiv\delta_{ij}\delta_{jk},$}
\draw[thick] (3.8,0) -- (4.8,0);
\draw[thick,fill=white] (4.1,-0.2) rectangle (4.5,0.2);
\Text[x=3.6,fontsize=\footnotesize]{$j$}
\Text[x=4.3,fontsize=\footnotesize]{$H$}
\Text[x=5,fontsize=\footnotesize]{$k$}
\Text[x=5.75]{$\equiv
\omega^{jk}_d$}
\end{tikzpicture}
\end{equation}
under the eigenbasis of $X_j$'s, the building block of the $\mathbb{Z}_d\times\mathbb{Z}_d$ SPT MPU (\ref{USPT}) can be represented as
\begin{equation}
\begin{tikzpicture}
\draw[thick] (0,-0.5) -- (0,0.5) (-0.5,0) -- (0.5,0);
\Vertex[shape = rectangle,size=0.5,x=0,label=$\mathcal{U}$,fontsize=\normalsize,color=white]{A}
\Text[x=0.8]{$=$}
\draw[thick,dashed] (1.4,-0.55) rectangle (3.5,0.55);
\draw[thick] (1.2,0.25) -- (1.7,0.25) (3.7,0.25) -- (2.7,0.25) (1.7,-0.25) -- (2.7,-0.25);
\draw[thick] (1.7,0.75) -- (1.7,-0.75) (2.7,0.75) -- (2.7,-0.75);
\draw[thick,fill=white] (1.7,0.25) circle (0.05);
\draw[thick,fill=white] (1.7,-0.25) circle (0.05);
\draw[thick,fill=white] (2.7,0.25) circle (0.05);
\draw[thick,fill=white] (2.7,-0.25) circle (0.05);
\draw[thick,fill=white] (2,-0.45) rectangle (2.4,-0.05);
\draw[thick,fill=white] (3,0.05) rectangle (3.4,0.45);
\Text[x=2.2,y=-0.25,fontsize=\footnotesize]{$H$}
\Text[x=3.22,y=0.27,fontsize=\footnotesize]{$H^\dag$}
\Text[x=4,y=-0.25]{.}
\end{tikzpicture}
\label{Ubb}
\end{equation}
Note that the delta tensor and the generalized Hadamard matrix satisfy 
\begin{equation*}
\begin{tikzpicture}
\draw[thick] (0,-0.5) -- (0,0.9) (0,0) -- (0.5,0);
\draw[thick,fill=white] (0,0.5) circle (0.25);
\draw[thick,fill=white] (0,0) circle (0.05);
\Text[y=0.5,fontsize=\footnotesize]{$Z^n$}
\Text[x=0.8]{$=$}
\draw[thick] (1.2,-0.9) -- (1.2,0.5) (1.2,0) -- (2.1,0);
\draw[thick,fill=white] (1.2,0) circle (0.05);
\draw[thick,fill=white] (1.7,0) circle (0.25);
\draw[thick,fill=white] (1.2,-0.5) circle (0.25);
\Text[x=1.7,fontsize=\footnotesize]{$Z^n$}
\Text[x=1.2,y=-0.5,fontsize=\footnotesize]{$Z^n$}
\Text[x=2.25,y=-0.2]{,}
\draw[thick] (3.5,-0.5) -- (3.5,0.9) (3.5,0) -- (3,0);
\draw[thick,fill=white] (3.5,0.5) circle (0.25);
\draw[thick,fill=white] (3.5,0) circle (0.05);
\Text[x=3.5,y=0.5,fontsize=\footnotesize]{$Z^n$}
\Text[x=3.8]{$=$}
\draw[thick] (5.1,-0.9) -- (5.1,0.5) (5.1,0) -- (4.1,0);
\draw[thick,fill=white] (5.1,0) circle (0.05);
\draw[thick,fill=white] (4.6,0) ellipse (0.3 and 0.2);
\draw[thick,fill=white] (5.1,-0.5) circle (0.25);
\Text[x=4.65,fontsize=\tiny]{$Z^{-n}$}
\Text[x=5.1,y=-0.5,fontsize=\footnotesize]{$Z^n$}
\Text[x=5.5,y=-0.2]{,}
\end{tikzpicture}
\end{equation*}
\begin{equation}
\begin{tikzpicture}
\draw[thick] (6.5,-0.5) -- (6.5,0.9) (6.5,0) -- (7,0);
\draw[thick,fill=white] (6.5,0.5) circle (0.25);
\draw[thick,fill=white] (6.5,0) circle (0.05);
\Text[x=6.52,y=0.5,fontsize=\footnotesize]{$X^n$}
\Text[x=7.3]{$=$}
\draw[thick] (7.7,-0.5) -- (7.7,0.5) (7.7,0) -- (8.6,0);
\draw[thick,fill=white] (7.7,0) circle (0.05);
\draw[thick,fill=white] (8.2,0) circle (0.25);
\Text[x=8.22,fontsize=\footnotesize]{$X^n$}
\Text[x=8.9]{$=$}
\draw[thick] (9.3,-0.9) -- (9.3,0.5) (9.3,0) -- (9.8,0);
\draw[thick,fill=white] (9.3,0) circle (0.05);
\draw[thick,fill=white] (9.3,-0.5) circle (0.25);
\Text[x=9.32,y=-0.5,fontsize=\footnotesize]{$X^n$}
\Text[x=10,y=-0.2]{,}
\end{tikzpicture}
\label{ZXH}
\end{equation}
\begin{equation*}
\begin{tikzpicture}
\draw[thick] (-0.6,0) -- (1,0);
\draw[thick,fill=white] (0.3,-0.2) rectangle (0.7,0.2);
\draw[thick,fill=white] (-0.15,0) circle (0.25);
\Text[x=0.5,fontsize=\footnotesize]{$H$}
\Text[x=-0.15,fontsize=\footnotesize]{$Z^n$}
\Text[x=1.3]{$=$}
\draw[thick] (1.6,0) -- (3.2,0);
\draw[thick,fill=white] (1.9,-0.2) rectangle (2.3,0.2);
\draw[thick,fill=white] (2.75,0) ellipse (0.3 and 0.2);
\Text[x=2.1,fontsize=\footnotesize]{$H$}
\Text[x=2.78,fontsize=\tiny]{$X^{-n}$}
\Text[x=3.4,y=-0.2]{,}
\end{tikzpicture}
\end{equation*}
\begin{equation*}
\begin{tikzpicture}
\draw[thick] (4.4,0) -- (6,0);
\draw[thick,fill=white] (5.3,-0.2) rectangle (5.7,0.2);
\draw[thick,fill=white] (4.85,0) circle (0.25);
\Text[x=5.5,fontsize=\footnotesize]{$H$}
\Text[x=4.87,fontsize=\footnotesize]{$X^n$}
\Text[x=6.3]{$=$}
\draw[thick] (6.6,0) -- (8.2,0);
\draw[thick,fill=white] (6.9,-0.2) rectangle (7.3,0.2);
\draw[thick,fill=white] (7.75,0) circle (0.25);
\Text[x=7.1,fontsize=\footnotesize]{$H$}
\Text[x=7.75,fontsize=\footnotesize]{$Z^n$}
\Text[x=8.4,y=-0.2]{.}
\end{tikzpicture}
\end{equation*}
Using these relations (\ref{ZXH}), for $\forall \rho_{(m,n)}=Z^m\otimes Z^n$, we have
\begin{widetext}
\begin{equation}
\begin{tikzpicture}
\draw[thick] (1.2,0.25) -- (1.7,0.25) (3.7,0.25) -- (2.7,0.25) (1.7,-0.25) -- (2.7,-0.25);
\draw[thick] (1.7,1.1) -- (1.7,-0.75) (2.7,1.1) -- (2.7,-0.75);
\draw[thick,fill=white] (1.7,0.25) circle (0.05);
\draw[thick,fill=white] (1.7,-0.25) circle (0.05);
\draw[thick,fill=white] (2.7,0.25) circle (0.05);
\draw[thick,fill=white] (2.7,-0.25) circle (0.05);
\draw[thick,fill=white] (2,-0.45) rectangle (2.4,-0.05);
\draw[thick,fill=white] (3,0.05) rectangle (3.4,0.45);
\draw[thick,fill=white] (1.7,0.7) circle (0.25);
\draw[thick,fill=white] (2.7,0.7) circle (0.25);
\Text[x=2.2,y=-0.25,fontsize=\footnotesize]{$H$}
\Text[x=3.22,y=0.27,fontsize=\footnotesize]{$H^\dag$}
\Text[x=1.72,y=0.7,fontsize=\footnotesize]{$Z^m$}
\Text[x=2.7,y=0.7,fontsize=\footnotesize]{$Z^n$}
\Text[x=4]{$=$}
\draw[thick] (4.5,0.25) -- (5.5,0.25) (7.8,0.25) -- (9.4,0.25) (5.5,-0.25) -- (7.8,-0.25);
\draw[thick] (5.5,0.75) -- (5.5,-1.2) (7.8,0.75) -- (7.8,-1.2);
\draw[thick,fill=white] (5,0.25) ellipse (0.3 and 0.2);
\draw[thick,fill=white] (5.5,0.25) circle (0.05);
\draw[thick,fill=white] (5.5,-0.25) circle (0.05);
\draw[thick,fill=white] (6,-0.25) circle (0.25);
\draw[thick,fill=white] (6.45,-0.45) rectangle (6.85,-0.05);
\draw[thick,fill=white] (7.3,-0.25) ellipse (0.3 and 0.2);
\draw[thick,fill=white] (7.8,0.25) circle (0.05);
\draw[thick,fill=white] (7.8,-0.25) circle (0.05);
\draw[thick,fill=white] (8.3,0.25) circle (0.25);
\draw[thick,fill=white] (8.75,0.05) rectangle (9.15,0.45);
\draw[thick,fill=white] (5.5,-0.75) circle (0.25);
\draw[thick,fill=white] (7.8,-0.75) circle (0.25);
\Text[x=6.65,y=-0.25,fontsize=\footnotesize]{$H$}
\Text[x=8.97,y=0.27,fontsize=\footnotesize]{$H^\dag$}
\Text[x=5.02,y=0.25,fontsize=\tiny]{$Z^{-m}$}
\Text[x=6.02,y=-0.25,fontsize=\footnotesize]{$Z^m$}
\Text[x=5.52,y=-0.75,fontsize=\footnotesize]{$Z^m$}
\Text[x=7.8,y=-0.75,fontsize=\footnotesize]{$Z^n$}
\Text[x=8.3,y=0.25,fontsize=\footnotesize]{$Z^n$}
\Text[x=7.32,y=-0.25,fontsize=\tiny]{$Z^{-n}$}
\Text[x=9.65]{$=$}
\Text[x=10.2,fontsize=\footnotesize]{$\omega^{mn}_d$}
\draw[thick] (10.5,0.25) -- (11.5,0.25) (13.8,0.25) -- (15.4,0.25) (11.5,-0.25) -- (13.8,-0.25);
\draw[thick] (11.5,0.75) -- (11.5,-1.2) (13.8,0.75) -- (13.8,-1.2);
\draw[thick,fill=white] (11,0.25) ellipse (0.3 and 0.2);
\draw[thick,fill=white] (11.5,0.25) circle (0.05);
\draw[thick,fill=white] (11.5,-0.25) circle (0.05);
\draw[thick,fill=white] (12,-0.25) ellipse (0.3 and 0.2);
\draw[thick,fill=white] (12.45,-0.45) rectangle (12.85,-0.05);
\draw[thick,fill=white] (13.3,-0.25) ellipse (0.3 and 0.2);
\draw[thick,fill=white] (13.8,0.25) circle (0.05);
\draw[thick,fill=white] (13.8,-0.25) circle (0.05);
\draw[thick,fill=white] (14.95,0.25) circle (0.25);
\draw[thick,fill=white] (14.1,0.05) rectangle (14.5,0.45);
\draw[thick,fill=white] (11.5,-0.75) circle (0.25);
\draw[thick,fill=white] (13.8,-0.75) circle (0.25);
\Text[x=12.65,y=-0.25,fontsize=\footnotesize]{$H$}
\Text[x=14.97,y=0.27,fontsize=\footnotesize]{$X^n$}
\Text[x=11.02,y=0.25,fontsize=\tiny]{$Z^{-m}$}
\Text[x=12.02,y=-0.25,fontsize=\tiny]{$X^{-n}$}
\Text[x=11.52,y=-0.75,fontsize=\footnotesize]{$Z^m$}
\Text[x=13.8,y=-0.75,fontsize=\footnotesize]{$Z^n$}
\Text[x=14.32,y=0.27,fontsize=\footnotesize]{$H^\dag$}
\Text[x=13.32,y=-0.25,fontsize=\tiny]{$X^{-m}$}
\end{tikzpicture}
\end{equation}
\begin{align*}
\begin{tikzpicture}
\Text[x=-1]{$=$}
\Text[x=-0.45,fontsize=\footnotesize]{$\omega^{mn}_d$}
\draw[thick] (-0.1,0.25) -- (1.7,0.25) (4.95,0.25) -- (2.7,0.25) (1.7,-0.25) -- (2.7,-0.25);
\draw[thick] (1.7,-1.1) -- (1.7,0.75) (2.7,-1.1) -- (2.7,0.75);
\draw[thick,fill=white] (0.8,0.25) ellipse (0.65 and 0.25);
\draw[thick,fill=white] (1.7,0.25) circle (0.05);
\draw[thick,fill=white] (1.7,-0.25) circle (0.05);
\draw[thick,fill=white] (2.7,0.25) circle (0.05);
\draw[thick,fill=white] (2.7,-0.25) circle (0.05);
\draw[thick,fill=white] (2,-0.45) rectangle (2.4,-0.05);
\draw[thick,fill=white] (3.2,0.25) ellipse (0.3 and 0.2);
\draw[thick,fill=white] (4.5,0.25) circle (0.25);
\draw[thick,fill=white] (3.65,0.05) rectangle (4.05,0.45);
\draw[thick,fill=white] (1.7,-0.7) circle (0.25);
\draw[thick,fill=white] (2.7,-0.7) circle (0.25);
\Text[x=2.2,y=-0.25,fontsize=\footnotesize]{$H$}
\Text[x=3.22,y=0.25,fontsize=\tiny]{$X^{-m}$}
\Text[x=3.87,y=0.27,fontsize=\footnotesize]{$H^\dag$}
\Text[x=1.72,y=-0.7,fontsize=\footnotesize]{$Z^m$}
\Text[x=2.7,y=-0.7,fontsize=\footnotesize]{$Z^n$}
\Text[x=0.8,y=0.27,fontsize=\tiny]{$Z^{-m}X^{-n}$}
\Text[x=4.52,y=0.25,fontsize=\footnotesize]{$X^n$}
\Text[x=5.25]{$=$}
\draw[thick] (5.6,0.25) -- (7.4,0.25) (10.6,0.25) -- (8.4,0.25) (7.4,-0.25) -- (8.4,-0.25);
\draw[thick] (7.4,-1.1) -- (7.4,0.75) (8.4,-1.1) -- (8.4,0.75);
\draw[thick,fill=white] (6.5,0.25) ellipse (0.65 and 0.25);
\draw[thick,fill=white] (7.4,0.25) circle (0.05);
\draw[thick,fill=white] (7.4,-0.25) circle (0.05);
\draw[thick,fill=white] (8.4,0.25) circle (0.05);
\draw[thick,fill=white] (8.4,-0.25) circle (0.05);
\draw[thick,fill=white] (7.7,-0.45) rectangle (8.1,-0.05);
\draw[thick,fill=white] (9.8,0.25) ellipse (0.5 and 0.2);
\draw[thick,fill=white] (8.7,0.05) rectangle (9.1,0.45);
\draw[thick,fill=white] (7.4,-0.7) circle (0.25);
\draw[thick,fill=white] (8.4,-0.7) circle (0.25);
\Text[x=7.9,y=-0.25,fontsize=\footnotesize]{$H$}
\Text[x=8.92,y=0.27,fontsize=\footnotesize]{$H^\dag$}
\Text[x=7.42,y=-0.7,fontsize=\footnotesize]{$Z^m$}
\Text[x=8.4,y=-0.7,fontsize=\footnotesize]{$Z^n$}
\Text[x=6.5,y=0.27,fontsize=\tiny]{$X^{-n}Z^{-m}$}
\Text[x=9.8,y=0.27,fontsize=\tiny]{$Z^mX^n$}
\Text[x=10.75]{$.$}
\end{tikzpicture}
\end{align*}
\end{widetext}
Therefore, the projective representation on the virtual level reads
\begin{equation}
z_{(m,n)}=Z^mX^n.
\end{equation}

According to the building block (\ref{Ubb}), we can write down the standard form as follows:
\begin{equation}
\begin{tikzpicture}
\draw[thick] (-2,0) rectangle (-1,0.5);
\draw[thick] (-1.8,-0.25) -- (-1.8,0) (-1.2,-0.25) -- (-1.2,0) (-1.8,0.5) -- (-1.8,0.75) (-1.2,0.5) -- (-1.2,0.75);
\Text[x=-1.5,y=0.25]{$u$}
\Text[x=-0.55,y=0.25]{$=$}
\draw[thick] (-0.1,-0.05) -- (0.5,-0.05) (0.5,0.55) -- (1.3,0.55) (1.3,-0.05) -- (1.9,-0.05);
\draw[thick,fill=white] (0,-0.25) rectangle (0.4,0.15);
\draw[thick,fill=white] (1.4,-0.25) rectangle (1.8,0.15);
\draw[thick,fill=white] (0.7,0.35) rectangle (1.1,0.75);
\draw[thick] (-0.1,1) -- (-0.1,-0.4) (0.5,1) -- (0.5,-0.4); 
\draw[thick] (1.3,1) -- (1.3,-0.4) (1.9,-0.4) -- (1.9,1);
\draw[thick,fill=white] (-0.1,-0.05) circle (0.05);
\draw[thick,fill=white] (0.5,-0.05) circle (0.05);
\draw[thick,fill=white] (0.5,0.55) circle (0.05);
\draw[thick,fill=white] (1.3,-0.05) circle (0.05);
\draw[thick,fill=white] (1.3,0.55) circle (0.05);
\draw[thick,fill=white] (1.9,-0.05) circle (0.05);
\Text[x=0.2,y=-0.05,fontsize=\footnotesize]{$H$}
\Text[x=1.6,y=-0.05,fontsize=\footnotesize]{$H$}
\Text[x=0.92,y=0.55,fontsize=\footnotesize]{$H^\dag$}
\Text[x=2.15,y=0]{,}
\end{tikzpicture}
\label{uv}
\end{equation}
\begin{equation*}
\begin{tikzpicture}
\draw[thick] (3,0) rectangle (4,0.5);
\draw[thick] (3.8,0.5) -- (3.8,0.75) (3.2,0.5) -- (3.2,0.75) (3.8,-0.25) -- (3.8,0) (3.2,-0.25) -- (3.2,0);
\Text[x=3.5,y=0.25]{$v$}
\Text[x=4.45,y=0.25]{$=$}
\draw[thick] (5.5,0.25) -- (6.3,0.25);
\draw[thick,fill=white] (5.7,0.05) rectangle (6.1,0.45);
\draw[thick] (4.9,-0.25) -- (4.9,0.75) (5.5,-0.25) -- (5.5,0.75) (6.3,-0.25) -- (6.3,0.75) (6.9,-0.25) -- (6.9,0.75);
\draw[thick,fill=white] (5.5,0.25) circle (0.05);
\draw[thick,fill=white] (6.3,0.25) circle (0.05);
\Text[x=5.92,y=0.25,fontsize=\footnotesize]{$H^\dag$}
\Text[x=7.15,y=0]{.}
\end{tikzpicture}
\end{equation*}
Using Eq.~(\ref{ZXH}), we can determine the projective representations $x_g$ and $y_g$ on the virtual level of the standard form as
\begin{equation}
\begin{split}
x_{(m,n)}&=\omega^{-mn}_d X^nZ^m\otimes Z^n,\\
y_{(m,n)}&=
Z^m\otimes X^mZ^n.
\end{split}
\end{equation}
It is easy to check $\omega_x((m,n),(m',n'))=\omega^{m'n}_d=\omega_z((m,n),(m',n'))$, which is consistent with the general relation that $x_g$ and $z_g$ belong to the same cohomology class.

\begin{figure}[!b]
\begin{center}
       \begin{tikzpicture}[thick,scale=0.7, every node/.style={scale=0.7}]
\fill[gray!20] (-0.8,0.8) rectangle (0,1.6) (0.8,0.8) rectangle (1.6,1.6) (0,0) rectangle (0.8,0.8) (-1.6,0) rectangle (-0.8,0.8) (-0.8,-0.8) rectangle (0,0) (0.8,-0.8) rectangle (1.6,0) (-1.6,-1.6) rectangle (-0.8,-0.8) (0,-1.6) rectangle (0.8,-0.8);
\draw[ultra thick,red!50!black] (-0.75,-1.55) -- (-0.05,-1.55) (0.85,-1.55) -- (1.55,-1.55);
\draw[ultra thick,blue!50!black] (-1.55,-1.55) -- (-0.85,-1.55) (0.05,-1.55) -- (0.75,-1.55);
\draw[ultra thick,red!50!black] (-1.55,-0.75) -- (-0.85,-0.75) (-0.75,-0.85) -- (-0.05,-0.85) (0.05,-0.75) -- (0.75,-0.75) (0.85,-0.85) -- (1.55,-0.85);
\draw[ultra thick,blue!50!black] (-1.55,-0.85) -- (-0.85,-0.85) (-0.75,-0.75) -- (-0.05,-0.75) (0.05,-0.85) -- (0.75,-0.85) (0.85,-0.75) -- (1.55,-0.75);
\draw[ultra thick,red!50!black] (-1.55,-0.05) -- (-0.85,-0.05) (-0.75,0.05) -- (-0.05,0.05) (0.05,-0.05) -- (0.75,-0.05) (0.85,0.05) -- (1.55,0.05);
\draw[ultra thick,blue!50!black] (-1.55,0.05) -- (-0.85,0.05) (-0.75,-0.05) -- (-0.05,-0.05) (0.05,0.05) -- (0.75,0.05) (0.85,-0.05) -- (1.55,-0.05);
\draw[ultra thick,red!50!black] (-1.55,0.85) -- (-0.85,0.85) (-0.75,0.75) -- (-0.05,0.75) (0.05,0.85) -- (0.75,0.85) (0.85,0.75) -- (1.55,0.75);
\draw[ultra thick,blue!50!black] (-1.55,0.75) -- (-0.85,0.75) (-0.75,0.85) -- (-0.05,0.85) (0.05,0.75) -- (0.75,0.75) (0.85,0.85) -- (1.55,0.85);
\draw[ultra thick,blue!50!black] (-0.75,1.55) -- (-0.05,1.55) (0.85,1.55) -- (1.55,1.55);
\draw[ultra thick,red!50!black] (-1.55,1.55) -- (-0.85,1.55) (0.05,1.55) -- (0.75,1.55);
\draw[ultra thick,blue!50!black] (-1.55,-0.75) -- (-1.55,-0.05) (-1.55,0.85) -- (-1.55,1.55);
\draw[ultra thick,red!50!black] (-1.55,-1.55) -- (-1.55,-0.85) (-1.55,0.05) -- (-1.55,0.75);
\draw[ultra thick,blue!50!black] (-0.75,-1.55) -- (-0.75,-0.85) (-0.85,-0.75) -- (-0.85,-0.05) (-0.75,0.05) -- (-0.75,0.75) (-0.85,0.85) -- (-0.85,1.55);
\draw[ultra thick,red!50!black] (-0.85,-1.55) -- (-0.85,-0.85) (-0.75,-0.75) -- (-0.75,-0.05) (-0.85,0.05) -- (-0.85,0.75) (-0.75,0.85) -- (-0.75,1.55);
\draw[ultra thick,blue!50!black] (-0.05,-1.55) -- (-0.05,-0.85) (0.05,-0.75) -- (0.05,-0.05) (-0.05,0.05) -- (-0.05,0.75) (0.05,0.85) -- (0.05,1.55);
\draw[ultra thick,red!50!black] (0.05,-1.55) -- (0.05,-0.85) (-0.05,-0.75) -- (-0.05,-0.05) (0.05,0.05) -- (0.05,0.75) (-0.05,0.85) -- (-0.05,1.55);
\draw[ultra thick,blue!50!black] (0.85,-1.55) -- (0.85,-0.85) (0.75,-0.75) -- (0.75,-0.05) (0.85,0.05) -- (0.85,0.75) (0.75,0.85) -- (0.75,1.55);
\draw[ultra thick,red!50!black] (0.75,-1.55) -- (0.75,-0.85) (0.85,-0.75) -- (0.85,-0.05) (0.75,0.05) -- (0.75,0.75) (0.85,0.85) -- (0.85,1.55);
\draw[ultra thick,red!50!black] (1.55,-0.75) -- (1.55,-0.05) (1.55,0.85) -- (1.55,1.55);
\draw[ultra thick,blue!50!black] (1.55,-1.55) -- (1.55,-0.85) (1.55,0.05) -- (1.55,0.75);
\draw[thick,fill=white] (-1.6,0) circle (0.12) (0,0) circle (0.12) (1.6,0) circle (0.12) (-1.6,1.6) circle (0.12) (0,1.6) circle (0.12) (1.6,1.6) circle (0.12) (-1.6,-1.6) circle (0.12) (0,-1.6) circle (0.12) (1.6,-1.6) circle (0.12);
\draw[thick,fill=white] (-0.8,0.8) circle (0.12) (0.8,0.8) circle (0.12) (-0.8,-0.8) circle (0.12) (0.8,-0.8) circle (0.12);
\draw[thick,fill=gray] (-1.6,-0.8) circle (0.12) (0,-0.8) circle (0.12) (1.6,-0.8) circle (0.12) (-1.6,0.8) circle (0.12) (0,0.8) circle (0.12) (1.6,0.8) circle (0.12);
\draw[thick,fill=gray] (-0.8,-1.6) circle (0.12) (0.8,-1.6) circle (0.12) (-0.8,0) circle (0.12) (0.8,0) circle (0.12) (-0.8,1.6) circle (0.12) (0.8,1.6) circle (0.12);
\draw[thick,fill=violet!50] (2.5,0.1) -- (3.2,0.1) -- (3.2,0.2) -- (3.5,0) -- (3.2,-0.2) -- (3.2,-0.1) -- (2.5,-0.1) -- cycle;
\fill[gray!20] (5.2,0.8) rectangle (6,1.6) (6.8,0.8) rectangle (7.6,1.6) (6,0) rectangle (6.8,0.8) (4.4,0) rectangle (5.2,0.8) (5.2,-0.8) rectangle (6,0) (6.8,-0.8) rectangle (7.6,0) (4.4,-1.6) rectangle (5.2,-0.8) (6,-1.6) rectangle (6.8,-0.8);
\draw[ultra thick,blue!50!black] (5.2,1.6) -- (6,1.6) (6.8,1.6) -- (7.6,1.6);
\draw[ultra thick,red!50!black] (5.2,-1.6) -- (6,-1.6) (6.8,-1.6) -- (7.6,-1.6);
\draw[ultra thick,blue!50!black] (4.4,-1.6) -- (5.2,-1.6) (6,-1.6) -- (6.8,-1.6);
\draw[ultra thick,red!50!black] (4.4,1.6) -- (5.2,1.6) (6,1.6) -- (6.8,1.6);
\draw[ultra thick,blue!50!black] (4.4,-0.8) -- (4.4,0) (4.4,0.8) -- (4.4,1.6);
\draw[ultra thick,red!50!black] (7.6,-0.8) -- (7.6,0) (7.6,0.8) -- (7.6,1.6);
\draw[ultra thick,blue!50!black] (7.6,-1.6) -- (7.6,-0.8) (7.6,0) -- (7.6,0.8);
\draw[ultra thick,red!50!black] (4.4,-1.6) -- (4.4,-0.8) (4.4,0) -- (4.4,0.8);
\draw[thick,fill=white] (4.4,0) circle (0.12) (6,0) circle (0.12) (7.6,0) circle (0.12) (4.4,1.6) circle (0.12) (6,1.6) circle (0.12) (7.6,1.6) circle (0.12) (4.4,-1.6) circle (0.12) (6,-1.6) circle (0.12) (7.6,-1.6) circle (0.12);
\draw[thick,fill=white] (5.2,0.8) circle (0.12) (6.8,0.8) circle (0.12) (5.2,-0.8) circle (0.12) (6.8,-0.8) circle (0.12);
\draw[thick,fill=gray] (4.4,-0.8) circle (0.12) (6,-0.8) circle (0.12) (7.6,-0.8) circle (0.12) (4.4,0.8) circle (0.12) (6,0.8) circle (0.12) (7.6,0.8) circle (0.12);
\draw[thick,fill=gray] (5.2,-1.6) circle (0.12) (6.8,-1.6) circle (0.12) (5.2,0) circle (0.12) (6.8,0) circle (0.12) (5.2,1.6) circle (0.12) (6.8,1.6) circle (0.12);
\draw[ultra thick,blue!50!black] (8.6,0.4) -- (9.2,0.4);
\Text[x=9.5,y=0.4]{$=$}
\draw[thick] (9.8,0.4) -- (10.6,0.4); 
\draw[thick,fill=white] (9.8,0.4) circle (0.05) (10.6,0.4) circle (0.05);
\draw[thick,fill=white] (10,0.2) rectangle (10.4,0.6);
\Text[x=10.2,y=0.4,fontsize=\footnotesize]{$H$}
\draw[ultra thick,red!50!black] (8.6,-0.4) -- (9.2,-0.4);
\Text[x=9.5,y=-0.4]{$=$}
\draw[thick] (9.8,-0.4) -- (10.6,-0.4); 
\draw[thick,fill=white] (9.8,-0.4) circle (0.05) (10.6,-0.4) circle (0.05);
\draw[thick,fill=white] (10,-0.2) rectangle (10.4,-0.6);
\Text[x=10.22,y=-0.4,fontsize=\footnotesize]{$H^\dag$}
       \end{tikzpicture}
       \end{center}
   \caption{2D Floquet SPT model with a trivial bulk and an anomalous edge dynamics governed by the $\mathbb{Z}_d\times\mathbb{Z}_d$ SPT MPU (\ref{USPT}).}
      \label{fig1}
\end{figure}

The MPU representation (\ref{Ubb}) in turn gives us an elegant picture of the associated 2D Floquet system with a trivial bulk. As shown in Fig.~\ref{fig1},  the 2D Floquet unitary is a product of commutative local unitaries \begin{tikzpicture}\draw[ultra thick,blue!50!black] (0,0) -- (0.25,0) (0,0.25) -- (0.25,0.25);\draw[ultra thick,red!50!black] (0,0) -- (0,0.25) (0.25,0) -- (0.25,0.25);\end{tikzpicture} and \begin{tikzpicture}\draw[ultra thick,red!50!black] (0,0) -- (0.25,0) (0,0.25) -- (0.25,0.25);\draw[ultra thick,blue!50!black] (0,0) -- (0,0.25) (0.25,0) -- (0.25,0.25);\end{tikzpicture} acting on gray and white plaquettes, respectively. Since \begin{tikzpicture}\draw[ultra thick,blue!50!black] (0,0) -- (0.25,0) (0,0.25) -- (0.25,0.25);\draw[ultra thick,red!50!black] (0,0) -- (0,0.25) (0.25,0) -- (0.25,0.25);\end{tikzpicture} and \begin{tikzpicture}\draw[ultra thick,red!50!black] (0,0) -- (0.25,0) (0,0.25) -- (0.25,0.25);\draw[ultra thick,blue!50!black] (0,0) -- (0,0.25) (0.25,0) -- (0.25,0.25);\end{tikzpicture} are actually small SPT MPU with $2$ unit cells, the entire 2D Floquet unitary is locally implemented under the symmetry constraint. While the bulk turns out to be trivial due to that \begin{tikzpicture}\draw[ultra thick,blue!50!black] (0,0) -- (0.25,0);\end{tikzpicture} and \begin{tikzpicture}\draw[ultra thick,red!50!black] (0,0) -- (0.25,0);\end{tikzpicture} cancel out, we obtain an anomalous edge as Eq.~(\ref{USPT}), which is not locally implementable under the symmetry constraint \cite{Potter2017}.

\subsection{General constraint on the entanglement-spectrum dynamics}
Starting from a trivial symmetric MPS $|\Psi_0\rangle$, after $t$ steps of time evolution by a $\mathbb{Z}_d\times\mathbb{Z}_d$ SPT MPU with the same cohomology class as Eq.~(\ref{USPT}), the entanglement spectrum (under the periodic boundary condition) must be at least $(d/{\rm gcd}(d,t))^2$-fold degenerate. To show this, we first point out a useful property that the spectrum of the transfer matrix of an MPS is conserved during the time evolution by an MPU. This result comes from the unitary nature of time evolution, which implies $\Tr(\sum_jM_j\otimes\bar M_j)^L=\Tr(\sum_j M'_j\otimes\bar M'_j)^L$ for arbitrary $L\in\mathbb{Z}^+$, $M_j\in
\mathbb{C}^{D\times D}$ and $M'_j=\sum_{j'}\mathcal{U}_{jj'}M_{j'}$. In particular, if the transfer matrix of $|\Psi_0\rangle$ has a unique fixed point, the uniqueness is preserved for $|\Psi_t\rangle$. Let $\Lambda$ be the unique fixed point point of the transfer matrix of $|\Psi_t\rangle$, it is known that the entanglement spectrum is given by $\{\lambda_\alpha\lambda_\beta\}^D_{\alpha,\beta=1}$, where $\lambda_\alpha$'s are the eigenvalues of $\Lambda$ \cite{Perez2007}. Moreover, following the same analysis for the composition of two MPUs in Sec.~\ref{cohoSMPU}, we know that the cohomology-class sum rule applies equally to an MPS evolved by an MPU.

Denoting the projective representation on the virtual level of $|\Psi_t\rangle\equiv U^t_{\rm SPT}|\Psi_0\rangle$ as $V_{(m,n)}$, we know from the additivity of the cohomology class that
\begin{equation}
V_{(m,n)}V_{(m',n')}=\omega^{tm'n}_dV_{(m+m',n+n')}
\label{projt}
\end{equation}
for some phase gauge. Moreover, due to the uniqueness of the fixed point, we have $[V_{(m,n)},\Lambda]=0$ for $\forall (m,n)\in\mathbb{Z}_d\times\mathbb{Z}_d$. 
Noting $V^d_{(1,0)}=\mathbb{1}_{\rm v}$, the identity on the virtual level, we can define a set of projectors $P_{n\in\mathbb{Z}_d}\equiv d^{-1}\sum_{j\in\mathbb{Z}_d}\omega^{-jn}_dV^j_{(1,0)}$ satisfying $P_nP_m=\delta_{mn}P_n$, $\sum_{n\in\mathbb{Z}_d}P_n=\mathbb{1}_{\rm v}$ and $P_nV_{(1,0)}=V_{(1,0)}P_n=\omega^n_dP_n$. Now consider an arbitrary eigenstate $|\lambda\rangle_{\rm v}$ of $\Lambda$ with eigenvalue $\lambda$, there should be at least one $P_{n_0}$ such that $P_{n_0}|\lambda\rangle_{\rm v}\neq0$ and thus $|\lambda_{n_0}\rangle_{\rm v}\equiv P_{n_0}|\lambda\rangle_{\rm v}/\|P_{n_0}|\lambda\rangle_{\rm v}\|$ is a common eigenstate of $\Lambda$ and $V_{(1,0)}$ with eigenvalues $\lambda$ and $\omega^{n_0}_d$, respectively.
Using Eq.~(\ref{projt}), we have
\begin{equation}
V_{(1,0)}V^n_{(0,1)}|\lambda_{n_0}\rangle_{\rm v}=\omega^{-tn+n_0}_dV^n_{(0,1)}|\lambda_{n_0}\rangle_{\rm v},
\end{equation}
so $V^n_{(0,1)}|\lambda_{n_0}\rangle_{\rm v}$ are also  common eigenstate of $\Lambda$ and $V_{(1,0)}$ with eigenvalues $\lambda$ and $\omega^{-tn+n_0}_d$, respectively. Since $\omega^{-tn+n_0}_d$ takes $d/{\rm gcd}(d,t)$ different values, the degeneracy in the entanglement spectrum should be at least $(d/{\rm gcd}(d,t))^2$.

\subsection{Analytical and numerical examples}
We consider an analytically solvable case in which $|\Psi_0\rangle=|0_Z\rangle^{\otimes 2L}$ ($|j_Z\rangle$ defined from $Z|j_Z\rangle=\omega^j_d|j_Z\rangle$) is evolved by the MPU in Eq.~(\ref{USPT}). With the local basis chosen as the eigenstates of $X$, the MPS building block for $|\Psi_t\rangle$ is given by
\begin{equation}
[M^X_{(a,b)}]_{\alpha\beta}=d^{-1}\delta_{\alpha a}\omega^{tb(\alpha-\beta)}_d,
\end{equation}
where we have used $\langle j_X|j'_Z\rangle=\omega^{-j'j}_d/\sqrt{d}$ with $j'=0$ ($|j_X\rangle$ defined from $X|j_X\rangle=\omega^j_d|j_X\rangle$). To make the representation matrix of $\mathbb{Z}_d\times\mathbb{Z}_d$ symmetry diagonal, we prefer to use the $Z$-basis, under which
\begin{equation}
\begin{split}
[M^Z_{(a,b)}]_{\alpha\beta}&=d^{-1}\sum_{a',b'\in\mathbb{Z}_d}\omega^{a'a+b'b}_d[M^X_{(a',b')}]_{\alpha\beta}\\
&=d^{-1}\omega^{\alpha a}_d\delta_{(\beta-\alpha)t,b},
\end{split}
\end{equation}
which is zero unless ${\rm gcd}(d,t)\equiv q|b$. When $q|b$, defining $\tilde b\equiv b/q$, $\tilde d\equiv d/q$ and $\tilde t\equiv t/q$, we have 
\begin{equation}
M^Z_{(a,b)}=\tilde d^{-1}Z^aX^{\tilde b\tilde t^{-1}}P, 
\end{equation}
where $\tilde t^{-1}$ is the well-defined inverse (due to ${\rm gcd}(\tilde d,\tilde t)=1$) of $\tilde t$ on $\mathbb{Z}_{\tilde d}$ and $P\equiv q^{-1}\sum_{j\in\mathbb{Z}_q}X^{j\tilde d}$ is a projector with rank $\Tr P=d/q=\tilde d$. Since $P^2=P$, $\tilde M^Z_{(a,b)}\equiv PM^Z_{(a,b)}$ gives the same MPS and the bond dimension is $\tilde d$. This result already implies that the ``least" case of $(d/{\rm gcd}(d,t))^2=\tilde d^2$-fold degeneracy is achievable. Moreover, defining $X_P\equiv PXP$ and $Z_P\equiv PZ^qP$, we can check that $X^{\tilde d}_P=Z^{\tilde d}_P=P$, $X_PZ_P=\omega_{\tilde d}Z_PX_P$ and $PZ^aP=0$ unless $a=q\tilde a$, in which case $PZ^aP=Z^{\tilde a}_PP=PZ^{\tilde a}_P$. Therefore, we obtain $\tilde M^Z_{(a,b)}=0$ except for
\begin{equation}
\tilde M^Z_{q(\tilde a,\tilde b)}=\tilde d^{-1}Z^{\tilde a}_PX^{\tilde b\tilde t^{-1}}_P,
\end{equation}
Under the $Z$-basis, the action of $Z^m\otimes Z^n$ on $|a_Zb_Z\rangle$ is just a phase $\omega^{am+bn}_d$, so the projective representation on the virtual level can be determined as
 \begin{equation}
 V_{(m,n)}=Z^{n\tilde t}_PX^{-m}_P,
 \end{equation}
 which satisfies
 \begin{equation}
 V_{(m,n)}V_{(m',n')}=\omega^{-mn'\tilde t}_{\tilde d}V_{(m+m',n+n')}.
 \label{anaprojt}
 \end{equation}
 Since $\omega^{-mn'\tilde t}_{\tilde d}=\omega^{-mn't}_d=\omega^{m'nt}_d\omega^{-t(mn'+m'n)}_d$ and $\omega^{-t(mn'+m'n)}_d=\omega^{-(m+m')(n+n')t}_d/(\omega^{-mnt}_d\omega^{-m'n't}_d)$, Eq.~(\ref{anaprojt}) can be related to Eq.~(\ref{projt}) by a gauge transformation.

\begin{figure}
\begin{center}
       \includegraphics[width=8.5cm, clip]{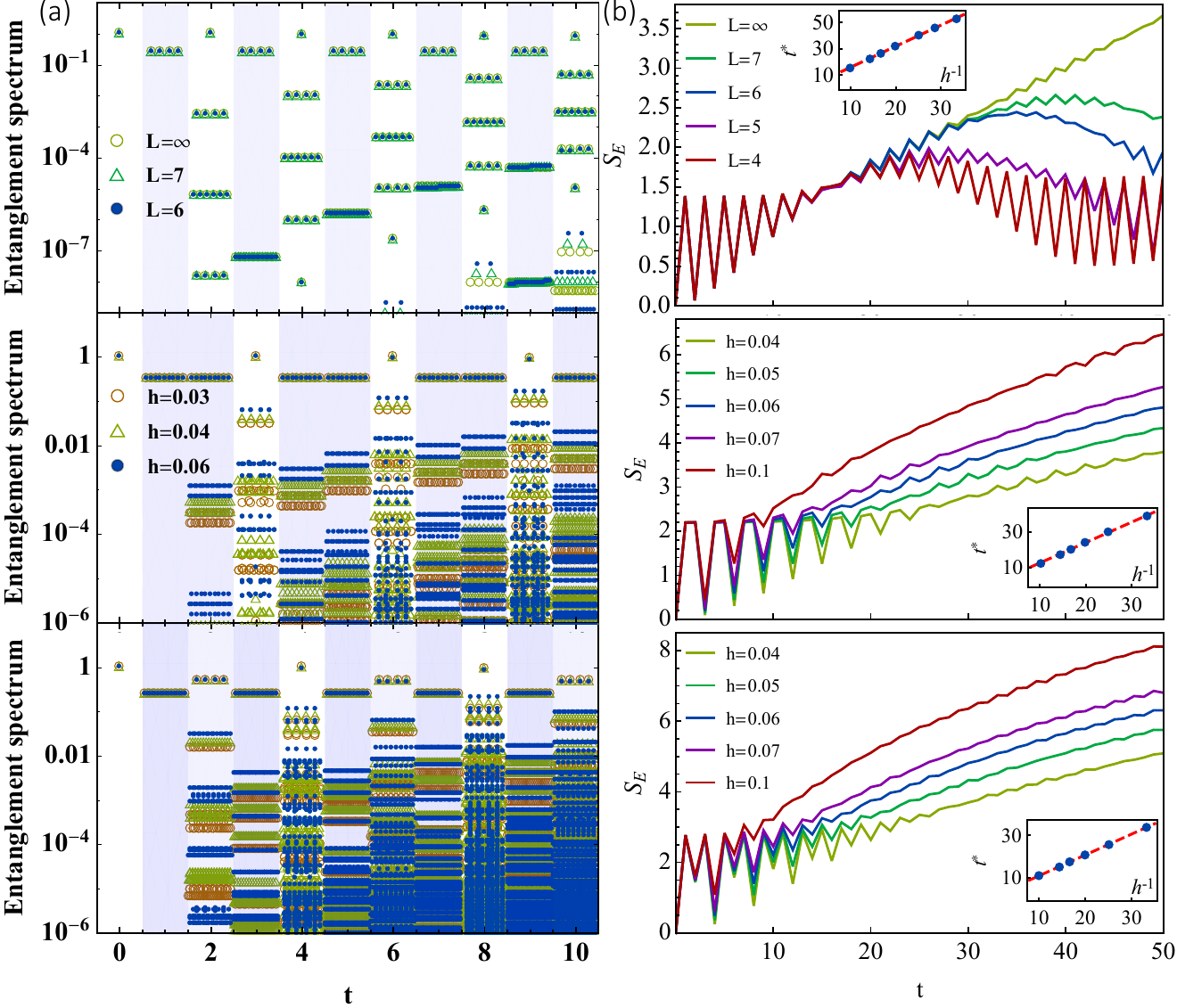}
       \end{center}
   \caption{Stroboscopic dynamics of (a) the entanglement spectrum and (b) the entanglement entropy governed by the Floquet unitary in Eq.~(\ref{UF}) starting from $|\Psi_0\rangle=|0_Z\rangle^{\otimes 2L}$. The top, middle and bottom rows correspond to the cases of $d=2$ (qubit), $d=3$ (qutrit) and $d=4$ (ququard), respectively. Times steps with SPT-enforced entanglement-spectrum degeneracy are shaded in (a). Inserts in (b) show the $h$ dependence of the crossover time $t^*$ is determined from Eq.~(\ref{tcr}). The entanglement dynamics in finite qubit systems are obtained by exact diagonalization while those in infinite qutrit and ququard systems are obtained by iTEBD with varying bond dimension $\chi=100\sim1500$ such that the numerical results converge.}
   \label{figS2}
\end{figure}

If $U_{\rm SPT}$ in Eq.~(\ref{USPT}) is perturbed in a symmetry-preserving manner, the strict $d$-fold period multiplication generally disappears but the constraint on the entanglement-spectrum degeneracy should still be valid. For simplicity, we consider the following $\mathbb{Z}_d\times\mathbb{Z}_d$-symmetric MPU which belongs to the same phase as $U_{\rm SPT}$:
\begin{equation}
U_{\rm F}=e^{-ih\sum_jN_Z}U_{\rm SPT},
\label{UF}
\end{equation}
where $N_Z\equiv\sum^{d-1}_{j=1}\frac{Z^j-\mathbb{1}}{\omega^{-j}_d-1}$. We perform numerical studies on the stroboscopic dynamics of both entanglement spectrum and entropy for $d=2,3$ and $4$ with varying $h$. In particular, we study the finite-size effect using exact diagonalization for $d=2$ (with $h=0.1$ only) while we focus on infinite systems for $d=3$ and $4$ using iTEBD \cite{Vidal2007}. As shown in Fig.~\ref{figS2}, while the growing (with oscillation) entanglement entropy indicates that the system will eventually thermalize, we do observe the (at least) $(d/{\rm gcd}(d,t))^2$-fold degeneracy in the entanglement spectrum. Even in a qubit system with $L=6$, the (at least) $4$-fold degeneracy at odd time steps is only slightly lifted (not visible in Fig.~\ref{figS2}(a)), implying that the SPT discrete time-crystalline oscillation is, in principle, observable in small systems. 

It is also worth mentioning that the entanglement entropy $S_{\rm E}$  exhibits a crossover between two dynamical regimes with large and small oscillations accompanied by the growth. The crossover time $t^*$, which we determine from
\begin{equation}
t^*=\frac{1}{d-1}\sum^{d-1}_{s=1}t^*_s,\;\;
t^*_s\equiv\min\{t:S^{(s)}_{\rm E}(t)=S^{(0)}_{\rm E}(t)\},
\label{tcr}
\end{equation}
where $S^{(s)}_{\rm E}(t)$ is a continuation of $S_{\rm E}(t=nd+s)$ ($s\in\mathbb{Z}_d$), turns out to be proportional to $h^{-1}$ for small $h$. We can understand the crossover from the SPT-enforced entanglement-spectrum degeneracy, which implies a lower bound $2\ln(d/{\rm gcd}(d,s))$ ($t\equiv s\;{\rm mod}\;d$) on the entanglement entropy. When $h$ is small, according to the Magnus expansion as used in Ref.~\cite{Potter2017}, the continuation $S^{(s)}_{\rm E}(t)$ is expected to be well approximated by the entanglement entropy of $|\Psi^{(s)}_{\lambda}\rangle=e^{-i\lambda H_{\rm eff}}U^s_{\rm SPT}|\Psi_0\rangle$ with $\lambda=ht$ and
\begin{equation}
H_{\rm eff}\simeq\frac{1}{d}\sum_j\sum_{a\in\mathbb{Z}_d}U^a_{\rm SPT}N_{j,Z}U^{-a}_{\rm SPT}.
\end{equation}
The initial strong oscillation and the $h^{-1}$ scaling follows.


\subsection{Generalization to arbitrary finite groups}
From the perspective of topological discrete time crystalline oscillation, we can seek for MPUs by considering a bilayer unitary circuit that transforms a trivial state, such as a product state, into an SPT MPS. Indeed, the $\mathbb{Z}_d\times\mathbb{Z}_d$ MPU transforms $|0_Z\rangle^{\otimes L}$ into a 1D cluster state of qudits.

In general, an SPT state can be constructed from a path integral of a cocycle action on a triangulated spacetime manifold \cite{Chen2013}. This construction might be regarded as a sort of the matter-field dual of the Dijkgraaf-Witten theory for gauge fields \cite{Dijkgraaf1990}. In 1D, given a symmetry group $G$ and a $2$-cocyle $e^{i\theta(g,h)}$ representing an element in $H^2(G,{\rm U}(1))$, we can easily identify the underlying MPU as \cite{Harper2017c}
\begin{equation}
U=\sum_{\{g_j\}^L_{j=1}}e^{-i\sum^L_{j=1}\theta(g^{-1}_jg_{j+1},g_{j+1}^{-1})}|g_1g_2...g_L\rangle\langle g_1g_2...g_L|.
\label{SPTMPU}
\end{equation}
Here the symmetry representation on a local Hilbert space $\mathbb{C}^{|G|}$ is assumed to be regular, i.e., $\rho_g|h\rangle=|gh\rangle$ for $\forall g,h\in G$, and the sum of each $g_j$ runs over $G$. To confirm that Eq.~(\ref{SPTMPU}) is topologically nontrivial, we first identify $u,v$ in the standard form as
\begin{equation}
u=v=\sum_{g_l,g_r}e^{-i\theta(g^{-1}_lg_r,g_r^{-1})}|g_l g_r\rangle\langle g_l g_r|.
\end{equation}
Accordingly, we can identify $x_g$ and $y_g$ from
\begin{equation}
\begin{split}
&(\rho_g\otimes\rho_g) v \\
=&\sum_{g_l,g_r}e^{-i\theta(g^{-1}_lg_r,g_r^{-1})}|gg_l, gg_r\rangle\langle g_l g_r| \\
=&\sum_{g_l,g_r}e^{-i\theta(g^{-1}_lg_r,g_r^{-1}g)}|g_lg_r\rangle\langle g^{-1}g_l,g^{-1}g_r| \\
=&\sum_{g_l,g_r}e^{i[\theta(g_r^{-1},g)-\theta(g_l^{-1}g_r,g_r^{-1})-\theta(g_l^{-1},g)]}|g_lg_r\rangle\langle g^{-1}g_l,g^{-1}g_r| \\
=&v(y_g\otimes x_g),
\end{split}
\end{equation}
where we have used the definition property of cocycles:
\begin{equation}
\theta(g_l^{-1}g_r,g_r^{-1})+\theta(g_l^{-1},g) = \theta(g_l^{-1}g_r,g_r^{-1}g)+\theta(g_r^{-1},g).
\end{equation}
Similarly, we can show that $\rho^{\otimes 2}_gu^\dag= u^\dag (x_g\otimes y_g)$. The expressions of $x_g$ and $y_g$ are given by
\begin{equation}
\begin{split}
x_g=y^*_g&=\sum_{k} e^{i\theta(k^{-1},g)} |k\rangle\langle g^{-1}k| \\
&=\sum_{k} e^{i\theta(k^{-1}g^{-1},g)} |gk\rangle\langle k|,
\end{split}
\label{x_g}
\end{equation}
according to which we obtain
\begin{equation}
\begin{split}
x_g x_h&=\sum_{k} e^{i[\theta(k^{-1}h^{-1}g^{-1},g)+\theta(k^{-1}h^{-1},h)]} |ghk\rangle\langle k| \\
&=\sum_{k} e^{i[\theta(k^{-1}h^{-1}g^{-1},gh)+\theta(g,h)]} |ghk\rangle\langle k| \\
&=e^{i\theta(g,h)}x_{gh},
\end{split}
\end{equation}
implying that $x_g$'s form a projective representation with factor set $e^{i\theta(g,h)}$. Since $y_g=x^*_g$, the factor set for $y_g$'s is $e^{-i\theta(g,h)}$.

The nontrivial nature can alternatively be confirmed by looking at the building-block tensor $\mathcal{U}$, which is given by
\begin{equation}
\mathcal{U}=\sum_{j,r}e^{-i\theta(j^{-1}r, r^{-1})}|j\rangle|j)(r|\langle j|.
\end{equation}
Here, we use $|\cdot\rangle$ and $|\cdot)$ to refer to physical and virtual states, respectively. We can identify the projective representation on the virtual level through
\begin{equation}
\begin{split}
&\rho_g\mathcal{U}\rho_g^\dag \\
=&\sum_{j,r}e^{-i\theta(j^{-1}r, r^{-1})}|gj\rangle|j)(r|\langle gj| \\
=&\sum_{j,r}e^{-i\theta(j^{-1}r, r^{-1}g)}|j\rangle |g^{-1}j)(g^{-1}r|\langle j| \\
=&\sum_{j,r}e^{i[\theta(r^{-1},g)-\theta(j^{-1}r, r^{-1})-\theta(j^{-1},g)]}|j\rangle|g^{-1}j)(g^{-1}r|\langle j| \\
=&z^\dag_g\mathcal{U} z_g,
\end{split}
\end{equation}
where we have again used the cocycle property and $z_g$ on the virtual level turns out to be
\begin{equation}
\begin{split}
z_g&=\sum_{h}e^{i\theta(h^{-1},g)}|h)(g^{-1}h| \\
&=\sum_{h}e^{i\theta(h^{-1}g^{-1},g)}|gh)(h|,
\end{split}
\end{equation}
whose entries are the same as those in $x_g$ (\ref{x_g}). This result is consistent with the fact that $z_g$, $x_g$ and $y^*_g$ belong to the same cohomology class.

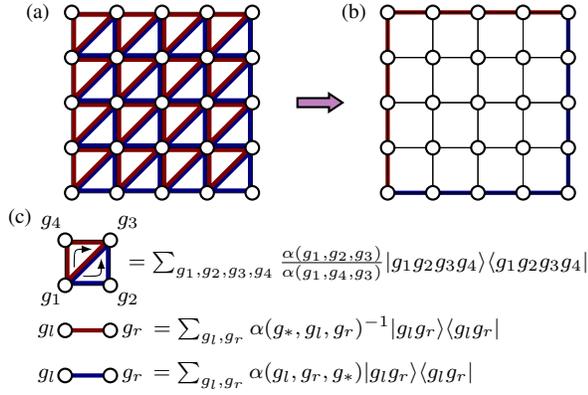
\begin{figure}
\begin{center}
       \begin{tikzpicture}[scale=0.6]
       \Text[x=-1.25,y=3.5,fontsize=\footnotesize]{(a)}
       \foreach \x in {0,...,3}
       \foreach \y in {0,...,3}
       {
           \draw (\x-0.5,\y-0.5) rectangle (\x+0.5,\y+0.5);
           \draw[ultra thick,blue!50!black] (\x-0.4,\y-0.45) -- (\x+0.45,\y-0.45) -- (\x+0.45,\y+0.4) -- cycle;
           \draw[ultra thick,red!50!black] (\x-0.45,\y-0.4) -- (\x-0.45,\y+0.45) -- (\x+0.4,\y+0.45) -- cycle;
       }
       \foreach \x in {0,...,4}
       \foreach \y in {0,...,4}
       {
                  \draw[fill=white,thick] (\x-0.5,\y-0.5) circle (0.15);
       }
       \draw[thick,fill=violet!50] (4.5,1.6) -- (5.2,1.6) -- (5.2,1.7) -- (5.5,1.5) -- (5.2,1.3) -- (5.2,1.4) -- (4.5,1.4) -- cycle;
       \Text[x=5.75,y=3.5,fontsize=\footnotesize]{(b)}
       \draw[ultra thick,blue!50!black] (6.5,-0.5) -- (10.5,-0.5) -- (10.5,3.5);
       \draw[ultra thick,red!50!black] (6.5,-0.5) -- (6.5,3.5) -- (10.5,3.5);
       \foreach \x in {7,...,10}
       \foreach \y in {0,...,3}
       {
           \draw (\x-0.5,\y-0.5) rectangle (\x+0.5,\y+0.5);
       }
       \foreach \x in {7,...,11}
       \foreach \y in {0,...,4}
       {
                  \draw[fill=white,thick] (\x-0.5,\y-0.5) circle (0.15);
       }
       \end{tikzpicture}
       \begin{tikzpicture}[scale=0.6]
       \Text[x=11.5,y=3.5,fontsize=\footnotesize]{(c)}
        \draw[ultra thick,blue!50!black] (12.6,2.05) -- (13.45,2.05) -- (13.45,2.9) -- cycle;
        \draw[ultra thick,red!50!black] (12.55,2.1) -- (12.55,2.95) -- (13.4,2.95) -- cycle;        
        \draw (12.5,2) rectangle (13.5,3);
        \begin{scope}[>=latex]
        \draw[->] (12.9,2.2) .. controls (13.3,2.2) .. (13.3,2.6);
        \draw[->] (12.7,2.4) .. controls (12.7,2.8) .. (13.1,2.8);
        \end{scope}
        \draw[thick,fill=white] (12.5,2) circle (0.15) (13.5,2) circle (0.15) (12.5,3) circle (0.15) (13.5,3) circle (0.15);
       \Text[x=12.2,y=1.7,fontsize=\footnotesize]{$g_1$};
       \Text[x=13.9,y=1.7,fontsize=\footnotesize]{$g_2$};
       \Text[x=13.9,y=3.4,fontsize=\footnotesize]{$g_3$};
       \Text[x=12.2,y=3.4,fontsize=\footnotesize]{$g_4$};
       \Text[x=19,y=2.5,fontsize=\footnotesize]{$=\sum_{g_1,g_2,g_3,g_4}\frac{\alpha(g_1,g_2,g_3)}{\alpha(g_1,g_4,g_3)}|g_1g_2g_3g_4\rangle\langle g_1g_2g_3g_4|$}
       \draw[ultra thick,red!50!black] (12.5,1) -- (13.5,1);
       \draw[thick,fill=white] (12.5,1) circle (0.15) (13.5,1) circle (0.15);
       \Text[x=12.1,y=1,fontsize=\footnotesize]{$g_l$};
       \Text[x=14,y=1,fontsize=\footnotesize]{$g_r$};
       \Text[x=18.3,y=1,fontsize=\footnotesize]{$=\sum_{g_l,g_r} \alpha(g_*,g_l,g_r)^{-1}|g_lg_r\rangle\langle g_l g_r|$}
       \draw[ultra thick,blue!50!black] (12.5,0) -- (13.5,0);
       \draw[thick,fill=white] (12.5,0) circle (0.15) (13.5,0) circle (0.15);
       \Text[x=12.1,y=0,fontsize=\footnotesize]{$g_l$};
       \Text[x=14,y=0,fontsize=\footnotesize]{$g_r$};
       \Text[x=18,y=0,fontsize=\footnotesize]{$=\sum_{g_l,g_r} \alpha(g_l,g_r,g_*)|g_lg_r\rangle\langle g_l g_r|$}
              \end{tikzpicture}
       \end{center}
   \caption{(a) 2D Floquet SPT phase built from a 2-cocycle (\ref{UF2D}). (b) Such a Floquet system has a trivial bulk and a nontrivial edge dynamics described by an SPT MPU (\ref{SPTMPU}). (c) Building blocks of the 2D Floquet unitary and the SPT MPU as its edge dynamics.}
   \label{2DFSPT}
\end{figure}

Finally, let us construct the parent 2D Floquet system of the general SPT MPU (\ref{SPTMPU}). In fact, such a construction naturally extends to arbitrary dimensions and realizes all the intrinsic Floquet SPT phases in the cohomology classification \cite{Harper2017c}. To make the notations elegant, we introduce $\alpha(g_1,g_2,g_3)\equiv e^{-i\theta(g_1^{-1}g_2,g_2^{-1}g_3)}$, which satisfies $\alpha(gg_1,gg_2,gg_3)=\alpha(g_1,g_2,g_3)$ for $\forall g\in G$ and the cocycle property
\begin{equation}
\frac{\alpha(g_2,g_3,g_4)\alpha(g_1,g_2,g_4)}{\alpha(g_1,g_3,g_4)\alpha(g_1,g_2,g_3)}=1.
\label{2co}
\end{equation}
As shown in Fig.~\ref{2DFSPT}, the Floquet unitary is simply a product of commutative four-site unitaries acting on each plaquettes: 
\begin{equation}
\begin{split}
U_{\rm F}&=\prod_{\begin{tikzpicture} \draw[thick,blue!50!black] (0.02,0) -- (0.2,0) -- (0.2,0.18) -- cycle; \draw[thick,red!50!black] (0,0.02) -- (0,0.2) -- (0.18,0.2) -- cycle; \end{tikzpicture}}U_{\begin{tikzpicture} \draw[thick,blue!50!black] (0.02,0) -- (0.2,0) -- (0.2,0.18) -- cycle; \draw[thick,red!50!black] (0,0.02) -- (0,0.2) -- (0.18,0.2) -- cycle; \end{tikzpicture}}, \\
U_{\begin{tikzpicture} \draw[thick,blue!50!black] (0.02,0) -- (0.2,0) -- (0.2,0.18) -- cycle; \draw[thick,red!50!black] (0,0.02) -- (0,0.2) -- (0.18,0.2) -- cycle; \end{tikzpicture}}&=\sum_{\{g_j\}^4_{j=1}}\frac{\alpha(g_1,g_2,g_3)}{\alpha(g_1,g_4,g_3)}|g_1g_2g_3g_4\rangle\langle g_1g_2g_3g_4|.
\end{split}
\label{UF2D}
\end{equation}
Due to the topological nature of the action (especially the cocycle property given in Eq.~(\ref{2co})), the ${\rm U}(1)$ phase before any $|\{g_j\}\rangle\langle\{g_j\}|$ (eigenstate of $U_{\rm F}$) should not depend on the explicit way of triangulation \cite{Dijkgraaf1990,Chen2013}. In particular, we can put a single point into the bulk and assign an arbitrary group element $g_*\in G$ to determine the phase. This observation implies a trivial bulk dynamics, i.e., $U_{\rm F}=\mathbb{I}_{\rm bulk}\otimes U_{\rm edge}$ with $\mathbb{I}_{\rm bulk}$ being the bulk identity \cite{Harper2017c}. For $g_*=e$, the bottom edge dynamics is described by the SPT MPU given in Eq.~(\ref{SPTMPU}).

\section{SPIs for MPUs}
In this section, we prove in details that the SPIs are topological invariants for symmetric MPUs. Regarding the physical implication and experimental probing, we give the explicit expressions of $L_g$ and $R_g$ (see Fig.~2 in the main text) and the derivation of Eq.~(7) in the main text.

\subsection{Basic properties}
\begin{figure*}
\begin{center}
\begin{tikzpicture}[scale=0.9]
\Text[x=-1,y=1.7]{(a)};
\draw[thick,blue,dashed,fill=blue!5] (-0.1,-0.15) -- (-0.1,0.65) -- (0.5,0.65) -- (0.5,1.45) -- (3.7,1.45) -- (3.7,0.65) -- (4.3,0.65) -- (4.3,-0.15) -- cycle;
\draw[thick,blue,dashed,fill=blue!5] (4.3,-0.15) -- (4.3,0.65) -- (4.9,0.65) -- (4.9,1.45) -- (8.1,1.45) -- (8.1,0.65) -- (8.7,0.65) -- (8.7,-0.15) -- cycle;
\draw[thick,blue,dashed] (-0.1,1.65) -- (8.7,1.65) (2.1,1.45) -- (2.1,1.65) (6.5,1.45) -- (6.5,1.65);
\draw[thick] (0.2,-0.25) -- (0.2,0) (0.8,-0.25) -- (0.8,0) (1.4,-0.25) -- (1.4,0) (2.8,-0.25) -- (2.8,0) (3.4,-0.25) -- (3.4,0) (4,-0.25) -- (4,0) (4.6,-0.25) -- (4.6,0) (5.2,-0.25) -- (5.2,0) (5.8,-0.25) -- (5.8,0) (7.2,-0.25) -- (7.2,0) (7.8,-0.25) -- (7.8,0) (8.4,-0.25) -- (8.4,0);
\draw[thick] (0.2,1.75) -- (0.2,1.3) (0.8,1.75) -- (0.8,1.3) (1.4,1.75) -- (1.4,1.3) (2.8,1.75) -- (2.8,1.3) (3.4,1.75) -- (3.4,1.3) (4,1.75) -- (4,1.3) (4.6,1.75) -- (4.6,1.3) (5.2,1.75) -- (5.2,1.3) (5.8,1.75) -- (5.8,1.3) (7.2,1.75) -- (7.2,1.3) (7.8,1.75) -- (7.8,1.3) (8.4,1.75) -- (8.4,1.3);
\draw[ultra thick] (0.2,0.5) -- (0.2,0.8) (1.4,0.5) -- (1.4,0.8) (3.4,0.5) -- (3.4,0.8) (4.6,0.5) -- (4.6,0.8) (5.8,0.5) -- (5.8,0.8) (7.8,0.5) -- (7.8,0.8);
\draw[ultra thick,dotted] (0.8,0.5) -- (0.8,0.8) (2.8,0.5) -- (2.8,0.8) (4,0.5) -- (4,0.8) (5.2,0.5) -- (5.2,0.8) (7.2,0.5) -- (7.2,0.8) (8.4,0.5) -- (8.4,0.8);
\Text[x=-0.5,y=0.65]{$\cdots$}
\draw[thick] (0,0) rectangle (1,0.5);
\draw[thick] (1.7,0.5) -- (1.2,0.5) -- (1.2,0) -- (1.7,0);
\draw[thick] (0.6,0.8) rectangle (1.6,1.3);
\draw[thick] (-0.1,0.8) -- (0.4,0.8) -- (0.4,1.3) -- (-0.1,1.3);
\Text[x=2.1,y=0.65]{$\cdots$}
\draw[thick] (2.5,0.5) -- (3,0.5) -- (3,0) -- (2.5,0);
\draw[thick] (3.2,0) rectangle (4.2,0.5) (4.4,0) rectangle (5.4,0.5);
\draw[thick] (6.1,0) -- (5.6,0) -- (5.6,0.5) -- (6.1,0.5);
\draw[thick] (2.6,0.8) rectangle (3.6,1.3) (3.8,0.8) rectangle (4.8,1.3) (5,0.8) rectangle (6,1.3);
\Text[x=6.5,y=0.65]{$\cdots$}
\draw[thick] (6.9,0) -- (7.4,0) -- (7.4,0.5) -- (6.9,0.5);
\draw[thick] (7.6,0) rectangle (8.6,0.5);
\draw[thick] (8.7,1.3) -- (8.2,1.3) -- (8.2,0.8) -- (8.7,0.8);
\draw[thick] (7,0.8) rectangle (8,1.3);
\Text[x=9.1,y=0.65]{$\cdots$}
\Text[x=0.5,y=0.25]{$u_k$}
\Text[x=3.7,y=0.25]{$u_k$}
\Text[x=4.9,y=0.25]{$u_k$}
\Text[x=8.1,y=0.25]{$u_k$}
\Text[x=1.1,y=1.05]{$v_k$}
\Text[x=3.1,y=1.05]{$v_k$}
\Text[x=4.3,y=1.05]{$v_k$}
\Text[x=5.5,y=1.05]{$v_k$}
\Text[x=7.5,y=1.05]{$v_k$}
\draw (0.5,-0.3) .. controls (0.5,-0.35) and (2.1,-0.25) .. (2.1,-0.45);
\draw (3.7,-0.3) .. controls (3.7,-0.35) and (2.1,-0.25) .. (2.1,-0.45);
\draw (4.9,-0.3) .. controls (4.9,-0.35) and (6.5,-0.25) .. (6.5,-0.45);
\draw (8.1,-0.3) .. controls (8.1,-0.35) and (6.5,-0.25) .. (6.5,-0.45);
\Text[x=2.1,y=-0.6]{$m$}
\Text[x=6.5,y=-0.6]{$m$}
\Text[x=9.6,y=0.65]{$=$}
\draw[thick] (10.8,-0.25) -- (10.8,0) (11.4,-0.25) -- (11.4,0) (12,-0.25) -- (12,0) (12.6,-0.25) -- (12.6,0);
\draw[thick] (10.8,1.55) -- (10.8,1.3) (11.4,1.55) -- (11.4,1.3) (12,1.55) -- (12,1.3) (12.6,1.55) -- (12.6,1.3);
\draw[ultra thick] (10.8,0.5) -- (10.8,0.8) (12,0.5) -- (12,0.8);
\draw[ultra thick,dotted] (11.4,0.5) -- (11.4,0.8) (12.6,0.5) -- (12.6,0.8);
\Text[x=10.1,y=0.65]{$\cdots$}
\draw[thick,fill=blue!5] (10.6,0) rectangle (11.6,0.5) (11.8,0) rectangle (12.8,0.5);
\draw[thick] (11.2,0.8) rectangle (12.2,1.3);
\draw[thick] (10.5,0.8) -- (11,0.8) -- (11,1.3) -- (10.5,1.3);
\draw[thick] (12.9,0.8) -- (12.4,0.8) -- (12.4,1.3) -- (12.9,1.3);
\Text[x=13.3,y=0.65]{$\cdots$}
\Text[x=11.1,y=0.25]{$u_K$}
\Text[x=12.3,y=0.25]{$u_K$}
\Text[x=11.7,y=1.05]{$v_K$}
\end{tikzpicture}
\begin{tikzpicture}
\fill[white] (0,0) rectangle (10,0.01);
\end{tikzpicture}
\begin{tikzpicture}[scale=0.9]
\Text[x=-4,y=2.9]{(b)};
\Text[x=10.6,y=2.9]{};
\draw[thick,orange,dashed,fill=orange!5] (-2.5,-0.15) -- (-2.5,0.65) -- (-1.9,0.65) -- (-1.9,1.45) -- (-1.3,1.45) -- (-1.3,2.25) -- (-0.1,2.25) -- (-0.1,1.45) -- (0.5,1.45) -- (0.5,0.65) -- (1.1,0.65) -- (1.1,-0.15) -- cycle;
\draw[thick,orange,dashed,fill=orange!5] (1.1,-0.15) -- (1.1,0.65) -- (1.7,0.65) -- (1.7,1.45) -- (2.3,1.45) -- (2.3,2.25) -- (3.5,2.25) -- (3.5,1.45) -- (4.1,1.45) -- (4.1,0.65) -- (4.7,0.65) -- (4.7,-0.15) -- cycle;
\draw[thick,orange,dashed] (-2.5,3) -- (4.7,3) (-0.7,2.25) -- (-0.7,3) (2.9,2.25) -- (2.9,3);
\draw[ultra thick] (-2.2,0.5) -- (-2.2,0.8) (-1,0.5) -- (-1,0.8) (0.2,0.5) -- (0.2,0.8) (1.4,0.5) -- (1.4,0.8) (2.6,0.5) -- (2.6,0.8) (3.8,0.5) -- (3.8,0.8) (-2.2,2.1) -- (-2.2,2.4) (-1,2.1) -- (-1,2.4) (0.2,2.1) -- (0.2,2.4) (1.4,2.1) -- (1.4,2.4) (2.6,2.1) -- (2.6,2.4) (3.8,2.1) -- (3.8,2.4);
\draw[ultra thick,dotted] (-1.6,0.5) -- (-1.6,0.8) (-0.4,0.5) -- (-0.4,0.8) (0.8,0.5) -- (0.8,0.8) (2,0.5) -- (2,0.8) (3.2,0.5) -- (3.2,0.8) (4.4,0.5) -- (4.4,0.8) (-1.6,2.1) -- (-1.6,2.4) (-0.4,2.1) -- (-0.4,2.4) (0.8,2.1) -- (0.8,2.4) (2,2.1) -- (2,2.4) (3.2,2.1) -- (3.2,2.4) (4.4,2.1) -- (4.4,2.4);
\draw[thick] (-2.2,-0.25) -- (-2.2,0) (-2.2,1.3) -- (-2.2,1.6) (-2.2,2.9) -- (-2.2,3.15) (-1.6,-0.25) -- (-1.6,0) (-1.6,1.3) -- (-1.6,1.6) (-1.6,2.9) -- (-1.6,3.15) (-1,-0.25) -- (-1,0) (-1,1.3) -- (-1,1.6) (-1,2.9) -- (-1,3.15) (-0.4,-0.25) -- (-0.4,0) (-0.4,1.3) -- (-0.4,1.6) (-0.4,2.9) -- (-0.4,3.15) (0.2,-0.25) -- (0.2,0) (0.2,1.3) -- (0.2,1.6) (0.2,2.9) -- (0.2,3.15) (0.8,-0.25) -- (0.8,0) (0.8,1.3) -- (0.8,1.6) (0.8,2.9) -- (0.8,3.15) (1.4,-0.25) -- (1.4,0) (1.4,1.3) -- (1.4,1.6) (1.4,2.9) -- (1.4,3.15) (2,-0.25) -- (2,0) (2,1.3) -- (2,1.6) (2,2.9) -- (2,3.15) (2.6,-0.25) -- (2.6,0) (2.6,1.3) -- (2.6,1.6) (2.6,2.9) -- (2.6,3.15) (3.2,-0.25) -- (3.2,0) (3.2,1.3) -- (3.2,1.6) (3.2,2.9) -- (3.2,3.15) (3.8,-0.25) -- (3.8,0) (3.8,1.3) -- (3.8,1.6) (3.8,2.9) -- (3.8,3.15) (4.4,-0.25) -- (4.4,0) (4.4,1.3) -- (4.4,1.6) (4.4,2.9) -- (4.4,3.15);
\draw[thick] (-2.4,0) rectangle (-1.4,0.5);
\draw[thick] (-1.2,0) rectangle (-0.2,0.5);
\draw[thick] (0,0) rectangle (1,0.5);
\draw[thick] (1.2,0) rectangle (2.2,0.5);
\draw[thick] (2.4,0) rectangle (3.4,0.5);
\draw[thick] (3.6,0) rectangle (4.6,0.5);
\draw[thick] (-2.4,1.6) rectangle (-1.4,2.1);
\draw[thick] (-1.2,1.6) rectangle (-0.2,2.1);
\draw[thick] (0,1.6) rectangle (1,2.1);
\draw[thick] (1.2,1.6) rectangle (2.2,2.1);
\draw[thick] (2.4,1.6) rectangle (3.4,2.1);
\draw[thick] (3.6,1.6) rectangle (4.6,2.1);
\draw[thick] (-2.5,0.8) -- (-2,0.8) -- (-2,1.3) -- (-2.5,1.3);
\draw[thick] (-2.5,2.4) -- (-2,2.4) -- (-2,2.9) -- (-2.5,2.9);
\draw[thick] (-1.8,0.8) rectangle (-0.8,1.3);
\draw[thick] (-0.6,0.8) rectangle (0.4,1.3);
\draw[thick] (0.6,0.8) rectangle (1.6,1.3);
\draw[thick] (1.8,0.8) rectangle (2.8,1.3);
\draw[thick] (3,0.8) rectangle (4,1.3);
\draw[thick] (-1.8,2.4) rectangle (-0.8,2.9);
\draw[thick] (-0.6,2.4) rectangle (0.4,2.9);
\draw[thick] (0.6,2.4) rectangle (1.6,2.9);
\draw[thick] (1.8,2.4) rectangle (2.8,2.9);
\draw[thick] (3,2.4) rectangle (4,2.9);
\draw[thick] (4.7,0.8) -- (4.2,0.8) -- (4.2,1.3) -- (4.7,1.3);
\draw[thick] (4.7,2.4) -- (4.2,2.4) -- (4.2,2.9) -- (4.7,2.9);
\Text[x=-1.9,y=0.25]{$u_0$}
\Text[x=-0.7,y=0.25]{$u_0$}
\Text[x=0.5,y=0.25]{$u_0$}
\Text[x=1.7,y=0.25]{$u_0$}
\Text[x=2.9,y=0.25]{$u_0$}
\Text[x=4.1,y=0.25]{$u_0$}
\Text[x=-1.3,y=1.05]{$v_0$}
\Text[x=-0.1,y=1.05]{$v_0$}
\Text[x=1.1,y=1.05]{$v_0$}
\Text[x=2.3,y=1.05]{$v_0$}
\Text[x=3.5,y=1.05]{$v_0$}
\Text[x=-1.3,y=2.65]{$v_1$}
\Text[x=-0.1,y=2.65]{$v_1$}
\Text[x=1.1,y=2.65]{$v_1$}
\Text[x=2.3,y=2.65]{$v_1$}
\Text[x=3.5,y=2.65]{$v_1$}
\Text[x=-1.9,y=1.85]{$u_1$}
\Text[x=-0.7,y=1.85]{$u_1$}
\Text[x=0.5,y=1.85]{$u_1$}
\Text[x=1.7,y=1.85]{$u_1$}
\Text[x=2.9,y=1.85]{$u_1$}
\Text[x=4.1,y=1.85]{$u_1$}
\Text[x=-2.9,y=1.45]{$\cdots$}
\Text[x=5.1,y=1.45]{$\cdots$}
\Text[x=5.6,y=1.45]{$=$}
\Text[x=6.1,y=1.45]{$\cdots$}
\Text[x=9.2,y=1.45]{$\cdots$}
\draw[thick,fill=orange!5] (6.5,0.8) rectangle (7.5,1.3);
\draw[thick,fill=orange!5] (7.7,0.8) rectangle (8.7,1.3);
\draw[thick] (7.1,1.6) rectangle (8.1,2.1);
\draw[thick] (6.4,1.6) -- (6.9,1.6) -- (6.9,2.1) -- (6.4,2.1);
\draw[thick] (8.8,1.6) -- (8.3,1.6) -- (8.3,2.1) -- (8.8,2.1);
\draw[thick] (6.7,0.55) -- (6.7,0.8)  (7.3,0.55) -- (7.3,0.8) (7.9,0.55) -- (7.9,0.8) (8.5,0.55) -- (8.5,0.8) 
(6.7,2.1) -- (6.7,2.35)  (7.3,2.1) -- (7.3,2.35) (7.9,2.1) -- (7.9,2.35) (8.5,2.1) -- (8.5,2.35);
\draw[ultra thick] (6.7,1.3) -- (6.7,1.6)  (7.9,1.3) -- (7.9,1.6);
\draw[ultra thick,dotted] (7.3,1.3) -- (7.3,1.6) (8.5,1.3) -- (8.5,1.6);
\Text[x=7,y=1.05]{$u$}
\Text[x=8.2,y=1.05]{$u$}
\Text[x=7.6,y=1.85]{$v$}
\end{tikzpicture}
       \end{center}
   \caption{Blocking protocols for showing (a) the blocking independence and (b) the additivity of the SPIs.}
   \label{figBlocking}
\end{figure*}
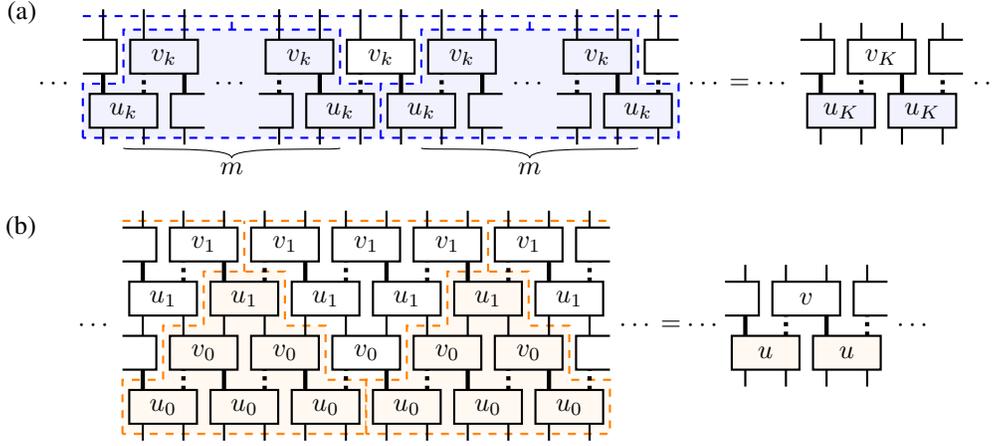

We start with proving
\begin{proposition}
Given a $G$-symmetric MPU generated by $\mathcal{U}$ and $g\in G$ with $\chi_g\neq0$, the SPI 
\begin{equation}
\ind_g=\frac{1}{2}\log\left|\frac{\Tr y_g}{\Tr x_g}\right|
\end{equation}
is well-defined, although the standard form is not unique. 
\end{proposition}
\emph{Proof:} In the main text, we already know that the SPI is well-defined for a given blocking number $k\ge k_0$, so we only have to prove that the SPI does not depend on $k$. For $\forall k>k_0$, we can always find $m,m_0\in\mathbb{Z}^+$ such that $km=k_0m_0\equiv K$. Denoting the standard form of $\mathcal{U}_{k_0}$ and $\mathcal{U}_k$ as $u_{k_0},v_{k_0}$ and $u_k,v_k$, respectively, by further blocking the former $m_0$ times (see Fig.~\ref{figBlocking}(a)) or the latter $m$ times, we obtain two equivalent standard forms for $\mathcal{U}_K$:
\begin{equation}
\begin{split}
u_K&=(\mathbb{1}^{\otimes k}\otimes v^{\otimes(m-1)}_k\otimes \mathbb{1}^{\otimes k})u^{\otimes m}_k,\\
v_K&=\mathbb{1}^{\otimes k(m-1)}\otimes v_k\otimes \mathbb{1}^{\otimes k(m-1)},\;{\rm and}\;\\
u'_K&=(\mathbb{1}^{\otimes k_0}\otimes v^{\otimes(m_0-1)}_{k_0}\otimes \mathbb{1}^{\otimes k_0})u^{\otimes m_0}_{k_0},\\
v'_K&=\mathbb{1}^{\otimes k_0(m_0-1)}\otimes v_{k_0}\otimes \mathbb{1}^{\otimes k_0(m_0-1)},
\end{split}
\end{equation}
which should be related by a gauge transformation \cite{Cirac2017}. Accordingly, $x_g$ and $y_g$ for $\mathcal{U}_K$ can be obtained to be
\begin{equation}
\begin{split}
&x_{k,g}\otimes\rho_g^{\otimes k(m-1)},\;\rho_g^{\otimes k(m-1)}\otimes y_{k,g},\;{\rm and}\;\;\\
&x_{k_0,g}\otimes\rho_g^{\otimes k_0(m_0-1)},\;\rho_g^{\otimes k_0(m_0-1)}\otimes y_{k_0,g},
\end{split}
\end{equation}
where $x_{k,g},y_{k,g}$ are the projective representations on the virtual level of the standard form of $\mathcal{U}_k$. 
Recalling that $\ind_g$ is well-defined for a given blocking number, we have
\begin{equation}
\begin{split}
{\rm ind}_g(\mathcal{U}_K)=\frac{1}{2}\log\frac{|\Tr y_{k,g}||\chi_g|^{k(m-1)}}{|\Tr x_{k,g}||\chi_g|^{k(m-1)}}&={\rm ind}_g(\mathcal{U}_k)\\
=\frac{1}{2}\log\frac{|\Tr y_{k_0,g}||\chi_g|^{k_0(m_0-1)}}{|\Tr x_{k_0,g}||\chi_g|^{k_0(m_0-1)}}&={\rm ind}_g(\mathcal{U}_{k_0}).
\end{split}
\end{equation}
So far, we have confirmed that the SPI is well-defined. \hfill$\Box$

We move on to prove that the SPI is a topological invariant.
\begin{theorem}
Given a continuous path of $\mathcal{U}_\lambda$ generating $G$-symmetric MPUs and $g\in G$ with $\chi_g\neq0$, then $\ind_g$ stays unchanged along the path. 
\label{indgtopo}
\end{theorem}
\emph{Proof:} According to Corollary 4.7 in Ref.~\cite{Cirac2017}, there exists $k_0\le D^4_{\rm m}$ ($D_{\rm m}$: largest bond dimension of $\mathcal{U}_\lambda$) such that the MPU $U^{(2k_0L)}_\lambda$ generated by $\mathcal{U}_\lambda$ has a standard form $u_\lambda$, $v_\lambda$ which are continuous with respect to $\lambda$. Using $u_\lambda\rho^{\otimes2k_0}_gu^\dag_\lambda=x_{\lambda,g}\otimes y_{\lambda,g}\equiv w_{\lambda,g}$, we can rewrite the SPI for $\mathcal{U}_\lambda$ into
\begin{equation}
\ind_g(\mathcal{U}_\lambda)=
\frac{1}{2}\log\frac{\Tr_{\rm l}[\Tr_{\rm r} w_{\lambda,g}\Tr_{\rm r} w^\dag_{\lambda,g}]}{l|\chi_g|^{2k_0}},
\end{equation}
where $\Tr_{\rm r}$ and $\Tr_{\rm l}$ are the partial traces over $\mathbb{C}^r$ and $\mathbb{C}^l$ on the virtual level, with $l$ and $r$ being the dimensions of $x_{\lambda,g}$ and $y_{\lambda,g}$, respectively. Since $w_{\lambda,g}\equiv u_\lambda\rho^{\otimes2k_0}_gu^\dag_\lambda$ and $w^\dag_\lambda$ are both continuous with respect to $\lambda$, the continuity of $\ind_g(\mathcal{U}_\lambda)$ follows. On the other hand, due to the fact that $x_{\lambda,g}\otimes y_{\lambda,g}$ is a linear representation, we have $(x_{\lambda,g}\otimes y_{\lambda,g})^{d_g}=\mathbb{1}^{\otimes2k_0}$, implying $x^{d_g}_{\lambda,g}=\mathbb{1}_l$ and $y^{d_g}_{\lambda,g}=\mathbb{1}_r$ for certain phase gauge. Accordingly, $\ind_g(\mathcal{U}_\lambda)$ only takes values over a finite set
\begin{equation}
\left\{\log\frac{|\sum_{j\in\mathbb{Z}_{d_g}} n_j\omega^j_{d_g}|}{|\chi_g|^{k_0}}:
\sum_{j\in\mathbb{Z}_{d_g}} n_j=r\right\}
\end{equation} 
in $\log|\mathbb{Q}(\omega_{d_g})\backslash\{0\}|$. Combining the discrete image of $\ind_g(\mathcal{U}_\lambda)$ with its continuity, we conclude that $\ind_g$ is a quantized constant, 
which is a logarithm of the absolute value of a cyclotomic number, along the continuous path. \hfill$\Box$

Finally, we turn to prove the additivity of the SPI.
\begin{theorem}
The SPI is additive by tensoring and composition.
\end{theorem}
\emph{Proof:} The case of tensoring is almost trivial --- on the virtual level of the standard form, the projective representations of a tensored MPU $U=U_1\otimes U_0$ are given by the tensor product of those of $U_1$ and $U_0$. 
Using the trace identity $\Tr x_1\otimes x_0=\Tr x_1\Tr x_0$, we immediately obtain the additivity of the SPI. 

For the case of composition, we consider two $G$-symmetric standard-form (after blocking $k_0$ sites into one) MPUs $U^{(6k_0L)}_0$ and $U^{(6k_0L)}_1$ generated by $u_0,v_0$ and $u_1,v_1$
and their composition $U^{(6k_0L)}=U^{(6k_0L)}_1U^{(6k_0L)}_0$. By further blocking three sites into one, the building blocks $u,v$ for the standard form of $U^{(6k_0L)}$ can be related to those for $U^{(6k_0L)}_0$ and $U^{(6k_0L)}_1$ via (see Fig.~\ref{figBlocking}(b))
\begin{equation}
\begin{split}
u&=\{x_{0,e}
\otimes[(\mathbb{1}^{\otimes k_0}\otimes u_1\otimes \mathbb{1}^{\otimes k_0})v^{\otimes2}_0]\otimes y_{0,e}
\}u^{\otimes3}_0,\\
v&=v^{\otimes3}_1\{y_{1,e}
\otimes [u^{\otimes2}_1(\mathbb{1}^{\otimes k_0}\otimes v_0\otimes\mathbb{1}^{\otimes k_0})]\otimes x_{1,e}
\}.
\end{split}
\end{equation}
Therefore, on the virtual level of the standard form of $U^{(6k_0L)}$, the projective representations read
\begin{equation}
x=x_0\otimes\rho^{\otimes k_0}\otimes x_1,\;\;
y=y_1\otimes\rho^{\otimes k_0}\otimes y_0,
\end{equation}
again implying that the SPIs sum up. \hfill$\Box$ 

The SPI $\ind_g$ for $g\in G$ is only well-defined for $\Tr\rho_g\neq0$. A necessary condition for this is the \emph{$c$-regularity} of $g$. For the cohomology class represented by a $2$-cocyle $c$: $G\times G\to{\rm U}(1)$, $g\in G$ is said to be $c$-regular if $c(g,h)=c(h,g)$ for $\forall h\in N_g\equiv\{h\in G:gh=hg\}$, which is the stabilizer group of $g$. We can show that $c$-regular is actually a property for conjugacy classes, since $N_{kgk^{-1}}=kN_gk^{-1}$ and $c(kgk^{-1},khk^{-1})$ is related to $c(g,h)$ via a $k$-dependent $2$-coboundary \cite{Cheng2015}:
\begin{equation}
c(kgk^{-1},khk^{-1})=c(g,h)\frac{\beta_k(gh)}{\beta_k(g)\beta_k(h)},
\end{equation}
where $\beta_k(g)=\frac{c(k,gk^{-1})c(g,k^{-1})}{c(k^{-1},k)}$ is a $1$-cochain. Now let us consider an arbitrary projective representation $z$ with factor set $c$ together with a group element $g$ that is not $c$-regular. By assumption, there exists $h\in N_g$ such that $c(g,h)\neq c(h,g)$. From the definition of projective representation $z_gz_h=c(g,h)z_{gh}$ and $z_h z_g=c(h,g)z_{hg}$, we obtain
\begin{equation}
\begin{split}
\Tr z_g&=c(g,h)\Tr (z_{gh}z_h^{-1})=c(h,g)\Tr (z_h^{-1}z_{hg}) \\
&=c(h,g)\Tr (z_{gh}z_h^{-1}),
\end{split}
\end{equation}
where we have used $\Tr (AB)=\Tr (BA)$ and $gh=hg$ in the second line. Since $c(g,h)\neq c(h,g)$, the only possibility is $\Tr z_g=\Tr (z_{gh}z_h^{-1})=0$. Similarly, we have $\Tr z_h=\Tr z_{gh}=0$. This completes the proof that the SPI cannot be defined for $g$ whenever $g$ is not $c$-regular.

\subsection{Expressions of $L_g$ and $R_g$ for MPUs}
In order to experimentally measure the SPI, we rewrite it in terms of the operators $L_g$ and $R_g$ arising in the evolution of a symmetry string operator (c.f.~Eq.~(6) and Fig.~2(a) in the main text).

Without loss of generality, we first assume $\mathcal{U}$ to be simple (otherwise block and redefine $\mathcal{U}_k$ as $\mathcal{U}$ and correspondingly $\rho^{\otimes k}_g\to\rho_g$) so that \cite{Cirac2017}
\begin{equation}
\begin{tikzpicture}
\draw[thick] (-0.5,0) -- (1.5,0) (-0.5,1) -- (1.5,1) (0,-0.5) -- (0,1.5) (1,-0.5) -- (1,1.5); 
\Vertex[shape = rectangle,size=0.5,x=0,label=$\mathcal{U}$,fontsize=\normalsize,color=white]{A}
\Vertex[shape = rectangle,size=0.5,x=1,label=$\mathcal{U}$,fontsize=\normalsize,color=white]{B}
\Vertex[shape = rectangle,size=0.5,x=0,y=1,label=$\overline{\mathcal{U}}$,fontsize=\normalsize,color=white]{C}
\Vertex[shape = rectangle,size=0.5,x=1,y=1,label=$\overline{\mathcal{U}}$,fontsize=\normalsize,color=white]{D}
\Text[x=2,y=0.5]{$=$}
\draw[thick] (2.5,0) -- (3.5,0) -- (3.5,1) -- (2.5,1);
\draw[thick] (5,0) -- (4,0) -- (4,1) -- (5,1);
\draw[thick] (3,-0.5) -- (3,1.5) (4.5,-0.5) -- (4.5,1.5);
\Vertex[shape = rectangle,size=0.5,x=3,label=$\mathcal{U}$,fontsize=\normalsize,color=white]{a}
\Vertex[shape = rectangle,size=0.5,x=4.5,label=$\mathcal{U}$,fontsize=\normalsize,color=white]{b}
\Vertex[shape = rectangle,size=0.5,x=3,y=1,label=$\overline{\mathcal{U}}$,fontsize=\normalsize,color=white]{c}
\Vertex[shape = rectangle,size=0.5,x=4.5,y=1,label=$\overline{\mathcal{U}}$,fontsize=\normalsize,color=white]{d}
\Vertex[shape = circle,size=0.4,x=3.5,y=0.5,label=$\Sigma$,fontsize=\footnotesize,color=white]{e}
\Text[x=5.25,y=0.25]{,}
\end{tikzpicture}
\label{simple}
\end{equation}
where $\overline{\mathcal{U}}_{ij}\equiv\mathcal{U}^*_{ji}$ and $\Sigma$ is the unique fixed point of the quantum channel
\begin{equation}
\begin{tikzpicture}
\Text[x=-1.5,y=0.625]{$\mathcal{E}\equiv d^{-1}$}
\draw[thick] (-0.75,1) -- (0.75,1);
\draw[thick] (-0.75,0) -- (0.75,0);
\draw[thick] (0,-0.35) -- (0,1.35);
\draw[thick] (0,1.35) .. controls (-0,1.55) and (-0.1,1.55) .. (-0.1,1.35);
\draw[thick] (0,-0.35) .. controls (-0,-0.55) and (-0.1,-0.55) .. (-0.1,-0.35);
\Vertex[shape = rectangle,size=0.5,x=0,label=$\mathcal{U}$,fontsize=\normalsize,color=white]{A}
\Vertex[shape = rectangle,size=0.5,x=0,y=1,label=$\overline{\mathcal{U}}$,fontsize=\normalsize,color=white]{B}
\Text[x=1,y=0.35]{.}
\end{tikzpicture}
\end{equation}
Applying Eqs.~(\ref{simple}) and (\ref{proj MPS}) to $U^{(L)\dag}\rho^{\otimes N}_gU^{(L)}$, we obtain
\begin{equation}
\begin{tikzpicture}
\Text[x=-1.1,y=0.5]{$L_g=$}
\draw[thick] (0,-0.9) -- (0,1.5) (1.2,-0.9) -- (1.2,1.5);
\draw[thick] (-0.5,0) rectangle (1.7,1);
\Vertex[shape = rectangle,size=0.5,x=0,label=$\mathcal{U}$,fontsize=\normalsize,color=white]{A}
\Vertex[shape = rectangle,size=0.5,x=1.2,label=$\mathcal{U}$,fontsize=\normalsize,color=white]{B}
\Vertex[shape = rectangle,size=0.5,x=0,y=1,label=$\overline{\mathcal{U}}$,fontsize=\normalsize,color=white]{C}
\Vertex[shape = rectangle,size=0.5,x=1.2,y=1,label=$\overline{\mathcal{U}}$,fontsize=\normalsize,color=white]{D}
\Vertex[shape = circle,size=0.4,x=0.6,y=0,label=$z^\dag_g$,fontsize=\footnotesize,color=white]{E}
\Vertex[shape = circle,size=0.4,x=1.7,y=0.5,label=$\Sigma$,fontsize=\footnotesize,color=white]{F}
\Vertex[shape = circle,size=0.4,x=1.2,y=-0.55,label=$\rho_g$,fontsize=\footnotesize,color=white]{G}
\Text[x=2,y=0.25]{,}
\draw[thick] (4,-0.9) -- (4,1.5) (5.2,-0.9) -- (5.2,1.5);
\Text[x=2.9,y=0.5]{$R_g=$}
\draw[thick] (3.5,0) rectangle (5.7,1);
\Vertex[shape = rectangle,size=0.5,x=4,label=$\mathcal{U}$,fontsize=\normalsize,color=white]{a}
\Vertex[shape = rectangle,size=0.5,x=5.2,label=$\mathcal{U}$,fontsize=\normalsize,color=white]{b}
\Vertex[shape = rectangle,size=0.5,x=4,y=1,label=$\overline{\mathcal{U}}$,fontsize=\normalsize,color=white]{c}
\Vertex[shape = rectangle,size=0.5,x=5.2,y=1,label=$\overline{\mathcal{U}}$,fontsize=\normalsize,color=white]{d}
\Vertex[shape = circle,size=0.4,x=4.6,y=0,label=$z_g$,fontsize=\footnotesize,color=white]{e}
\Vertex[shape = circle,size=0.4,x=5.7,y=0.5,label=$\Sigma$,fontsize=\footnotesize,color=white]{f}
\Vertex[shape = circle,size=0.4,x=4,y=-0.55,label=$\rho_g$,fontsize=\footnotesize,color=white]{g}
\Text[x=6,y=0.25]{.}
\end{tikzpicture}
\label{LgRg}
\end{equation}
Substituting the singular-value decomposed forms
\begin{equation}
\begin{tikzpicture}
\draw[thick] (0,-0.5) -- (0,0.5) (-0.5,0) -- (0.5,0);
\Vertex[shape = rectangle,size=0.5,x=0,label=$\mathcal{U}$,fontsize=\normalsize,color=white]{A}
\Text[x=0.8,y=0]{$=$}
\draw[ultra thick,dotted] (1.6,-0.2) -- (1.6,0.2);
\draw[thick] (1.1,-0.2) -- (1.35,-0.2) (1.6,-0.45) -- (1.6,-0.7);
\draw[thick,fill=white] (1.35,-0.2) -- (1.6,-0.2) -- (1.6,-0.45) -- cycle;
\draw[thick] (1.6,0.45) -- (1.6,0.7) (1.85,0.45) -- (2.1,0.45);
\draw[thick,fill=white] (1.6,0.2) -- (1.6,0.45) -- (1.85,0.45) -- cycle;
\Text[x=2.4,y=0]{$=$}
\draw[ultra thick] (3.2,-0.2) -- (3.2,0.2);
\draw[thick] (2.7,0.45) -- (2.95,0.45) (3.2,0.45) -- (3.2,0.7);
\draw[thick,fill=lightgray] (2.95,0.45) -- (3.2,0.45) -- (3.2,0.2) -- cycle;
\draw[thick] (3.45,-0.2) -- (3.7,-0.2) (3.2,-0.45) -- (3.2,-0.7);
\draw[thick,fill=lightgray] (3.2,-0.45) -- (3.2,-0.2) -- (3.45,-0.2) -- cycle;
\end{tikzpicture}
\end{equation}
into Eq.~(\ref{LgRg}), we obtain
\begin{equation}
\begin{tikzpicture}
\Text[x=-0.5,y=1]{$L_g=$}
\filldraw[thick,fill=lightgray] (1,0) -- (1.25,0) -- (1,-0.25) -- cycle;
\filldraw[thick,fill=lightgray] (1,0.75) -- (0.75,0.75) -- (1,0.5) -- cycle;
\filldraw[thick,fill=lightgray] (1,1.25) -- (0.75,1.25) -- (1,1.5) -- cycle;
\filldraw[thick,fill=lightgray] (1,2) -- (1.25,2) -- (1,2.25) -- cycle;
\filldraw[thick,fill=white] (2.25,0) -- (2,0) -- (2.25,-0.25) -- cycle;
\filldraw[thick,fill=white] (2.25,0.75) -- (2.5,0.75) -- (2.25,0.5) -- cycle;
\filldraw[thick,fill=white] (2.25,1.25) -- (2.5,1.25) -- (2.25,1.5) -- cycle;
\filldraw[thick,fill=white] (2.25,2) -- (2,2) -- (2.25,2.25) -- cycle;
\draw[thick] (1,-0.85) -- (1,0) -- (2.25,0) -- (2.25,-0.85);
\draw[thick] (0.25,1.25) -- (1,1.25) -- (1,0.75) -- (0.25,0.75) -- cycle;
\draw[thick] (1,2.6) -- (1,2) -- (2.25,2) -- (2.25,2.6);
\draw[thick] (2.25,1.25) -- (3,1.25) -- (3,0.75) -- (2.25,0.75) -- cycle;
\draw[ultra thick] (1,0) -- (1,0.5) (1,1.5) -- (1,2);
\draw[ultra thick,dotted] (2.25,0) -- (2.25,0.5) (2.25,1.5) -- (2.25,2);
\Vertex[shape = circle,size=0.4,x=1.625,y=0,label=$z^\dag_g$,fontsize=\footnotesize,color=white]{A}
\Vertex[shape = circle,size=0.4,x=2.25,y=-0.55,label=$\rho_g$,fontsize=\footnotesize,color=white]{B}
\Vertex[shape = circle,size=0.35,x=3,y=1,label=$\Sigma$,fontsize=\footnotesize,color=white]{C}
\Text[x=3.5,y=1]{$=$}
\filldraw[thick,fill=lightgray] (4,0.5) -- (4.25,0.5) -- (4,0.25) -- cycle;
\filldraw[thick,fill=lightgray] (4,1.5) -- (4.25,1.5) -- (4,1.75) -- cycle;
\filldraw[thick,fill=white] (4.75,0.5) -- (5,0.5) -- (5,0.25) -- cycle;
\filldraw[thick,fill=white] (4.75,1.5) -- (5,1.5) -- (5,1.75) -- cycle;
\draw[ultra thick] (4,0.5) -- (4,1.5);
\draw[ultra thick,dotted] (5,0.5) -- (5,1.5);
\draw[thick] (4,-0.1) -- (4,0.5) -- (5,0.5) -- (5,-0.1);
\draw[thick] (4,2.1) -- (4,1.5) -- (5,1.5) -- (5,2.1);
\Vertex[shape = circle,size=0.4,x=5,y=1,label=$y_g$,fontsize=\footnotesize,color=white]{D}
\Text[x=5.4,y=0.8]{,}
\end{tikzpicture}
\label{LgMPU}
\end{equation}
and
\begin{equation}
\begin{tikzpicture}
\Text[x=-0.5,y=1]{$R_g=$}
\filldraw[thick,fill=lightgray] (1,0) -- (1.25,0) -- (1,-0.25) -- cycle;
\filldraw[thick,fill=lightgray] (1,0.75) -- (0.75,0.75) -- (1,0.5) -- cycle;
\filldraw[thick,fill=lightgray] (1,1.25) -- (0.75,1.25) -- (1,1.5) -- cycle;
\filldraw[thick,fill=lightgray] (1,2) -- (1.25,2) -- (1,2.25) -- cycle;
\filldraw[thick,fill=white] (2.25,0) -- (2,0) -- (2.25,-0.25) -- cycle;
\filldraw[thick,fill=white] (2.25,0.75) -- (2.5,0.75) -- (2.25,0.5) -- cycle;
\filldraw[thick,fill=white] (2.25,1.25) -- (2.5,1.25) -- (2.25,1.5) -- cycle;
\filldraw[thick,fill=white] (2.25,2) -- (2,2) -- (2.25,2.25) -- cycle;
\draw[thick] (1,-0.85) -- (1,0) -- (2.25,0) -- (2.25,-0.85);
\draw[thick] (0.25,1.25) -- (1,1.25) -- (1,0.75) -- (0.25,0.75) -- cycle;
\draw[thick] (1,2.6) -- (1,2) -- (2.25,2) -- (2.25,2.6);
\draw[thick] (2.25,1.25) -- (3,1.25) -- (3,0.75) -- (2.25,0.75) -- cycle;
\draw[ultra thick] (1,0) -- (1,0.5) (1,1.5) -- (1,2);
\draw[ultra thick,dotted] (2.25,0) -- (2.25,0.5) (2.25,1.5) -- (2.25,2);
\Vertex[shape = circle,size=0.4,x=1.625,y=0,label=$z_g$,fontsize=\footnotesize,color=white]{A}
\Vertex[shape = circle,size=0.4,x=1,y=-0.55,label=$\rho_g$,fontsize=\footnotesize,color=white]{B}
\Vertex[shape = circle,size=0.35,x=3,y=1,label=$\Sigma$,fontsize=\footnotesize,color=white]{C}
\Text[x=3.5,y=1]{$=$}
\filldraw[thick,fill=lightgray] (4,0.5) -- (4.25,0.5) -- (4,0.25) -- cycle;
\filldraw[thick,fill=lightgray] (4,1.5) -- (4.25,1.5) -- (4,1.75) -- cycle;
\filldraw[thick,fill=white] (4.75,0.5) -- (5,0.5) -- (5,0.25) -- cycle;
\filldraw[thick,fill=white] (4.75,1.5) -- (5,1.5) -- (5,1.75) -- cycle;
\draw[ultra thick] (4,0.5) -- (4,1.5);
\draw[ultra thick,dotted] (5,0.5) -- (5,1.5);
\draw[thick] (4,-0.1) -- (4,0.5) -- (5,0.5) -- (5,-0.1);
\draw[thick] (4,2.1) -- (4,1.5) -- (5,1.5) -- (5,2.1);
\Vertex[shape = circle,size=0.4,x=4,y=1,label=$x_g$,fontsize=\footnotesize,color=white]{D}
\Text[x=5.4,y=0.8]{.}
\end{tikzpicture}
\label{RgMPU}
\end{equation}
Here we have used the definition properties of singular-value decomposition
\begin{equation}
\begin{tikzpicture}
\draw[thick] (-0.6,-0.25) rectangle (0,0.25);
\draw[thick,fill=lightgray] (-0.25,-0.25) -- (0,-0.25) -- (0,-0.5) -- cycle;
\draw[thick,fill=lightgray] (-0.25,0.25) -- (0,0.25) -- (0,0.5) -- cycle;
\draw[ultra thick] (0,-0.8) -- (0,-0.5) (0,0.8) -- (0,0.5);
\Text[x=0.5]{$=$};
\draw[ultra thick] (1,-0.8) -- (1,0.8);
\Text[x=1.25,y=-0.2]{,}
\draw[thick] (2,-0.25) rectangle (2.6,0.25);
\Vertex[shape = circle,size=0.35,x=2.6,label=$\Sigma$,fontsize=\footnotesize,color=white]{A}
\draw[thick,fill=white] (2,0.25) -- (2.25,0.25) -- (2,0.5) -- cycle;
\draw[thick,fill=white] (2,-0.25) -- (2.25,-0.25) -- (2,-0.5) -- cycle;
\draw[ultra thick,dotted] (2,-0.5) -- (2,-0.8) (2,0.5) -- (2,0.8);
\Text[x=3.1]{$=$};
\draw[ultra thick,dotted] (3.6,-0.8) -- (3.6,0.8);
\end{tikzpicture}
\end{equation}
and the symmetry constraint (complementary to Eq.~(\ref{rhoxyz}))
\begin{equation}
\begin{tikzpicture}[scale=0.8]
\filldraw[thick,fill=white] (0,0) -- (-0.3,0) -- (0,-0.3) -- cycle;
\draw[thick] (-0.6,0) -- (-0.3,0) (0,-0.6) -- (0,-0.3);
\draw[ultra thick,dotted] (0,0) -- (0,0.7);
\draw[thick,fill=white] (0,0.34) circle (0.22);
\Text[y=0.34,fontsize=\footnotesize]{$y_g$}
\Text[x=0.58]{$=$}
\draw[thick] (0.98,0) -- (1.7,0) (2,-0.3) -- (2,-1);
\draw[ultra thick,dotted] (2,0) -- (2,0.3);
\filldraw[thick,fill=white] (2,0) -- (1.7,0) -- (2,-0.3) -- cycle;
\draw[thick,fill=white] (1.34,0) circle (0.22);
\draw[thick,fill=white] (2,-0.64) circle (0.22);
\Text[x=1.34,fontsize=\footnotesize]{$z^\dag_g$}
\Text[x=2,y=-0.64,fontsize=\footnotesize]{$\rho_g$}
\Text[x=2.5,y=-0.25]{,}
\filldraw[thick,fill=lightgray] (4,0) -- (4.3,0) -- (4,-0.3) -- cycle;
\draw[thick] (4.3,0) -- (4.6,0) (4,-0.6) -- (4,-0.3);
\draw[ultra thick] (4,0) -- (4,0.7);
\draw[thick,fill=white] (4,0.34) circle (0.22);
\Text[x=4,y=0.34,fontsize=\footnotesize]{$x_g$}
\Text[x=5]{$=$}
\draw[thick] (5.8,0) -- (6.5,0) (5.5,-0.3) -- (5.5,-1);
\draw[ultra thick] (5.5,0) -- (5.5,0.3);
\filldraw[thick,fill=lightgray] (5.5,0) -- (5.8,0) -- (5.5,-0.3) -- cycle;
\draw[thick,fill=white] (6.14,0) circle (0.22);
\draw[thick,fill=white] (5.5,-0.64) circle (0.22);
\Text[x=6.14,fontsize=\footnotesize]{$z_g$}
\Text[x=5.5,y=-0.64,fontsize=\footnotesize]{$\rho_g$}
\Text[x=6.75,y=-0.25]{.}
\end{tikzpicture}
\end{equation}
Note that Eqs.~(\ref{LgMPU}) and (\ref{RgMPU}) are nothing but $L_g=u^\dag(x_e\otimes y_g)u$ and $R_g=u^\dag(x_g\otimes y_e)u$. It is worth mentioning that, by evolving a more general string operator $g^{\otimes N}\otimes h^{\otimes L-N}$ with $g,h\in G$, we will obtain $D_{g,h}=u^\dag(x_g\otimes y_h)u$ near the domain wall where the left is $g$'s and the right is $h$'s. Note that $L_g=D_{e,g}$ and $R_g=D_{g,e}$.

As a byproduct of Eq.~(\ref{LgRg}), we can compute the relative SPI from a single $\mathcal{U}$, which is not necessarily simple. For $\forall g\in G$ with $\chi_g\neq0$, replacing $\mathcal{U}$ and $\rho_g$ in Eq.~(\ref{LgRg}) with $\mathcal{U}_k$ (which is simple) and $\rho^{\otimes k}_g$, respectively, followed by taking the trace and using
\begin{equation}
\begin{tikzpicture} 
\draw[thick] (0.5,0) -- (-0.5,0) -- (-0.5,1) -- (0.5,1);
\draw[thick] (0,-0.35) -- (0,1.35);
\draw[thick] (0,1.35) .. controls (-0,1.55) and (-0.1,1.55) .. (-0.1,1.35);
\draw[thick] (0,-0.35) .. controls (-0,-0.55) and (-0.1,-0.55) .. (-0.1,-0.35);
\Vertex[shape = rectangle,size=0.5,x=0,label=$\mathcal{U}$,fontsize=\normalsize,color=white]{A}
\Vertex[shape = rectangle,size=0.5,x=0,y=1,label=$\overline{\mathcal{U}}$,fontsize=\normalsize,color=white]{B}
\Text[x=0.95,y=0.5]{$=\;d$}
\draw[thick] (2,0) -- (1.5,0) -- (1.5,1) -- (2,1);
\end{tikzpicture}
\end{equation}
$k$ times, we obtain
\begin{equation}
\begin{split}
\Tr L_g&=d^k\chi^k_g\Tr\Sigma_g,
\end{split}
\label{LgSigma}
\end{equation}
where $\Sigma_g\equiv\mathcal{E}^k_g(z^\dag_g\Sigma)$ and
\begin{equation}
\begin{tikzpicture}
\Text[x=-1.5,y=0.625]{$\mathcal{E}_g\equiv\chi_g^{-1}$}
\draw[thick] (-0.75,1.25) -- (0.75,1.25);
\draw[thick] (-0.75,0) -- (0.75,0);
\draw[thick] (0,-0.35) -- (0,1.6);
\draw[thick] (0,1.6) .. controls (-0,1.8) and (-0.1,1.8) .. (-0.1,1.6);
\draw[thick] (0,-0.35) .. controls (-0,-0.55) and (-0.1,-0.55) .. (-0.1,-0.35);
\Vertex[shape = rectangle,size=0.5,x=0,label=$\mathcal{U}$,fontsize=\normalsize,color=white]{A}
\Vertex[shape = rectangle,size=0.5,x=0,y=1.25,label=$\overline{\mathcal{U}}$,fontsize=\normalsize,color=white]{B}
\Vertex[shape = circle,size=0.4,x=0,y=0.625,label=$\rho_g$,fontsize=\footnotesize,color=white]{C}
\Text[x=1,y=0.5]{.}
\end{tikzpicture}
\end{equation}
Using the fact that for $\forall k\ge k_0$ ($k_0$: smallest integer such that $\mathcal{U}_{k_0}$ is simple)
\begin{equation}
\begin{tikzpicture}
\draw[thick] (0,-0.5) -- (0,1.5);
\draw[thick] (-0.5,1) -- (0.5,1) -- (0.5,0) -- (-0.5,0);
\Vertex[shape = rectangle,size=0.5,x=0,label=$\mathcal{U}_k$,fontsize=\small,color=white]{A}
\Vertex[shape = rectangle,size=0.5,x=0,y=1,label=$\overline{\mathcal{U}}_k$,fontsize=\small,color=white]{B}
\Vertex[shape = circle,size=0.4,x=0.5,y=0.5,label=$\Sigma$,fontsize=\footnotesize,color=white]{C}
\Text[x=1,y=0.5]{$=$}
\draw[thick] (1.5,1) -- (2.5,1) -- (2.5,0) -- (1.5,0);
\draw[thick] (2,-0.5) -- (2,1.5);
\Vertex[shape = rectangle,size=0.5,x=2,label=$\;\mathcal{U}_{k_0}$,fontsize=\footnotesize,color=white]{a}
\Vertex[shape = rectangle,size=0.5,x=2,y=1,label=$\;\overline{\mathcal{U}}_{k_0}$,fontsize=\footnotesize,color=white]{b}
\Vertex[shape = circle,size=0.4,x=2.5,y=0.5,label=$\Sigma$,fontsize=\footnotesize,color=white]{c}
\draw[thick] (3,-0.5) -- (3,1.5);
\Text[x=3.2,y=1.7,fontsize=\tiny]{$d^{k-k_0}$}
\Text[x=3.4,y=0.25]{,}
\end{tikzpicture}
\end{equation}
 which can be obtained from contracting the rightmost $k_0$ identities in
 \begin{widetext}
\begin{equation}
\begin{tikzpicture}
\draw[thick] (1.5,1) -- (2.5,1) -- (2.5,0) -- (1.5,0);
\draw[thick] (2,-0.5) -- (2,1.5);
\Vertex[shape = rectangle,size=0.5,x=2,label=$\;\mathcal{U}_k$,fontsize=\small,color=white]{a}
\Vertex[shape = rectangle,size=0.5,x=2,y=1,label=$\;\overline{\mathcal{U}}_k$,fontsize=\small,color=white]{b}
\Vertex[shape = circle,size=0.4,x=2.5,y=0.5,label=$\Sigma$,fontsize=\footnotesize,color=white]{c}
\draw[thick] (3,-0.5) -- (3,1.5);
\Text[x=3.15,y=1.7,fontsize=\tiny]{$d^{k_0}$}
\end{tikzpicture}
\begin{tikzpicture}
\Text[x=-0.75,y=0.5]{$=$}
\draw[thick] (0,-0.5) -- (0,1.5);
\draw[thick] (-0.5,1) -- (0.5,1) -- (0.5,0) -- (-0.5,0);
\Vertex[shape = rectangle,size=0.5,x=0,label=$\mathcal{U}_k$,fontsize=\small,color=white]{A}
\Vertex[shape = rectangle,size=0.5,x=0,y=1,label=$\overline{\mathcal{U}}_k$,fontsize=\small,color=white]{B}
\Vertex[shape = circle,size=0.4,x=0.5,y=0.5,label=$\Sigma$,fontsize=\footnotesize,color=white]{C}
\draw[thick] (0.8,1) -- (1.8,1) -- (1.8,0) -- (0.8,0) -- cycle;
\draw[thick] (1.3,-0.5) -- (1.3,1.5);
\Vertex[shape = rectangle,size=0.5,x=1.3,label=$\;\mathcal{U}_{k_0}$,fontsize=\footnotesize,color=white]{a}
\Vertex[shape = rectangle,size=0.5,x=1.3,y=1,label=$\;\overline{\mathcal{U}}_{k_0}$,fontsize=\footnotesize,color=white]{b}
\Vertex[shape = circle,size=0.4,x=1.8,y=0.5,label=$\Sigma$,fontsize=\footnotesize,color=white]{c}
\end{tikzpicture}
\begin{tikzpicture}
\Text[x=-0.75,y=0.5]{$=$}
\draw[thick] (0,-0.5) -- (0,1.5);
\draw[thick] (-0.5,1) -- (0.5,1) -- (0.5,0) -- (-0.5,0);
\Vertex[shape = rectangle,size=0.5,x=0,label=$\mathcal{U}_{k'}$,fontsize=\footnotesize,color=white]{A}
\Vertex[shape = rectangle,size=0.5,x=0,y=1,label=$\overline{\mathcal{U}}_{k'}$,fontsize=\footnotesize,color=white]{B}
\Vertex[shape = circle,size=0.4,x=0.5,y=0.5,label=$\Sigma$,fontsize=\footnotesize,color=white]{C}
\end{tikzpicture}
\begin{tikzpicture}
\Text[x=-0.75,y=0.5]{$=$}
\draw[thick] (0,-0.5) -- (0,1.5);
\draw[thick] (-0.5,1) -- (0.5,1) -- (0.5,0) -- (-0.5,0);
\Vertex[shape = rectangle,size=0.5,x=0,label=$\mathcal{U}_{k_0}$,fontsize=\footnotesize,color=white]{A}
\Vertex[shape = rectangle,size=0.5,x=0,y=1,label=$\overline{\mathcal{U}}_{k_0}$,fontsize=\footnotesize,color=white]{B}
\Vertex[shape = circle,size=0.4,x=0.5,y=0.5,label=$\Sigma$,fontsize=\footnotesize,color=white]{C}
\draw[thick] (0.8,1) -- (1.8,1) -- (1.8,0) -- (0.8,0) -- cycle;
\draw[thick] (1.3,-0.5) -- (1.3,1.5);
\Vertex[shape = rectangle,size=0.5,x=1.3,label=$\;\mathcal{U}_k$,fontsize=\small,color=white]{a}
\Vertex[shape = rectangle,size=0.5,x=1.3,y=1,label=$\;\overline{\mathcal{U}}_k$,fontsize=\small,color=white]{b}
\Vertex[shape = circle,size=0.4,x=1.8,y=0.5,label=$\Sigma$,fontsize=\footnotesize,color=white]{c}
\end{tikzpicture}
\begin{tikzpicture}
\Text[x=1.25,y=0.5]{$=$}
\draw[thick] (1.5,1) -- (2.5,1) -- (2.5,0) -- (1.5,0);
\draw[thick] (2,-0.5) -- (2,1.5);
\Vertex[shape = rectangle,size=0.5,x=2,label=$\;\mathcal{U}_{k_0}$,fontsize=\footnotesize,color=white]{a}
\Vertex[shape = rectangle,size=0.5,x=2,y=1,label=$\;\overline{\mathcal{U}}_{k_0}$,fontsize=\footnotesize,color=white]{b}
\Vertex[shape = circle,size=0.4,x=2.5,y=0.5,label=$\Sigma$,fontsize=\footnotesize,color=white]{c}
\draw[thick] (3,-0.5) -- (3,1.5);
\Text[x=3.1,y=1.7,fontsize=\tiny]{$d^k$}
\end{tikzpicture}
\end{equation}
\end{widetext}
with $k'=k+k_0$, as well as the symmetry requirement, we know that $\Sigma_g=\mathcal{E}^{k_0}_g(z^\dag_g\Sigma)$ and satisfies $\mathcal{E}_g(\Sigma_g)=\Sigma_g$. On the other hand, due to $\Tr \mathcal{E}^N_g=1$ for $\forall N\in\mathbb{Z}^+$, the fixed point of $\mathcal{E}_g$ is unique (just like $\mathcal{E}=\mathcal{E}_e$). Therefore, we can determine $\Sigma_g$ by solving $\mathcal{E}_g(\Sigma_g)=\Sigma_g$ subject to $\Tr z_g\Sigma_g=1$, where both $\mathcal{E}_g$ and $z_g$ can be obtained from a single $\mathcal{U}$. According to Eq.~(\ref{LgSigma}) and $\Tr L_g\Tr R_g=d^{2k}\chi^{2k}_g$, the relative SPI is directly related to $\Sigma_g$ by
\begin{equation}
\ind_g-\ind=\log|\Tr\Sigma_g|.
\label{indgSigma}
\end{equation}
Recalling that $\ind$ does not rely on blocking, according to Eq.~(\ref{indgSigma}), we again confirm the blocking independence of $\ind_g$.

\subsection{Derivation of Eq.~(7)}
We first figure out the explicit expression of $\langle X\rangle$ in Fig.~\ref{figS3} for a general bipartite unitary $U$. Recalling that the controlled-SWAP gate reads
\begin{equation}
U_{\rm CS}=|0\rangle\langle0|\otimes \mathbb{1}_{A'A}+|1\rangle\langle1|\otimes\mathbb{S},
\end{equation}
where $\mathbb{S}\equiv\sum^{d_A}_{j,j'=1}|j_Aj'_{A'}\rangle\langle j'_Aj_{A'}|=\mathbb{S}^\dag$, we obtain the unitary evolution of the entire system, including the bipartite physical system $A\bigcup B$ with interest, a copy $A'$ and an auxiliary qubit, as
\begin{equation}
\begin{split}
U_{\rm tot}&=
U_{\rm CS}(X\otimes U)U_{\rm CS}H\\
&=
|1\rangle\langle+|\otimes\mathbb{S}U+|0\rangle\langle-|\otimes U\mathbb{S},
\end{split}
\label{Utot}
\end{equation}
where $|\pm\rangle\equiv\frac{1}{\sqrt{2}}(|0\rangle\pm|1\rangle)$, $X\equiv|1\rangle\langle0|+|0\rangle\langle1|$, and $H\equiv\frac{1}{\sqrt{2}}(X+Z)$ (Hadamard gate) with $Z\equiv|0\rangle\langle0|-|1\rangle\langle1|$. At the initial time, we prepare the qubit in the pure state $|0\rangle$ and the remaining systems in the infinite-temperature state $\rho_\infty\equiv d^{-2}_Ad^{-1}_B\mathbb{1}_{A'AB}$. After evolving the entire system by $U_{\rm tot}$ (\ref{Utot}), we measure the qubit under the $X$ basis, so the expectation value $\langle X\rangle$ should be
\begin{equation}
\begin{split}
\langle X\rangle&=\Tr[XU_{\rm tot}(|0\rangle\langle0|\otimes\rho_\infty)U^\dag_{\rm tot}]\\
&={\rm Re}\Tr[XU_{\rm tot}(|+\rangle\langle-|\otimes\rho_\infty)U^\dag_{\rm tot}]\\
&=d^{-2}_Ad^{-1}_B\Tr[U\mathbb{S}U^\dag\mathbb{S}],
\end{split}
\label{Xexp}
\end{equation}
where we have used $|0\rangle\langle0|=\frac{1}{2}(\mathbb{1}_{\rm qubit}+|-\rangle\langle+|+|+\rangle\langle-|)$ and $\Tr[\mathbb{S}U\mathbb{S}U^\dag]=\Tr[U\mathbb{S}U^\dag\mathbb{S}]$. Combining Eq.~(\ref{Xexp}) with $\Tr_B[\Tr_A U\Tr_A U^\dag]=\Tr[U\mathbb{S}U^\dag\mathbb{S}]$ and $U=U_A\otimes U_B$, we end up with
\begin{equation}
|\Tr U_A|=d_A\sqrt{\langle X\rangle}.
\end{equation}

\begin{figure}
\begin{center}
\begin{tikzpicture}[scale=0.9]
\draw[thick] (1,0) -- (7,0) (1,0.7) -- (7,0.7) (1,1.4) -- (7,1.4) (1,2.1) -- (7,2.1);
\fill[black] (2.9,2.1) circle (0.075) (5.6,2.1) circle (0.075);
\draw[thick] (2.9,2.1) -- (2.9,1.7) (5.6,2.1) -- (5.6,1.7);
\draw[thick,fill=green!5] (1.5,1.8) rectangle (2.1,2.4);
\draw[thick,fill=green!5] (3.95,1.8) rectangle (4.55,2.4);
\draw[thick,fill=green!20] (2.6,0.4) rectangle (3.2,1.7);
\draw[thick,fill=green!20] (5.3,0.4) rectangle (5.9,1.7);
\draw[thick,fill=blue!15] (3.6,-0.3) rectangle (4.9,1);
\draw[thick,fill=gray!10] (6.2,1.8) rectangle (7.2,2.4);
\draw[thick] (6.5,2) .. controls (6.6,2.2) and (6.8,2.2).. (6.9,2);
\draw[thick] (6.7,2) -- (6.8,2.2);
\Text[x=1.8,y=2.1,fontsize=\large]{$H$}
\Text[x=4.25,y=2.1,fontsize=\large]{$X$}
\Text[x=2.9,y=1.05,fontsize=\large]{$\mathbb{S}$}
\Text[x=5.6,y=1.05,fontsize=\large]{$\mathbb{S}$}
\Text[x=4.25,y=0.35,fontsize=\large]{$U$}
\Text[x=7.6,y=2.1]{$\langle X\rangle$}
\Text[x=0.65,y=2.1]{$|0\rangle$}
\Text[x=0.15,y=1.4]{$\rho_{A'}=\frac{\mathbb{1}_{A'}}{d_A}$}
\Text[x=0.2,y=0.7]{$\rho_A=\frac{\mathbb{1}_A}{d_A}$}
\Text[x=0.2,y=0]{$\rho_B=\frac{\mathbb{1}_B}{d_B}$}
\end{tikzpicture}
       \end{center}
   \caption{Interferometric scheme for measuring $\Tr_B[\Tr_AU\Tr_AU^\dag]$.}
   \label{figS3}
\end{figure}

To measure the relative SPI, we should implement $U$ as $U^\dag_{\rm MPU}\rho^{\otimes N}_gU_{\rm MPU}$. Thanks to the locality-preserving property of MPUs, we can choose $A$ to be as small as $2k_0\sim O(1)$ sites ($k_0$ is the smallest $k$ such that $\mathcal{U}_k$ is simple) across the left domain wall between $\rho_g$ and $\mathbb{1}$. The controlled-SWAP gate between $A$ and its copy can be decomposed into $2k_0$ individual two-site controlled-SWAP gates acting only on the $j$th site of $A$ and that of the copy. Once we succeed in measuring $\langle X\rangle$, we obtain 
\begin{equation}
|\Tr L_g|=d^{2k_0}\sqrt{\langle X\rangle}, 
\label{Lgk0}
\end{equation}
so the relative SPI reads
\begin{equation}
{\rm ind}_g-\ind=\log\frac{|\Tr L_g|}{d^{k_0}|\chi_g|^{k_0}}=\frac{1}{2}\log\langle X\rangle+k_0\log\frac{d}{|\chi_g|}.
\label{indgk0}
\end{equation}
By choosing $A$ as $2k$ sites ($k\ge k_0$) across the left edge of the symmetry string operator, following a similar analysis, we can obtain Eqs.~(\ref{Lgk0}) and (\ref{indgk0}) with $k_0$ replaced by $k$ (note that $\langle X\rangle$ also depends on $k$). This is why ${\rm ind}_g$ is always measurable from linear fitting even if $d$ and $\chi_g$ are unknown.


\section{SPIs for inhomogeneous locality-preserving unitaries}
In this section, we rigorously derive the factorization relation for the evolved symmetry string operator solely from the symmetry and the (strict) locality-preserving requirement. In analogy to the SPIs for MPUs, we prove that the (relative) SPIs defined from $L_g$ and $R_g$ in the factorization relation 
are topological invariants. The same technique also allows the generalization of the cohomology class to inhomogeneous locality-preserving unitaries.

\subsection{
Factorization relation}
We first introduce two useful lemmas:
\begin{lemma}
Given a unitary $U$ acting on a bipartite system $A\bigcup B$, then $U=U_A\otimes U_B$ for some subsystem unitaries $U_A$ and $U_B$ if and only if $[U^\dag(O_A\otimes\mathbb{1}_B)U,\mathbb{1}_A\otimes O_B]=[O_A\otimes\mathbb{1}_B,U^\dag(\mathbb{1}_A\otimes O_B)U]=0$ for any subsystem operators $O_A$ and $O_B$.
\label{comfac}
\end{lemma}
\emph{Proof:} ``Only if" is trivial. To show ``if", we first note that $[U^\dag(O_A\otimes\mathbb{1}_B)U,\mathbb{1}_A\otimes O_B]=0$ for arbitrary $O_B$ acting on $B$ is equivalent to $U^\dag(O_A\otimes\mathbb{1}_B)U=f_A(O_A)\otimes\mathbb{1}_B$ (see, e.g., Lemma 1.5 in Ref.~\cite{Christandl2006}). Moreover, $f_A$ is a ring automorphism on ${\rm M}_{d_A}(\mathbb{C})$, 
since it is a ring homomorphism and is bijective due to $U(O_A\otimes\mathbb{1}_B)U^\dag=f^{-1}_A(O_A)\otimes\mathbb{1}_B$. According to the Skolem-Noether theorem, we must have $f_A(O_A)=V^{-1}_AO_AV_A$ for some invertible $V_A\in{\rm M}_{d_A}(\mathbb{C})$. Similarly, we must have 
$U^\dag(\mathbb{1}_A\otimes O_B)U=\mathbb{1}_A\otimes V^{-1}_BO_BV_B$. Therefore, for an arbitrary operator $O=\sum_j O_{A,j}\otimes O_{B,j}$ acting on the entire system, we have $U^\dag OU=\sum_jU^\dag(O_{A,j}\otimes\mathbb{1}_B)UU^\dag(\mathbb{1}_A\otimes O_{B,j})U=(V_A\otimes V_B)^{-1}O(V_A\otimes V_B)$ and thus $(V_A\otimes V_B)U^\dag=c\mathbb{1}_{AB}$ with $c\neq0$. Absorbing $c^{-1}$ into $V_A$ or $V_B$ followed by choosing a proper $\mathbb{C}^\times$ gauge, we end up with $U=U_A\otimes U_B$ where $U_A$ and $U_B$ are subsystem unitaries. \hfill$\Box$
\begin{lemma}
Given a unitary $U$ acting on an $M$-partite system $S=\bigcup^M_{m=1}S_j$, then $U=\bigotimes^M_{m=1}U_j$ if and only if $U=U_m\otimes U_{\bar m}$ for $m=1,2,...,M-1$, where $U_m$ and $U_{\bar m}$ are unitaries acting only on $S_m$ and $S_{\bar m}\equiv S\backslash S_m$, respectively.
\label{Ufac}
\end{lemma}
\emph{Proof:} ``Only if" is trivial. To show ``if", we start from $U=U_1\otimes U_{\bar 1}$ and prove $U_{\bar 1}=U_2\otimes U_{\overline{12}}$. Focusing on the singular-value decomposition with respect to the bipartition $S=S_2\bigcup S_{\bar 2}$, we know from $U=U_2\otimes U_{\bar 2}$ that the bond dimension is one, implying $U_{\bar 1}=V_2\otimes V_{\overline{12}}$. Moreover, from the fact that $\mathbb{1}_S=U^\dag U=U^\dag_2V_2\otimes[U^\dag_{\bar 2}(U_1\otimes V_{\overline{12}})]$, we can properly choose the $\mathbb{C}^\times$ gauge such that $V_2=U_2$, which in turn implies $U_{\overline{12}}\equiv V_{\overline{12}}$ is unitary. Following a similar analysis, we can factorize $U_{\overline{12}}$ into $U_3\otimes U_{\overline{123}}$ and 
so on, and end up with $U=\bigotimes^M_{m=1}U_m$. \hfill$\Box$

Before deriving the factorization relation, we list a few fundamental properties of locality-preserving unitaries: 
\begin{proposition}
For a locality-preserving unitary $U_{\rm LP}$ acting on a ring of $L$ spins and with Lieb-Robinson length $l_{\rm LR}$, we have \newline
(i) If $O$ 
acts nontrivially only on 
$[j_{\rm l},j_{\rm r}]\subset\mathbb{Z}_L$, then $U^\dag_{\rm LP}O
U_{\rm LP}$ is nontrivial at most on $[j_{\rm l}-l_{\rm LR},j_{\rm r}+l_{\rm LR}]$; \newline
(ii) If $U_{\rm LP}=U_{\rm LP,1}U_{\rm LP,2}$, where the Lieb-Robinson length of $U_{{\rm LP},a}$ is $l_{{\rm LR},a}$ ($a=1,2$), then $l_{\rm LR}\le l_{{\rm LR},1}+l_{{\rm LR},2}$;  \newline
(iii) $U^\dag_{\rm LP}$ has the same Lieb-Robinson length $l_{\rm LR}$.
\label{ULPpro}
\end{proposition}
\emph{Proof:} 
From the expansion $O
=\sum_{\{n_j\}^{j_{\rm r}}_{j=j_{\rm l}}}\bigotimes^{j_{\rm r}}_{j=j_{\rm l}}O^{[n_j]}_j$, where $O^{[n_j]}_j$ acts nontrivially only on the $j$th site, we obtain (i). Taking $O=U^\dag_{LP,1}O_jU_{LP,1}$ in (i) for an arbitrary operator $O_j$ acting nontrivially only on the $j$th site, we know that $U^\dag_{\rm LP,2}U^\dag_{\rm LP,1}O_jU_{\rm LP,1}U_{\rm LP,2}$ is nontrivial at most on $[j-l_{\rm LR,1}-l_{\rm LR,2},j+l_{\rm LR,1}+l_{\rm LR,2}]$ for $\forall j\in\mathbb{Z}_L$ and (ii) follows.To show (iii), we consider all the operators $O$ acting nontrivially only on $\mathbb{Z}_L\backslash[j-l_{\rm LR},j+l_{\rm LR}]$, so that $[U^\dag_{\rm LP}OU_{\rm LP},O_j]=[O,U_{\rm LP}O_jU^\dag_{\rm LP}]=0$ due to (i) ($O_j$ follows the previous notation). The arbitrariness of $O$ implies that $U_{\rm LP}O_jU^\dag_{\rm LP}$ acts nontrivially at most on $[j-l_{\rm LR},j+l_{\rm LR}]$. 
Hence, denoting the Lieb-Robinson length of $U^\dag_{\rm LP}$ as $\tilde l_{\rm LR}$, we have $\tilde l_{\rm LR}\le l_{\rm LR}$. Similarly, we can derive $l_{\rm LR}\le\tilde l_{\rm LR}$ from $(U^\dag_{\rm LP})^\dag=U_{\rm LP}$. Combining these two relations, we obtain (iii). \hfill$\Box$

With all the previous results in hand, we are ready to prove
\begin{theorem}[Factorization relation]
Given a $G$-symmetric locality-preserving unitary $U_{\rm LP}$ with Lieb-Robinson length $l_{\rm LR}$, for $\forall g\in G$ defining a $g$-string operator $\rho_{g[j_{\rm l},j_{\rm r}]}\equiv\bigotimes_{j\in[j_{\rm l},j_{\rm r}]}\rho_g
$ with $|j_{\rm l}-j_{\rm r}|\ge4l_{\rm LR}$, we have (with $\mathbb{1}$'s omitted for simplicity)
\begin{equation}
U^\dag_{\rm LP}\rho_{g[j_{\rm l},j_{\rm r}]}U_{\rm LP}=L_g(j_{\rm l})\otimes\rho_{g[j_{\rm l}+l_{\rm LR},j_{\rm r}-l_{\rm LR}]}\otimes R_g(j_{\rm r}),
\label{gfac}
\end{equation}
where $L_g(j_{\rm l})$ and $R_g(j_{\rm l})$ are unitary operators acting nontrivially (at most) on $I_{\rm L}=[j_{\rm l}-l_{\rm LR},j_{\rm l}+l_{\rm LR}-1]$ and $I_{\rm R}=[j_{\rm r}-l_{\rm LR}+1,j_{\rm r}+l_{\rm LR}]$, respectively.
\label{facre}
\end{theorem}
\emph{Proof:} We divide the lattice $\mathbb{Z}_L$ into three subsystems: $S_1\equiv\mathbb{Z}_L\backslash[j_{\rm l}-l_{\rm LR},j_{\rm r}+l_{\rm LR}]$, $S_2\equiv[j_{\rm l}+l_{\rm LR},j_{\rm r}-l_{\rm LR}]$ and $S_3\equiv I_{\rm L}\bigcup I_{\rm R}$. 
Applying Prop.~\ref{ULPpro}(i) directly to the LHS of Eq.~(\ref{gfac}), we obtain $U^\dag_{\rm LP}\rho_{g[j_{\rm l},j_{\rm r}]}U_{\rm LP}=U_1\otimes U_{\bar 1}$ with $U_1=\mathbb{1}_{S_1}$. Since $U_{\rm LP}$ is $G$-symmetric, we can rewrite the evolved $g$-string operator into 
\begin{equation}
U^\dag_{\rm LP}\rho_{g[j_{\rm l},j_{\rm r}]}U_{\rm LP}=\rho^{\otimes L}_gU^\dag_{\rm LP}\rho^\dag_{g\mathbb{Z}_L\backslash[j_{\rm l},j_{\rm r}]}U_{\rm LP}.
\label{facsym}
\end{equation}
Applying Prop.~\ref{ULPpro}(i) to the RHS of Eq.~(\ref{facsym}), we obtain $U^\dag_{\rm LP}\rho_{g[j_{\rm l},j_{\rm r}]}U_{\rm LP}=U_2\otimes U_{\bar 2}$ with $U_2=\rho^{\otimes(j_{\rm r}-j_{\rm l}-2l_{\rm LR}+1)}_g$. According to Lemma~\ref{Ufac}, there exists a unitary $U_3$ acting on $S_3$ such that $U^\dag_{\rm LP}\rho_{g[j_{\rm l},j_{\rm r}]}U_{\rm LP}=\bigotimes^3_{m=1}U_m$. 

We move on to show that $U_3$ can further be factorized. According to Prop.~\ref{ULPpro}(ii) and (iii), the Lieb-Robinson length of $U^\dag_{\rm LP}\rho_{g[j_{\rm l},j_{\rm r}]}U_{\rm LP}=\bigotimes^3_{m=1}U_m$ is no more than $2l_{\rm LR}$. Since $|j_{\rm l}-j_{\rm r}|\ge 4l_{\rm LR}$, given any two operators $O_{I_{\rm L}}$ and $O_{I_{\rm R}}$ acting nontrivially only on $I_{\rm L}$ and $I_{\rm R}$, respectively, we have $[U^\dag_3(O_{I_{\rm L}}\otimes\mathbb{1}_{I_{\rm R}})U_3,\mathbb{1}_{I_{\rm L}}\otimes O_{I_{\rm R}}]=[O_{I_{\rm L}}\otimes\mathbb{1}_{I_{\rm R}},U^\dag_3(\mathbb{1}_{I_{\rm L}}\otimes O_{I_{\rm R}})U_3]=0$. It follows from Lemma~\ref{comfac} that $U_3=L_g(j_{\rm l})\otimes R_g(j_{\rm r})$ for two unitaries $L_g(j_{\rm l})$ and $R_g(j_{\rm r})$ acting on $I_{\rm L}$ and $I_{\rm R}$, respectively. Substituting the expressions of $U_m$ ($m=1,2,3$) into $\bigotimes^3_{m=1}U_m$ yields the RHS of Eq.~(\ref{gfac}). \hfill$\Box$

\begin{figure}
\begin{center}
       \includegraphics[width=8.5cm, clip]{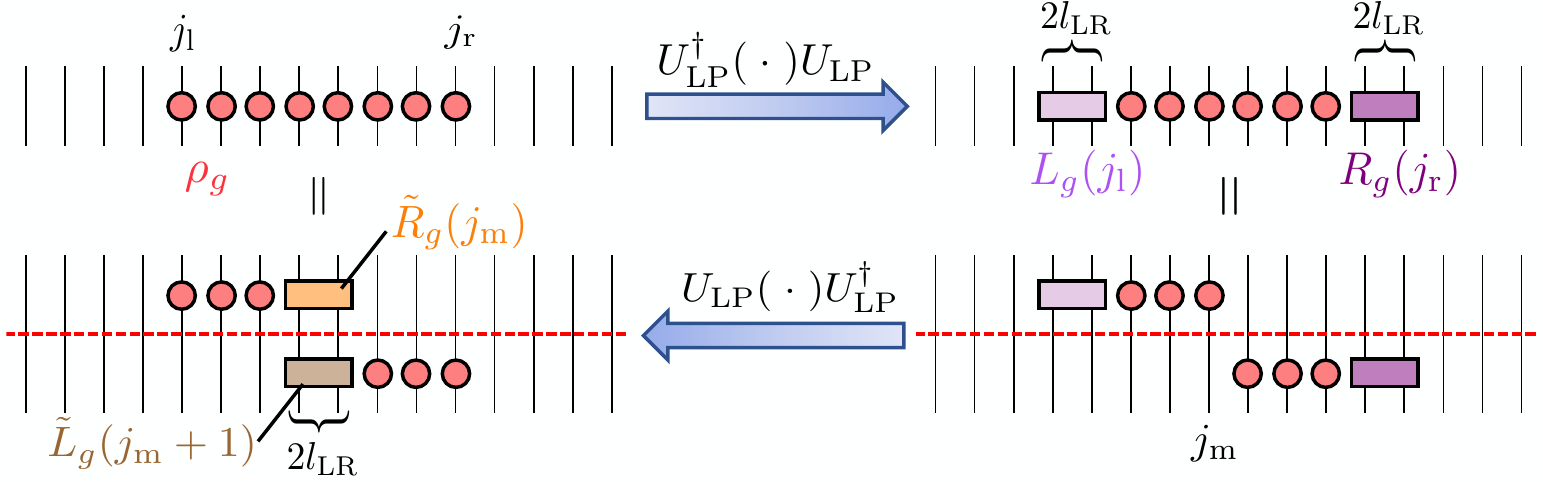}
       \end{center}
   \caption{Relations between $g$, $L_g$, $R_g$, $\tilde L_g$ and $\tilde R_g$ through a symmetric inhomogeneous locality-preserving unitary $U_{\rm LP}$ with $l_{\rm LR}=1$, which is always the case after sufficient blocking. Trace preservation under unitary transforms implies $\Tr L_g(j_{\rm l})=\Tr\tilde R_g(j_{\rm m})$ and $\Tr R_g(j_{\rm l})=\Tr\tilde L_g(j_{\rm m}+1)$. Note that the parts above and below the red dashed line are transformed separately.}
   \label{figS5}
\end{figure}

\subsection{Basic properties} 
We first mention that the robustness of $\ind$ against disorder and its additivity are well established in Ref.~\cite{Po2016}, so it is sufficient to focus on the relative SPI $\ind_g-\ind$. By sufficient we mean that once we know the relative SPI is an additive topological invariant, so is the SPI.
\begin{proposition}
The relative SPI
\begin{equation}
\ind_g-\ind\equiv\frac{1}{2}\log\left|\frac{\Tr L_g}{\Tr R_g}\right|
\label{RSPI}
\end{equation}
is a well-defined global character, although $L_g$ and $R_g$ are generally site-dependent.
\label{siteindep}
\end{proposition}
\emph{Proof:} 
To prove the site-independence of $|\Tr L_g|$ and $|\Tr R_g|$, we only have to show $\Tr L_g(j_{\rm l})=\Tr\tilde R_g(j_{\rm m})$ and $\Tr R_g(j_{\rm r})=\Tr\tilde L_g(j_{\rm m}+1)$ for $\forall j_{\rm m}\in(j_{\rm l}+l_{\rm LR},j_{\rm r}-l_{\rm LR})$, where $\tilde L_g$ and $\tilde R_g$ are determined from evolving the $g$-string operator by $U^\dag_{\rm LP}$. These two identities stem simply from the preservation of trace under unitary conjugation (see Fig.~\ref{figS5}). As $j_{\rm l}$ and $j_{\rm r}$ are variable for a fixed $j_{\rm m}$, $|\Tr L_g|$ and $|\Tr R_g|$ should be site-independent and ${\rm ind}_g-\ind$ in Eq.~(\ref{RSPI}) is a well-defined global character for $U_{\rm LP}$.

\begin{figure*}
\begin{center}
       \includegraphics[width=14cm, clip]{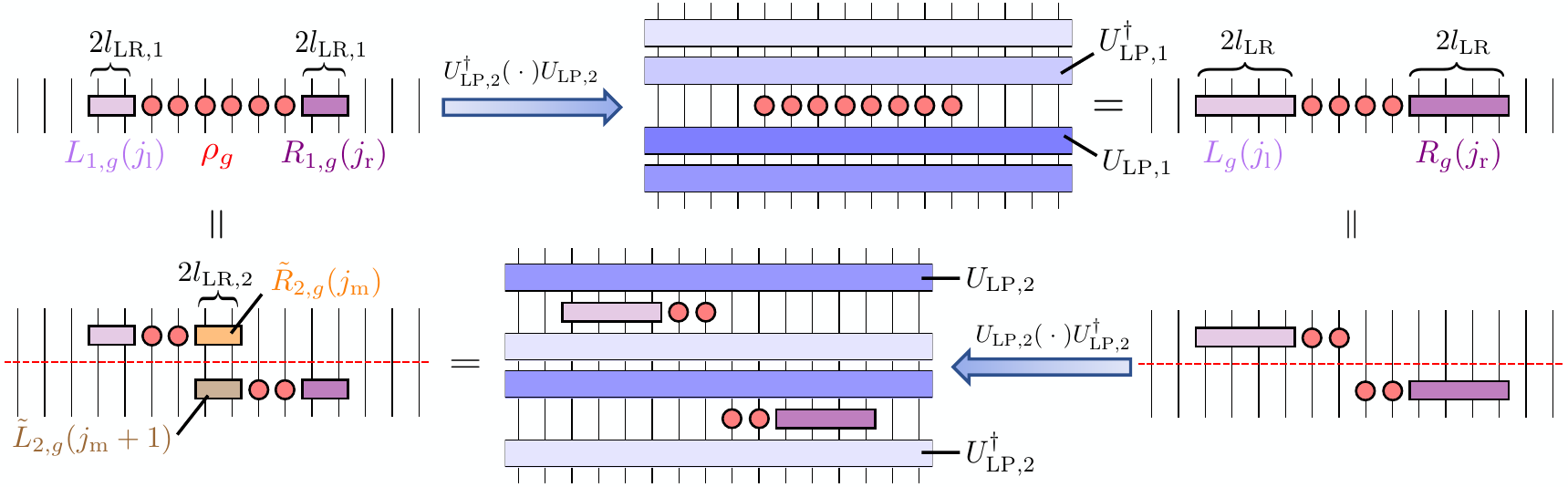}
       \end{center}
   \caption{Schematic illustration of the additivity of the relative SPI and the cohomology class. Up to tensoring with (equal numbers of) $\mathbb{1}$'s and $\rho_g$'s, $L_g(j_{\rm l})$ and $R_g(j_{\rm r})$ of $U_{\rm LP}=U_{\rm LP,1}U_{\rm LP,2}$ are unitarily equivalent to $L_{1,g}(j_{\rm l})\otimes\tilde R_{2,g}(j_{\rm m})$ and $\tilde L_{2,g}(j_{\rm m}+1)\otimes R_{1,g}(j_{\rm r})$, respectively.}
   \label{figS6}
\end{figure*}

It remains to derive the relations in Fig.~\ref{figS5}, where the upper half is nothing but Thm.~\ref{facre}, so we only have to rigorously derive the lower half. To show the relation above the red dashed line, which explicitly reads
\begin{equation}
U_{\rm LP}L_g(j_{\rm l})\otimes\rho_{g[j_{\rm l}+l_{\rm LR},j_{\rm m}]}U^\dag_{\rm LP}=\rho_{g[j_{\rm l},j_{\rm m}-l_{\rm LR}]}\otimes\tilde R_g(j_{\rm m}),
\label{ULUdag}
\end{equation} 
we divide $\mathbb{Z}_L$ into $S_1=\mathbb{Z}_L\backslash[j_{\rm l}-2l_{\rm LR},j_{\rm m}+l_{\rm LR}]$, $S_2=[j_{\rm l}-2l_{\rm LR},j_{\rm m}-l_{\rm LR}]$ and $S_3=[j_{\rm m}-l_{\rm LR}+1,j_{\rm m}+l_{\rm LR}]$. Applying Prop.~\ref{ULPpro}(i) directly to the LHS of Eq.~(\ref{ULUdag}), we obtain the factorization $U_1\otimes U_{\bar 1}$ with $U_1=\mathbb{1}_{S_1}$. Moreover, rewriting Eq.~(\ref{gfac}) into 
\begin{equation}
\begin{split}
&\;\;\;\;\;U_{\rm LP}L_g(j_{\rm l})\otimes\rho_{g[j_{\rm l}+l_{\rm LR},j_{\rm m}]}U^\dag_{\rm LP}\\
&=\rho_{g[j_{\rm l},j_{\rm r}]}U_{\rm LP}\rho^\dag_{g[j_{\rm m}+1,j_{\rm r}-l_{\rm LR}]}\otimes R_g(j_{\rm r})^\dag U^\dag_{\rm LP}
\end{split}
\end{equation} 
and applying Prop.~\ref{ULPpro}(i) directly to the RHS, we obtain the factorization $U_2\otimes U_{\bar 2}$ with $U_2=\rho_{g[j_{\rm l}-2l_{\rm LR},j_{\rm m}-l_{\rm LR}]}$. According to Lemma~\ref{Ufac}, Eq.~(\ref{ULUdag}) follows. Similarly, by choosing $S_1=\mathbb{Z}_L\backslash[j_{\rm m}-l_{\rm LR}+1,j_{\rm r}+2l_{\rm LR}]$, $S_2=[j_{\rm m}+l_{\rm LR}+1,j_{\rm r}+2l_{\rm LR}]$ and the same $S_3$, we can derive
 \begin{equation}
 \begin{split}
&\;\;\;\;\;U_{\rm LP}\rho_{g[j_{\rm m}+1,j_{\rm r}-l_{\rm LR}]}\otimes R_g(j_{\rm r})U^\dag_{\rm LP}\\
&=\tilde L_g(j_{\rm m}+1)\otimes\rho_{g[j_{\rm m}+l_{\rm LR}+1,j_{\rm r}]}.
\end{split}
\end{equation}
The independence of $\tilde R_g(j_{\rm m})$ and $\tilde L_g(j_{\rm m}+1)$ on $j_{\rm l}$ and $j_{\rm r}$ with $\min\{|j_{\rm l}-j_{\rm m}|,|j_{\rm r}-j_{\rm m}|\}\ge l_{\rm LR}$ is again a direct result of Prop.~\ref{ULPpro}(i), i.e., a local modification far from subsystem $S_3$ cannot propagate to $S_3$ through $U^\dag_{\rm LP}$. \hfill$\Box$

Similarly to the SPI, we can also generalize the cohomology character to inhomogeneous locality-preserving unitaries by realising that $L_g(j_{\rm l})\otimes R_g(j_{\rm r})$ is a linear representation, refer to Fig.~\ref{figS5} top.
\begin{proposition}
The cohomology class of $L_g$, which is opposite to that of $R_g$, is a well-defined global character.
\end{proposition}
To see this, note that $L_g(j_{\rm l})$ and $\tilde R_g(j_{\rm m})$ are equivalent projective representations (Fig.~\ref{figS5} bottom) and the cohomology class cannot depend on $j_{\rm l}$.
\hfill$\Box$

Just like the case of MPUs, we have
\begin{theorem}
The relative SPI 
is a topological invariant, which is additive by tensoring and composition.
\label{topoadd}
\end{theorem}
\emph{Proof:} We consider a continuous path of $G$-symmetric locality-preserving unitaries $U_{\rm LP}(\lambda)$,  along which the maximal Lieb-Robinson length is denoted as $l_{\rm max}\equiv\max_\lambda l_{\rm LR}(\lambda)$. For a fixed $g$-string operator $\rho_{g[j_{\rm l},j_{\rm r}]}$ with $|j_{\rm l}-j_{\rm r}|\ge 4l_{\rm max}$, we have the factorization relation
\begin{equation}
\begin{split}
&\;\;\;\;\;U^\dag_{\rm LP}(\lambda)\rho_{g[j_{\rm l},j_{\rm r}]}U_{\rm LP}(\lambda)\\
&=L_g(j_{\rm l};\lambda)\otimes\rho_{g[j_{\rm l}+l_{\rm max},j_{\rm r}-l_{\rm max}]}\otimes R_g(j_{\rm r};\lambda),
\end{split}
\label{faclam}
\end{equation}
where $L_g(j_{\rm l};\lambda)$ and $R_g(j_{\rm r};\lambda)$ act on $I_{\rm l}\equiv[j_{\rm l}-l_{\rm max},j_{\rm l}+l_{\rm max}-1]$ and $I_{\rm r}\equiv[j_{\rm r}-l_{\rm max}+1,j_{\rm r}+l_{\rm max}]$, respectively. Note that although the factorization in Eq.~(\ref{faclam}) may not be optimal in the sense that $L_g(j_{\rm l};\lambda)$ and $R_g(j_{\rm r};\lambda)$ can be reduced to $\mathbb{1}^{\otimes \Delta l}\otimes L^{\rm op}_g(j_{\rm l};\lambda)\otimes\rho^{\otimes \Delta l}_g$ and $\rho^{\otimes \Delta l}_g\otimes R^{\rm op}_g(j_{\rm l};\lambda)\otimes\mathbb{1}^{\otimes \Delta l}$ with $\Delta l=l_{\rm max}-l_{\rm LR}(\lambda)$. Nevertheless, the ratio $|\Tr L_g/\Tr R_g|$ always determines the relative SPI since additional factors $\chi_g$'s and $d$'s cancel out. Inspired by the experimental scheme, we can express the relative SPI in terms of $U_{\rm LP}(\lambda)$ via 
\begin{equation}
\frac{1}{4}\log\left|\frac{\Tr_{\bar I_{\rm l}}[\Tr_{I_{\rm l}}\rho^{\rm H}_{g[j_{\rm l},j_{\rm r}]}(\lambda)\Tr_{I_{\rm l}}\rho^{\rm H}_{g[j_{\rm l},j_{\rm r}]}(\lambda)^\dag]}{\Tr_{\bar I_{\rm r}}[\Tr_{I_{\rm r}}\rho^{\rm H}_{g[j_{\rm l},j_{\rm r}]}(\lambda)\Tr_{I_{\rm r}}\rho^{\rm H}_{g[j_{\rm l},j_{\rm r}]}(\lambda)^\dag]}
\right|,
\end{equation}
where $\rho^{\rm H}_{g[j_{\rm l},j_{\rm r}]}(\lambda)\equiv U^\dag_{\rm LP}(\lambda)\rho_{g[j_{\rm l},j_{\rm r}]}U_{\rm LP}(\lambda)$ is continuous with respect to $\lambda$. Accordingly, the relative SPI is a continuous function of $\lambda$. On the other hand, since $L_g\otimes R_g$ is a linear representation (see also the proof of Thm.~\ref{indgtopo}), 
we know that $\ind_g-\ind$ takes values over a finite set
\begin{equation}
\left\{\log\frac{|\sum_{j\in\mathbb{Z}_{d_g}} n_j\omega^j_{d_g}|}{d^{l_{\rm max}}|\chi_g|^{l_{\rm max}}}:\sum_{j\in\mathbb{Z}_{d_g}} n_j=d^{2l_{\rm max}}\right\}.
\end{equation} 
Therefore, the relative SPI must be a topological invariant. 

We move on to prove the additivity. Consider two $G$-symmetric locality-preserving unitaries $U_{\rm LP,1}$ and $U_{\rm LP,2}$ with Lieb-Robinson lengths $l_{\rm LR,1}$ and $l_{\rm LR,2}$, then their composition $U_{\rm LP}=U_{\rm LP,1}U_{\rm LP,2}$ is again $G$-symmetric and locality-preserving with $l_{\rm LR}=l_{\rm LR,1}+l_{\rm LR,2}-\Delta l_{12}$ ($\Delta l_{12}\ge0$). As shown in Fig.~\ref{figS6}, which is similar to Fig.~\ref{figS5}, we apply a ``pull-back" technique to $U^\dag_{\rm LP,1}\rho_{g[j_{\rm l},j_{\rm r}]}U_{\rm LP,1}$ to obtain (a rigorous derivation is parallel to the proof of Prop.~\ref{siteindep})
\begin{equation}
\begin{split}
(d\chi_g)^{\Delta l_{12}}\Tr L_g(j_{\rm l})&=\Tr L_{1,g}(j_{\rm l})\Tr\tilde R_{2,g}(j_{\rm m}),\\
(d\chi_g)^{\Delta l_{12}}\Tr R_g(j_{\rm r})&=\Tr R_{1,g}(j_{\rm r})\Tr\tilde L_{2,g}(j_{\rm m}+1),
\end{split}
\end{equation}
which implies 
\begin{equation}
\log\left|\frac{\Tr L_g(j_{\rm l})}{\Tr R_g(j_{\rm r})}\right|=\log\left|\frac{\Tr L_{1,g}(j_{\rm l})}{\Tr R_{1,g}(j_{\rm r})}\right|+\log\left|\frac{\Tr \tilde R_{2,g}(j_{\rm m})}{\Tr\tilde L_{2,g}(j_{\rm m}+1)}\right|
\end{equation}
and thus the additivity of the relative SPI. \hfill$\Box$ \newline

In fact, the same result (Thm.~\ref{topoadd}) holds for the cohomology class, to show which we need the following Lemma:
\begin{lemma}
Let $U(\lambda)=U_A(\lambda)\otimes U_B(\lambda)$ be a continuous path of unitaries acting on a bipartite system $A\bigcup B$, then we can always choose a proper ${\rm U}(1)$ gauge such that $U_A(\lambda)$ and $U_B(\lambda)$ are separately continuous paths. 
\label{bipcont}
\end{lemma}
\emph{Proof:} By assumption, we have
\begin{equation}
U_A(\lambda)\otimes\bar U_A(\lambda)=\langle\Phi_{B'B}|U(\lambda)\otimes\bar U(\lambda)|\Phi_{B'B}\rangle
\end{equation}
where $|\Phi_{BB'}\rangle=\frac{1}{\sqrt{d_B}}\sum^{d_B}_{j=1}|j_Bj_{B'}\rangle$ is the maximally entangled state of subsystem $B$ and its copy $B'$. Moreover, $U_A(\lambda)\otimes\bar U_A(\lambda)$ must be continuous with respect to $\lambda$ as $U(\lambda)$ is. 
Decomposing the entry of $U_A(\lambda)$ as $[U_A(\lambda)]_{mn}=r_{mn}(\lambda)\omega_{mn}(\lambda)$ with $r_{mn}(\lambda)\in\mathbb{R}^+\bigcup\{0\}$ and $\omega_{mn}(\lambda)\in{\rm U}(1)$, we can uniquely determine all $r_{mn}(\lambda)$ from $[U_A(\lambda)\otimes\bar U_A(\lambda)]_{mm,nn}=r_{mn}(\lambda)^2$ and all $\omega_{mn}(\lambda)/\omega_{m'n'}(\lambda)$ with $r_{mn}(\lambda)r_{m'n'}(\lambda)\neq0$ from $[U_A(\lambda)\otimes\bar U_A(\lambda)]_{mm',nn'}=r_{mn}(\lambda)r_{m'n'}(\lambda)\omega_{mn}(\lambda)/\omega_{m'n'}(\lambda)$, which are all continuous. Also, the unitarity $\sum_mr_{mn}(\lambda)^2=\sum_nr_{mn}(\lambda)^2=1$ implies at least $d_A$ nonzero entries. Imposing continuity to (the phase of) an arbitrary nonzero entry, the continuity of the others immediately follows from the continuity of $\omega_{mn}(\lambda)/\omega_{m'n'}(\lambda)$ and we obtain a continuous path of $U_A(\lambda)$. The corresponding $U_B(\lambda)$ can be determined from
\begin{equation}
U_B(\lambda)=d^{-1}_A\Tr_A[(U^\dag_A(\lambda)\otimes\mathbb{1}_B)U(\lambda)],
\end{equation}
which is also continuous with respect to $\lambda$. \hfill$\Box$ \newline
We are now ready to prove
\begin{theorem}
The cohomology class is a topological invariant, which is additive by tensoring and composition.
\end{theorem}
\emph{Proof:} According to the factorization relation (\ref{gfac}), using the same notations in the proof of Thm.~\ref{topoadd}, we have
\begin{equation}
\begin{split}
&\;\;\;\;\;L_g(j_{\rm l};\lambda)\otimes R_g(j_{\rm r};\lambda)\\
&=d^{4l_{\rm max}-L}\Tr_{\bar I_{\rm l}\bigcap\bar I_{\rm r}}[\rho^\dag_{g[j_{\rm l}+l_{\rm max},j_{\rm r}-l_{\rm max}]}\rho^{\rm H}_{g[j_{\rm l},j_{\rm r}]}(\lambda)],
\end{split}
\end{equation}
where $\rho^{\rm H}_{g[j_{\rm l},j_{\rm r}]}(\lambda)\equiv U^\dag_{\rm LP}(\lambda)\rho_{g[j_{\rm l},j_{\rm r}]}U_{\rm LP}(\lambda)$ is continuous with respect to $\lambda$. Therefore, we can apply Lemma~\ref{bipcont} and find two continuous paths $L_g(j_{\rm l};\lambda)$ and $R_g(j_{\rm r};\lambda)$, which are projective representations belonging to the opposite cohomology classes due to the fact that $L_g(j_{\rm l};\lambda)\otimes R_g(j_{\rm r};\lambda)$ is a linear representation.  According to Ref.~\cite{Schuch2011}, the cohomology class must stay unchanged along the continuous path. To prove the additivity, we only have to employ the pull-back technique shown in Fig.~\ref{figS6} and then use the additivity of cohomology class upon tensoring. \hfill$\Box$

Finally, we claim that, as a special case of inhomogeneous locality-preserving unitaries,
\begin{theorem}
An open-boundary locality-preserving unitary always has trivial index, SPIs and cohomology character.
\end{theorem}
\emph{Proof:} For the case of the (symmetry-irrelevant) index, see Ref.~\cite{Po2016}. In the presence of symmetry, 
we consider a $g$-string operator $\rho_{g[0,j_{\rm r}]}$ starting from the left edge of an open-boundary locality-preserving unitary $U_{\rm LP}$ with length $L$ and Lieb-Robinson length $l_{\rm LR}$ (we choose $j_{\rm r}>l_{\rm LR}$). Combining the locality-preserving property and the identity
\begin{equation}
U^\dag_{LP}\rho_{g[0,j_{\rm r}]}U_{LP}=\rho^{\otimes L}_gU^\dag_{LP}\rho^\dag_{g[j_{\rm r}+1,L-1]}U_{LP},
\end{equation}
we know that $L_g(0)=\mathbb{1}^{\otimes l_{\rm LR}}\otimes\rho^{\otimes l_{\rm LR}}_g$ (where $\mathbb{1}^{\otimes l_{\rm LR}}$ acts on the rightmost $l_{\rm LR}$ sites), implying that both the SPI and the cohomology class are trivial. \hfill$\Box$

\subsection{Implication for thermalization}
As pointed out in Ref.~\cite{Po2016}, ${\rm ind}\neq0$ implies Floquet thermalization. Here, we argue that even if ${\rm ind}=0$, thermalization will still be enforced whenever ${\rm ind}_g\neq0$. Combining Eq.~(\ref{RSPI}), $\Tr L_g\Tr R_g=d^{2l_{\rm LR}}\chi^{2l_{\rm LR}}_g$ with the fact that $|\Tr L_g|$ and $|\Tr R_g|$ are no more than $d^{2l_{\rm LR}}$, we can bound $l_{\rm LR}$ of $U_{\rm LP}$ from below by ($\ind=0$ used) 
\begin{equation}
l_{\rm LR}\ge\frac{|{\rm ind}_g|}{\log(d/|\chi_g|)}.
\end{equation}
It follows from the additivity of ${\rm ind}_g$ that $l_{\rm LR}$ of $U^t_{\rm LP}$ grows at least linearly with $t$, a universal feature of Floquet thermal phases. This expectation may be called SPT-enforced thermalization \footnote{The bound for $l_{\rm LR}$ is tight, as the edges $L_g,R_g$ of a $g$-string operator indeed grow in this manner. However, it does not prove spreading of every operator.}, and is to be contrasted with the case of nontrivial cohomology class, which implies either thermalization or many-body localization accompanied by spontaneous symmetry breaking \cite{Potter2017}. On the other hand, analogous to the fact that $U_{\rm LP}$ with ${\rm ind}\neq0$ cannot be generated by a finite-depth circuit of local unitaries \cite{Po2016}, $G$-symmetric $U_{\rm LP}$ with ${\rm ind}_g\neq0$ cannot be generated by a finite-depth circuit of $G$-symmetric local unitaries (with $\ind_g=0$ each), just like those in nontrivial cohomology classes \cite{Potter2017}.

\section{Towards a complete classification}
Similar to the case of equivalence, the question whether the SPIs together with the index and the cohomology class give a complete classification for strong equivalence is settled by whether an MPU with trivial SPIs, index and cohomology class is always strongly equivalent to the identity. The answer turns out to be \emph{no}, because the SPI can be further refined, at least for those MPUs with trivial cohomology.

\subsection{Refined SPI}
Let us first introduce the definition of \emph{refined SPI}:
\begin{definition}[Refined SPI]
Given a $G$-symmetric MPU in the trivial cohomology class, a refined index with respect to any $g\in G$ with $\chi_g\neq0$ is defined as
\begin{equation}
{\rm rind}_g=\left(\frac{\Tr y_g}{\Tr \rho_g}\right)^{d_g},
\label{rind}
\end{equation}
where $y_g$ is already lifted to a linear representation.
\end{definition}
As $y_g$ is a linear representation, the phase ambiguity is discretized as a 1D representation and thus killed by the power $d_g$, implying that the refined SPI is well-defined for a given standard form. To show 
\begin{proposition}
Given a $G$-symmetric MPU in the trivial cohomology class and $g\in G$ with $\chi_g\neq0$, the refined SPI given in Eq.~(\ref{rind}) is well-defined, although the standard form is not unique.
\label{RSPIpro}
\end{proposition}
we can use almost the same analysis for the blocking-independence of the SPI.

We still have to prove that the refined SPI is not only well-defined but also a topological invariant. To this end, we need the following Lemma: 
\begin{lemma}
Given a continuous path of projective representation $z_g(\lambda)$ of a finite group $G$ in the trivial cohomology class, there exists a continuous function $\omega_g:\lambda\to{\rm U}(1)$ such that $\omega_g(\lambda)z_g(\lambda)$ is a continuous path of linear representation.
\label{tricont}
\end{lemma}
\emph{Proof:} By assumption, 
there exists $z'_g(\lambda)=\omega'_g(\lambda)z_g(\lambda)$ which is a linear representation but not necessarily continuous with respect to $\lambda$. From $z'_g(\lambda)^{d_g}=\mathbb{1}$, we know that $\omega'_g(\lambda)^{d_g}=z_g(\lambda)^{-d_g}$ should be continuous, thus $\omega'_g(\lambda)=\omega^{\rm c}_g(\lambda)\omega^{n_g(\lambda)}_{d_g}$ with $\omega^{\rm c}_g(\lambda)$ being continuous and $n_g(\lambda)\in\mathbb{Z}_{d_g}$. Moreover, from $z'_g(\lambda)z'_h(\lambda)=z'_{gh}(\lambda)$, we know that $\omega'_g(\lambda)\omega'_h(\lambda)\omega'_{gh}(\lambda)^{-1}=[z_g(\lambda)z_h(\lambda)z_{gh}(\lambda)]^{-1}$ is continuous, implying the continuity of $\omega^{n_g(\lambda)}_{d_g}\omega^{n_h(\lambda)}_{d_h}\omega^{-n_{gh}(\lambda)}_{d_{gh}}$. On the other hand, $\omega^{n_g(\lambda)}_{d_g}\omega^{n_h(\lambda)}_{d_h}\omega^{-n_{gh}(\lambda)}_{d_{gh}}$ takes discrete values in $\{\omega^j_{d_G}:j\in\mathbb{Z}_{d_G}\}$ ($d_G$: order of group $G$), so $\omega^{n_g(\lambda)}_{d_g}\omega^{n_h(\lambda)}_{d_h}\omega^{-n_{gh}(\lambda)}_{d_{gh}}=\omega^{n_g(0)}_{d_g}\omega^{n_h(0)}_{d_h}\omega^{-n_{gh}(0)}_{d_{gh}}$ should be $\lambda$-independent. Now defining $z''_g(\lambda)\equiv\omega^{\rm c}_g(\lambda)\omega^{n_g(0)}_{d_g}z_g(\lambda)$ which is obviously continuous, we can check that it is a linear representation. \hfill$\Box$ \newline
Combining Prop.~\ref{RSPIpro} and Lemma~\ref{tricont}, we immediately obtain
\begin{theorem}
The refined SPI is a topological invariant, which is multiplicative by tensoring and composition.
\end{theorem}
The generalization of the refined SPI to inhomogeneous locality-preserving unitaries can be done in full analogy to the generalization of the SPI. 

We note that $\ind_g=d_g^{-1}\log|{\rm rind}_g|$, so the refinement comes from the \emph{phase} of ${\rm rind}_g$, whose absolute value contains the same information as $\ind_g$. On the other hand, due to the integer factor $d_g\neq1$ for $\forall g\neq e$ which is necessary for eliminating the ambiguity of the definition, $\log|{\rm rind}_g|$ or the real part of $\log{\rm rind}_g$ does not have a clear information-theoretic interpretation and is, in particular, not directly related to the edge imbalance of an evolved symmetry string operator. Therefore, we prefer not taking the logarithm when defining the refined SPIs.

Since ${\rm rind}_g$ is a topological invariant, we have 
\begin{theorem}
Given two symmetric, strongly equivalent MPUs $U_0$ and $U_1$, then $U^\dag_1U_0$ has trivial (unit) refined SPIs.
\label{thm3}
\end{theorem}
A minimal example that is not strongly equivalent to the identity as indicated by Thm.~\ref{thm3} but not captured by Thm.~2 in the main text is a $\mathbb{Z}_3$-symmetric MPU consisting of qutrits. Denoting $\rho^{(j)}$ ($j\in\mathbb{Z}_3$) as the irreducible (1D) representation of $\mathbb{Z}_3$ with $\rho^{(j)}_{1_{\mathbb{Z}_3}}=\omega^{j}_3$, for $\rho=\rho^{(0)}\oplus2\rho^{(1)}$, we can realize $x=y=\rho^{(0)}\oplus2\rho^{(2)}$ on the virtual level due to $\rho\otimes\rho=x\otimes y=\rho^{(0)}\oplus4\rho^{(1)}\oplus4\rho^{(2)}$. Substituting $\Tr y_{1_{\mathbb{Z}_3}}=1+2\omega^2_3$ and $\Tr \rho_{1_{\mathbb{Z}_3}}=1+2\omega_3$ into Eq.~(\ref{rind}), we obtain ${\rm rind}_{1_{\mathbb{Z}_3}}=-1\neq1$, implying that the MPU cannot be deformed into the identity without breaking the $\mathbb{Z}_3$ symmetry with such a fixed representation. We will see an explicit construction of such an MPU in the following subsection.

While Thm.~\ref{thm3} is undoubtedly an improved criterion for ruling out strong equivalence, we still do not know whether it gives the complete classification. It is also far from clear whether we can measure ${\rm rind}_g$ in experiments.

\subsection{Systematic construction of nontrivial MPUs in the trivial cohomology class} 
Given an MPU in the trivial cohomology class, all the projective representations $z_g$, $x_g$ and $y_g$ on the virtual level can be lifted to linear representations, which are all elements in the \emph{representation ring} \cite{Serre1977}:
\begin{definition}[Representation Ring]
For a finite group $G$ with in total $r$ different irreducible representations denoted by $\rho_1,\rho_2,...,\rho_r$, 
the representation ring $R(G)$ is defined as $\{\rho=\bigoplus^r_{\alpha=1}n_\alpha\rho_\alpha:n_\alpha\in\mathbb{Z}\}$, on which
the addition between two elements $\rho=\bigoplus^r_{\alpha=1}n_\alpha\rho_\alpha$ and $\rho'=\bigoplus^r_{\alpha=1}n'_\alpha\rho_\alpha$ is defined as
\begin{equation}
\rho\oplus\rho'\equiv\bigoplus^r_{\alpha=1}(n_\alpha+n'_\alpha)\rho_\alpha,
\end{equation}
while their multiplication is defined as
\begin{equation}
\rho\otimes\rho'\equiv\bigoplus^r_{\alpha=1}\left(\sum^r_{\beta,\gamma=1}N^\alpha_{\beta\gamma}n_\beta n'_\gamma\right)\rho_\alpha,
\end{equation}
where the nonnegative integer (Littlewood-Richardson) coefficients $N^\alpha_{\beta\gamma}$ are determined from $\rho_\beta\otimes\rho_\gamma=\bigoplus^r_{\alpha=1}N^\alpha_{\beta\gamma}\rho_\alpha$, the decomposition of the tensor-product representation. 
\end{definition}
According to the character theory, there is an injective map $\chi:R(G)\to\mathbb{C}^r$ ($r$ is also the number of conjugacy classes)
\begin{equation}
\chi(\rho)\equiv(\Tr\rho_{g_1},\Tr\rho_{g_2},...,\Tr\rho_{g_r}),
\end{equation}
where $\Tr\rho_{g_j}=\sum_in_i\chi_{ij}\in\mathbb{Z}[\omega_{d_{g_j}}]$ (Kummer ring), $g_j$ is a representative element of the $j$th conjugacy class and $\chi_{ij}\equiv\Tr\rho_{i,g_j}$ is an entry in the character table. From the injectivity of $\chi$, we know that $N^\alpha_{\beta\gamma}=N^\alpha_{\gamma\beta}$ and thus $R(G)$ is commutative. Note that 
$\mathbb{Z}$ can be regarded as the representation ring of the trivial group $G=\{e\}$.

With the notion of representation ring in mind, we can systematically construct MPUs with nontrivial SPIs through finding nontrivial solutions to 
\begin{equation}
\rho^{\otimes 2}=x\otimes y,\;\;\;\;
x,y,\rho\in R^+(G),
\label{PLR}
\end{equation}
where $R^+(G)$ is a subset of $R(G)$ consisting of all the linear representations (i.e., $n_\alpha\ge0$ and $\sum^r_{\alpha=1}n_\alpha\ge1$), $\rho$ is on the physical level and $x$ and $y$ are on the virtual level of the standard form (c.f. Eq.~(3) in the main text). By nontrivial, we mean that $x\neq \rho_{1\rm D}\otimes\rho$ for any one-dimensional representation $\rho_{1\rm D}$ of $G$, otherwise the MPU can be trivialized into the identity. 
In practice, we may alternatively focus on the nontrivial decomposition of the character vector, of which each component is decomposed on the Kummer ring.

For the simplest case $G=\mathbb{Z}_n$, we have $r=n$ and $N^\alpha_{\beta\gamma}=\delta_{\alpha,\beta+\gamma}$ and a minimal example of nontrivial decomposition
\begin{equation}
(2\rho_0\oplus 2\rho_1)^{\otimes2}=4\rho_0\otimes(\rho_0\oplus2\rho_1\oplus\rho_2) 
\label{swapdec}
\end{equation}
has already been mentioned in the main text and can be realized by the bilayer SWAP circuit on two qubits via $(2\rho_0 + 2 \rho_1) = (2\rho_0) \otimes (\rho_0 + \rho_1)$:
\begin{equation}
\begin{tikzpicture}
\draw[thick] (2.6,-0.85) -- (2.6,-0.6) (2.6,-0.1) -- (2.6,0.85) (3.2,-0.85) -- (3.2,0.85) (3.8,-0.85) -- (3.8,0.85);
\draw[thick] (4.4,0.85) -- (4.4,-0.1) (4.4,-0.6) -- (4.4,-0.85);
\draw[thick,fill=white] (2.4,0.1) rectangle (3.4,0.6) (3.6,0.1) rectangle (4.6,0.6);
\draw[thick,fill=white] (3,-0.1) rectangle (4,-0.6);
\draw[thick] (2.3,-0.1) -- (2.8,-0.1) -- (2.8,-0.6) -- (2.3,-0.6);
\draw[thick] (4.7,-0.1) -- (4.2,-0.1) -- (4.2,-0.6) -- (4.7,-0.6);
\Text[x=2.9,y=0.35,fontsize=\footnotesize]{SWAP}
\Text[x=4.1,y=0.35,fontsize=\footnotesize]{SWAP}
\Text[x=3.5,y=-0.35,fontsize=\footnotesize]{SWAP}
\Text[x=1.9]{$\cdots$}
\Text[x=5.1]{$\cdots$}
\Text[x=5.6]{$=$}
\Text[x=6.1]{$\cdots$}
\draw[dashed,fill=gray!10] (7.5,-0.75) rectangle (8.7,0.75); 
\draw[thick] (6.6,-0.85) -- (6.6,0.85) (7.2,-0.85) -- (7.2,0.85) (7.8,-0.85) -- (7.8,0.85) (8.4,-0.85) -- (8.4,0.85); 
\draw[thick] (6.6,-0.13) circle (0.08) (7.2,-0.35) circle (0.08) (6.6,-0.57) circle (0.08) (7.8,-0.13) circle (0.08) (8.4,-0.35) circle (0.08) (7.8,-0.57) circle (0.08); 
\draw[thick] (6.4,-0.13) -- (6.68,-0.13) (6.4,-0.35) -- (6.6,-0.35) (6.4,-0.57) -- (6.68,-0.57) 
(7.88,-0.13) -- (7.2,-0.13) (7.12,-0.35) -- (7.8,-0.35) (7.88,-0.57) -- (7.2,-0.57) 
(8.8,-0.13) -- (8.4,-0.13) (8.32,-0.35) -- (8.8,-0.35) (8.8,-0.57) -- (8.4,-0.57); 
\fill (7.2,-0.13) circle (0.05) (6.6,-0.35) circle (0.05) (7.2,-0.57) circle (0.05) (8.4,-0.13) circle (0.05) (7.8,-0.35) circle (0.05) (8.4,-0.57) circle (0.05); 
\Text[x=9.2]{$\cdots$}
\draw[thick] (6.6,0.13) circle (0.08) (7.2,0.35) circle (0.08) (6.6,0.57) circle (0.08) (7.8,0.13) circle (0.08) (8.4,0.35) circle (0.08) (7.8,0.57) circle (0.08); 
\fill (7.2,0.13) circle (0.05) (6.6,0.35) circle (0.05) (7.2,0.57) circle (0.05) (8.4,0.13) circle (0.05) (7.8,0.35) circle (0.05) (8.4,0.57) circle (0.05); 
\draw[thick] (6.52,0.13) -- (7.2,0.13) (6.6,0.35) -- (7.28,0.35) (6.52,0.57) -- (7.2,0.57) (7.72,0.13) -- (8.4,0.13) (7.8,0.35) -- (8.48,0.35) (7.72,0.57) -- (8.4,0.57); 
\Text[x=9.5,y=-0.2]{,}
\end{tikzpicture}
\label{biswap}
\end{equation}
where \begin{tikzpicture}\fill (0,0) circle (0.05); \draw[thick] (0,0) -- (0.48,0) (0.4,-0.08) -- (0.4,0.08); \draw[thick] (0.4,0) circle (0.08);\end{tikzpicture} is the CNOT gate $|0\rangle\langle0|\otimes\mathbb{1}+|1\rangle\langle1|\otimes X$ and the part marked in the gray rectangle is the building block $\mathcal{U}$. The equivalence between the left and right hand sides in Eq.~(\ref{biswap}) can be understood from 
\begin{equation}
\begin{tikzpicture}
\draw[thick] (0.2,-0.5) -- (0.2,0.5) (0.8,-0.5) -- (0.8,0.5);
\draw[thick,fill=white] (0,-0.25) rectangle (1,0.25);
\Text[x=0.5,fontsize=\footnotesize]{SWAP}
\Text[x=1.5]{$=$}
\draw[thick] (2.2,-0.5) -- (2.2,0.5) (2.8,-0.5) -- (2.8,0.5);
\draw[thick] (2.2,0) circle (0.08);
\draw[thick] (2.8,0.22) circle (0.08);
\draw[thick] (2.8,-0.22) circle (0.08);
\fill (2.8,0) circle (0.05) (2.2,0.22) circle (0.05) (2.2,-0.22) circle (0.05);
\draw[thick] (2.2,0.22) -- (2.88,0.22) (2.12,0) -- (2.8,0) (2.2,-0.22) -- (2.88,-0.22);
\Text[x=3.5]{$=$}
\draw[thick] (4.2,-0.5) -- (4.2,0.5) (4.8,-0.5) -- (4.8,0.5);
\draw[thick] (4.8,0) circle (0.08);
\draw[thick] (4.2,0.22) circle (0.08);
\draw[thick] (4.2,-0.22) circle (0.08);
\fill (4.2,0) circle (0.05) (4.8,0.22) circle (0.05) (4.8,-0.22) circle (0.05);
\draw[thick] (4.12,0.22) -- (4.8,0.22) (4.2,0) -- (4.88,0) (4.12,-0.22) -- (4.8,-0.22);
\Text[x=5.1,y=-0.2]{.}
\end{tikzpicture}
\end{equation}
For $n=2$, however, such a nontrivial decomposition (\ref{swapdec}) is not stable against blocking, i.e., $\rho\otimes x=\rho\otimes\rho$. Indeed, we can trivialize the bilayer SWAP circuit as follows:
\begin{widetext}
\begin{equation}
\begin{tikzpicture}
\Text[x=-0.4]{$\cdots$}
\draw[thin,dashed,fill=gray!20] (0.08,0.46) rectangle (0.92,0.68) (1.28,0.46) rectangle (2.12,0.68) (-0.1,-0.46) rectangle (0.32,-0.68) (0.68,-0.46) rectangle (1.52,-0.68) (1.88,-0.46) rectangle (2.3,-0.68);
\draw[thick] (0.2,-0.85) -- (0.2,0.85) (0.8,-0.85) -- (0.8,0.85) (1.4,-0.85) -- (1.4,0.85) (2,-0.85) -- (2,0.85); 
\draw[thick] (0.2,-0.13) circle (0.08) (0.2,-0.57) circle (0.08) (0.8,-0.35) circle (0.08) (1.4,-0.13) circle (0.08) (1.4,-0.57) circle (0.08) (2,-0.35) circle (0.08) 
(0.2,0.13) circle (0.08) (0.2,0.57) circle (0.08) (0.8,0.35) circle (0.08) (1.4,0.13) circle (0.08) (1.4,0.57) circle (0.08) (2,0.35) circle (0.08); 
\fill (0.2,-0.35) circle (0.05) (0.8,-0.13) circle (0.05) (0.8,-0.57) circle (0.05) (1.4,-0.35) circle (0.05) (2,-0.13) circle (0.05) (2,-0.57) circle (0.05) 
(0.2,0.35) circle (0.05) (0.8,0.13) circle (0.05) (0.8,0.57) circle (0.05) (1.4,0.35) circle (0.05) (2,0.13) circle (0.05) (2,0.57) circle (0.05); 
\draw[thick] (-0.1,-0.13) -- (0.28,-0.13) (-0.1,-0.35) -- (0.2,-0.35) (-0.1,-0.57) -- (0.28,-0.57) (0.8,-0.13) -- (1.48,-0.13) (0.72,-0.35) -- (1.4,-0.35) (0.8,-0.57) -- (1.48,-0.57) (2,-0.13) -- (2.3,-0.13) (1.92,-0.35) -- (2.3,-0.35) (2,-0.57) -- (2.3,-0.57) (0.12,0.13) -- (0.8,0.13) (0.2,0.35) -- (0.88,0.35) (0.12,0.57) -- (0.8,0.57) (1.32,0.13) -- (2.0,0.13) (1.4,0.35) -- (2.08,0.35) (1.32,0.57) -- (2.,0.57); 
\Text[x=2.7]{$\cdots$}
\Text[x=3.2]{$\to$}
\Text[x=3.7]{$\cdots$}
\draw[thin,dashed,fill=gray!20] (4,0.02) rectangle (4.42,0.24) (5.38,0.02) rectangle (6.4,0.24) (4.18,-0.02) rectangle (5.62,-0.24);
\draw[thick] (4.3,-0.85) -- (4.3,0.85) (4.9,-0.85) -- (4.9,0.85) (5.5,-0.85) -- (5.5,0.85) (6.1,-0.85) -- (6.1,0.85);
\draw[thick] (4.3,0.13) circle (0.08) (4.3,-0.13) circle (0.08) (4.9,-0.35) circle (0.08) (4.9,0.35) circle (0.08) (5.5,-0.13) circle (0.08) (5.5,0.13) circle (0.08) (6.1,-0.35) circle (0.08) (6.1,0.35) circle (0.08);
\fill (4.3,-0.35) circle (0.05) (4.3,0.35) circle (0.05) (4.9,-0.13) circle (0.05) 
(5.5,-0.35) circle (0.05) (5.5,0.35) circle (0.05) 
(6.1,0.13) circle (0.05);
\draw[thick] (4.3,0.35) -- (4.98,0.35) (5.5,0.35) -- (6.18,0.35) (3.92,-0.35) -- (4.3,-0.35) (4.82,-0.35) -- (5.5,-0.35) (6.02,-0.35) -- (6.4,-0.35) (4.22,-0.13) -- (5.58,-0.13) (4,0.13) -- (4.38,0.13) (5.42,0.13) -- (6.4,0.13);
\Text[x=6.8]{$\cdots$}
\Text[x=7.3]{$\to$}
\Text[x=7.8]{$\cdots$}
\draw[thin,dashed,fill=gray!20] (8.88,-0.46) rectangle (10.32,-0.24) (8.1,0.46) rectangle (9.12,0.24) (10.08,0.46) rectangle (10.5,0.24);
\draw[thick] (8.4,-0.85) -- (8.4,0.85) (9,-0.85) -- (9,0.85) (9.6,-0.85) -- (9.6,0.85) (10.2,-0.85) -- (10.2,0.85);
\fill (9.6,-0.35) circle (0.05) (8.4,0.35) circle (0.05);
\draw[thick] (9,-0.35) circle (0.08) (9,0.35) circle (0.08) (10.2,-0.35) circle (0.08) (10.2,0.35) circle (0.08);
\draw[thick] (8.92,-0.35) -- (10.28,-0.35) (8.1,0.35) -- (9.08,0.35) (10.12,0.35) -- (10.5,0.35);
\Text[x=10.9]{$\cdots$}
\Text[x=11.3,y=-0.2]{,}
\end{tikzpicture}
\end{equation}
\end{widetext}
where the local quantum gates in the gray boxes can continuously be deformed into local identities without breaking the $\mathbb{Z}_2$ symmetry (represented by $\mathbb{1}\otimes Z$). When $n\ge3$, the stability of the nontrivial MPU against blocking (and disorder) is ensured by the nontrivial SPI.

To construct a $\mathbb{Z}_2$-symmetric MPU with zero index and nonzero SPI, we first write down the general representation-decomposition relation
\begin{equation}
(d_0\rho_0\oplus d_1\rho_1)^{\otimes2}=(m_0\rho_0\oplus m_1\rho_1)\otimes(n_0\rho_0\oplus n_1\rho_1),
\end{equation}
which is equivalent to
\begin{equation}
(d_0\pm d_1)^2=(m_0\pm m_1)(n_0\pm n_1).
\end{equation}
By assumption, we would like to find a solution with $d=d_0+d_1=m_0+m_1=n_0+n_1$ and $m_0-m_1\neq n_0-n_1$, $d_0\neq d_1$. 
After some trials, a minimal solution with the smallest local Hilbert-space dimension $d=8$ is found to be
\begin{equation}
(6\rho_0\oplus2\rho_1)^{\otimes2}=(5\rho_0\oplus3\rho_1)\otimes8\rho_0.
\label{Z2min}
\end{equation}
Regarding $d=8=2^3$ as three qubits, we can implemented Eq.~(\ref{Z2min}) with
\begin{equation}
\begin{split}
\rho_{1_{\mathbb{Z}_2}}&=(|1\rangle\langle1|\otimes Z+|0\rangle\langle0|\otimes\mathbb{1})\otimes\mathbb{1},\\
x_{1_{\mathbb{Z}_2}}&=|00\rangle\langle00|\otimes\mathbb{1}+(\mathbb{1}^{\otimes2}-|00\rangle\langle00|)\otimes Z,\\
y_{1_{\mathbb{Z}_2}}&=\mathbb{1}^{\otimes3},
\end{split}
\end{equation}
which can be realized by
\begin{equation}
\begin{tikzpicture}[scale=0.8]
\Text[x=-2.25]{$u=$}
\draw[thick,dashed] (-1.8,-0.75) rectangle (1.8,0.75);
\draw[thick] (-1.5,-1) -- (-1.5,1) (1.5,-1) -- (1.5,1);
\draw[thick] (-0.5,-1) .. controls (-0.5,-0.5) .. (0,-0.45) (0,-0.45) .. controls (1,-0.4) .. (1,1); 
\draw[thick] (0.5,-1) .. controls (0.5,-0.5) .. (0,-0.45) (0,-0.45) .. controls (-1,-0.4) .. (-1,1); 
\draw[thick] (-1,-1) .. controls (-1,-0.5) .. (-0.75,-0.475) (-0.75,-0.475) .. controls (-0.5,-0.45) .. (-0.5,1); 
\draw[thick] (1,-1) .. controls (1,-0.5) .. (0.75,-0.475) (0.75,-0.475) .. controls (0.5,-0.45) .. (0.5,1); 
\draw[thick,fill=white] (-0.7,-0.1) rectangle (0.7,0.5);
\fill[black] (-1.5,0.1) circle (0.05) (-1,0.3) circle (0.05);
\draw[thick] (-1.5,0.1) -- (-0.7,0.1) (-1,0.3) -- (-0.7,0.3);
\Text[y=0.2]{$U$}
\Text[x=2.1,y=-0.1]{,}
\end{tikzpicture}
\end{equation}
\begin{equation*}
\begin{tikzpicture}[scale=0.8]
\Text[x=3.95]{$v=$}
\draw[thick,dashed] (4.4,-0.75) rectangle (8,0.75);
\draw[thick] (5.7,-1) -- (5.7,1) (6.7,-1) -- (6.7,1);
\draw[thick] (4.7,1) .. controls (4.7,0.5) .. (6.4,0.45) (6.4,0.45) .. controls (7.2,0.4) .. (7.2,-1);
\draw[thick] (7.7,1) .. controls (7.7,0.5) .. (6,0.45) (6,0.45) .. controls (5.2,0.4) .. (5.2,-1);
\draw[thick] (5.2,1) .. controls (5.2,0.35) .. (5.6,0.325) (5.6,0.325) .. controls (6,0.3) .. (6,0.1);
\draw[thick] (7.2,1) .. controls (7.2,0.35) .. (6.8,0.325) (6.8,0.325) .. controls (6.4,0.3) .. (6.4,0.1);
\draw[thick] (4.7,-1) .. controls (4.7,-0.65) .. (5.6,-0.625) (5.6,-0.625) .. controls (6,-0.6) .. (6,-0.5);
\draw[thick] (7.7,-1) .. controls (7.7,-0.65) .. (6.8,-0.625) (6.8,-0.625) .. controls (6.4,-0.6) .. (6.4,-0.5);
\draw[thick,fill=white] (5.85,0.1) rectangle (6.55,-0.5);
\fill[black] (6.7,-0.1) circle (0.05) (7.2,-0.3) circle (0.05);
\draw[thick] (6.7,-0.1) -- (6.55,-0.1) (7.2,-0.3) -- (6.55,-0.3);
\Text[x=6.2,y=-0.2]{$U'$}
\Text[x=8.3,y=-0.1]{,}
\end{tikzpicture}
\end{equation*}
where the two-qubit-controlled gate in $u$ is given by
\begin{equation}
\begin{split}
|11\rangle\langle11|
\otimes(X\otimes|0\rangle\langle0|+\mathbb{1}\otimes|1\rangle\langle1|)\\
+|01\rangle\langle01|
\otimes\mathbb{S}+\mathbb{1}\otimes|0\rangle\langle0|
\otimes\mathbb{1}^{\otimes2},
\end{split}
\end{equation}
and that in $v=\mathcal{S}u^\dag \mathcal{S}$ ($\mathcal{S}$ swaps two adjacent sites, either of which consists of three qubits) is given by
\begin{equation}
\begin{split}
(|0\rangle\langle0|\otimes X+|1\rangle\langle1|\otimes\mathbb{1})\otimes|11\rangle\langle11|\\
+\mathbb{S}\otimes|01\rangle\langle01|+\mathbb{1}^{\otimes 3}\otimes|0\rangle\langle0|,
\end{split}
\end{equation}

\begin{figure*}
\begin{center}
       \includegraphics[width=12cm, clip]{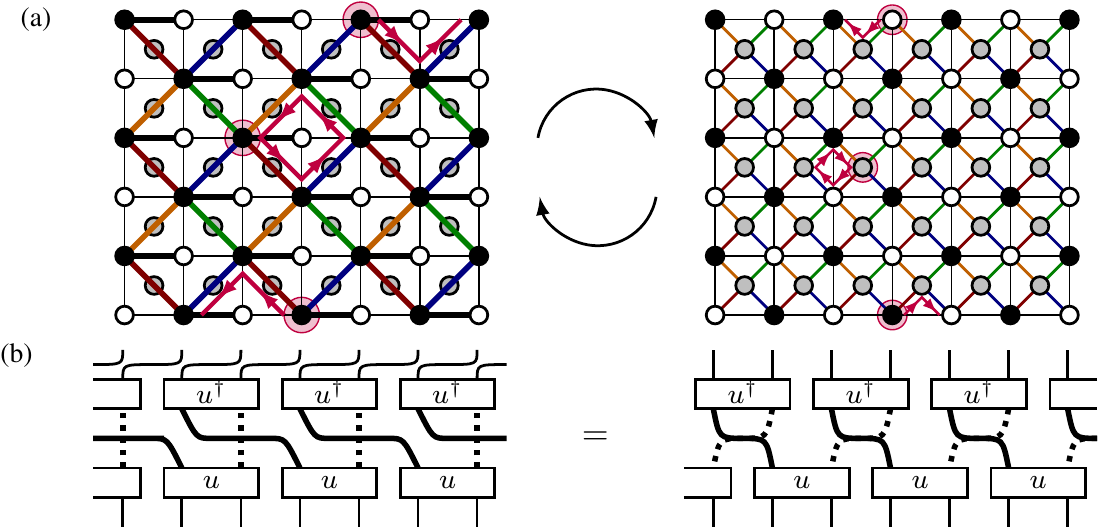}
       \end{center}
    \caption{(a) General construction of a 2D symmetric Floquet system whose edge dynamics is characterized by a nontrivial SPI. (Left) The thick black lines denote the conjugation by $u:\mathbb{C}^d\otimes\mathbb{C}^d\to\mathbb{C}^l\otimes\mathbb{C}^r$ and all the colored lines are the SWAP gates on two copies of $\mathbb{C}^l$ (left virtual Hilbert space). After $u$ conjugations, the SWAP gates are pulled back to the physical level. 
   (Right) Each colored line corresponds to a SWAP gate on two copies of $\mathbb{C}^d$ (physical Hilbert space). In both figures, the SWAP gates are switched on sequentially as \textcolor{red!50!black}{red}$\to$\textcolor{blue!50!black}{blue}$\to$\textcolor{green!50!black}{green}$\to$\textcolor{orange!75!black}{orange}. The bulk dynamics of the Floquet system is trivial, as indicated by the arrows forming loops. (b) Tensor-network representation (standard form) of the (bottom) edge dynamics. The left hand side corresponds directly to the action by (a) Left followed by (a) Right, while it is clear from the right hand side that $v=u^\dag\mathbb{S}_{\rm v}$, where $\mathbb{S}_{\rm v}$ is the SWAP gate $\mathbb{C}^l\otimes\mathbb{C}^r\to\mathbb{C}^r\otimes\mathbb{C}^l$ that exchanges the left and right virtual Hilbert spaces.}
   \label{figS8}
\end{figure*}

Finally, we present an explicit construction that realizes
\begin{equation}
(\rho_0\oplus2\rho_1)^{\otimes2}=(\rho_0\oplus2\rho_2)^{\otimes2},
\end{equation}
on $R(G=\mathbb{Z}_3)$, the minimal example with trivial SPIs but a nontrivial refined SPI, as mentioned in the previous subsection. We consider a qutrit system and specify the representations on the physical and virtual levels as
\begin{equation}
\begin{split}
\rho_{1_{\mathbb{Z}_3}}&=|0\rangle\langle0|+\omega_3(|1\rangle\langle1|+|2\rangle\langle2|),\\
x_{1_{\mathbb{Z}_3}}&=y_{1_{\mathbb{Z}_3}}=\rho^2_{1_{\mathbb{Z}_3}}=\rho_{2_{\mathbb{Z}_3}}.
\end{split}
\end{equation}
To transform $\rho\otimes\rho$ into $x\otimes y$, we can use
\begin{equation}
\begin{tikzpicture}
\draw[thick] (0,-0.4) -- (0,0.4) (0.6,-0.4) -- (0.6,0.4);
\draw[thick] (0,0.12) -- (0.68,0.12) (-0.08,-0.12) -- (0.6,-0.12); 
\fill[black] (0,0.12) circle (0.05) (0.6,-0.12) circle (0.05);
\draw[thick] (0,-0.12) circle (0.08) (0.6,0.12) circle (0.08);
\Text[x=-1,y=0]{$u=v=$}
\Text[x=1,y=-0.2]{,}
\end{tikzpicture}
\end{equation}
where \begin{tikzpicture}\fill (0,0) circle (0.05); \draw[thick] (0,0) -- (0.48,0) (0.4,-0.08) -- (0.4,0.08); \draw[thick] (0.4,0) circle (0.08);\end{tikzpicture} $\equiv\sum_{j\in\mathbb{Z}_3}|j\rangle\langle j|\otimes X^{-j}$ ($X\equiv\sum_{k\in\mathbb{Z}_3}|k-1\rangle\langle k|$) is a natural generalization of the CNOT gate. We can easily check that $v|00\rangle=|00\rangle$, $v(|01\rangle,|12\rangle,|02\rangle,|21\rangle)=(|12\rangle,|02\rangle,|21\rangle,|01\rangle)$ and $v(|10\rangle,|11\rangle,|20\rangle,|22\rangle)=(|11\rangle,|20\rangle,|22\rangle,|10\rangle)$, implying
\begin{equation}
\rho^{\otimes2}_{1_{\mathbb{Z}_3}}v=v\rho^{\otimes2}_{2_{\mathbb{Z}_3}},\;\;\;\;
\rho^{\otimes2}_{2_{\mathbb{Z}_3}}v=v\rho^{\otimes2}_{1_{\mathbb{Z}_3}}.
\end{equation}

\section{Details on the parent Floquet systems}
In this section, we show that given any (refined) SPI, we can construct a 2D Floquet system such that its bulk dynamics is trivial while its edge dynamics is an MPU characterized by the prescribed SPI (including the chiral index). In fact, this construction is equally applicable to nontrivial cohomology classes, thus provides a unified point of view for understanding topological MPUs with symmetries and their relations to 2D Floquet systems. In turn, this construction implies that 2D Floquet SPT phases are much richer than expected, in the sense of strong equivalence.

Given a set of well-defined (refined) SPIs and/or the cohomology class, we can find $u$ such that $x\otimes y=u\varrho^{\otimes2}u^\dag$ gives the desired decomposition of the symmetry representation $\varrho^{\otimes2}=\rho^{\otimes2k}$ (after blocking). The general construction is then schematically illustrated in Fig.~\ref{figS8}(a), where black and white sites live on the vertices of square plaquettes while gray sites live in the centers. Regardless of the color, each site is assigned with a local Hilbert space $\mathbb{C}^d$, where the symmetry is linearly represented as $\rho$. We impose the periodic (open) boundary condition to the horizontal (vertical) direction. In the first half of a Floquet period, we sequentially apply direct products of $u$-conjugated SWAP gates of black sites as follows (see Fig.~\ref{figS8}(a) Left):
\begin{equation}
U=
U_{\begin{tikzpicture}[scale=0.15] \draw[thick,orange!75!black] (1,1) -- (0,0); \draw[thick] (1,1) -- (2,1) (0,0) -- (1,0); \draw[fill=black] (1,1) circle (0.2); \draw[fill=black] (0,0) circle (0.2); \draw[fill=white] (2,1) circle (0.2); \draw[fill=white] (1,0) circle (0.2); \end{tikzpicture}}
U_{\begin{tikzpicture}[scale=0.15] \draw[thick,green!50!black] (-1,1) -- (0,0); \draw[thick] (-1,1) -- (0,1) (0,0) -- (1,0); \draw[fill=black] (-1,1) circle (0.2); \draw[fill=black] (0,0) circle (0.2); \draw[fill=white] (0,1) circle (0.2); \draw[fill=white] (1,0) circle (0.2); \end{tikzpicture}}
U_{\begin{tikzpicture}[scale=0.15] \draw[thick,blue!50!black] (1,1) -- (0,0); \draw[thick] (1,1) -- (2,1) (0,0) -- (1,0); \draw[fill=black] (1,1) circle (0.2); \draw[fill=black] (0,0) circle (0.2); \draw[fill=white] (2,1) circle (0.2); \draw[fill=white] (1,0) circle (0.2); \end{tikzpicture}}
U_{\begin{tikzpicture}[scale=0.15] \draw[thick,red!50!black] (-1,1) -- (0,0); \draw[thick] (-1,1) -- (0,1) (0,0) -- (1,0); \draw[fill=black] (-1,1) circle (0.2); \draw[fill=black] (0,0) circle (0.2); \draw[fill=white] (0,1) circle (0.2); \draw[fill=white] (1,0) circle (0.2); \end{tikzpicture}},
\label{Ufirst}
\end{equation}
where
\begin{equation}
U_{\begin{tikzpicture}[scale=0.15] \draw[thick,red!50!black] (-1,1) -- (0,0); \draw[thick] (-1,1) -- (0,1) (0,0) -- (1,0); \draw[fill=black] (-1,1) circle (0.2); \draw[fill=black] (0,0) circle (0.2); \draw[fill=white] (0,1) circle (0.2); \draw[fill=white] (1,0) circle (0.2); \end{tikzpicture}}=
\bigotimes_{\begin{tikzpicture}[scale=0.15] \draw[thick,red!50!black] (-1,1) -- (0,0); \draw[thick] (-1,1) -- (0,1) (0,0) -- (1,0); \draw[fill=black] (-1,1) circle (0.2); \draw[fill=black] (0,0) circle (0.2); \draw[fill=white] (0,1) circle (0.2); \draw[fill=white] (1,0) circle (0.2); \end{tikzpicture}}
u^{\dag\otimes2}_{\begin{tikzpicture}[scale=0.15] \draw[thick] (0,0) -- (1,0); \draw[fill=black] (0,0) circle (0.2); \draw[fill=white] (1,0) circle (0.2); \end{tikzpicture}}
\mathbb{S}_{\begin{tikzpicture}[scale=0.15] \draw[thick,red!50!black] (-1,1) -- (0,0); \draw[fill=black] (0,0) circle (0.2) (-1,1) circle (0.2); \end{tikzpicture}}u^{\otimes2}_{\begin{tikzpicture}[scale=0.15] \draw[thick] (0,0) -- (1,0); \draw[fill=black] (0,0) circle (0.2); \draw[fill=white] (1,0) circle (0.2); \end{tikzpicture}},
\end{equation}
with $\mathbb{S}_{\begin{tikzpicture}[scale=0.15] \draw[thick,red!50!black] (-1,1) -- (0,0); \draw[fill=black] (0,0) circle (0.2) (-1,1) circle (0.2); \end{tikzpicture}}$ acting on two copies of $\mathbb{C}^l$, on which the symmetry is represented as $x$. Here the identities on the virtual Hilbert space $\mathbb{C}^r$, where the symmetry representation is $y$, are omitted for simplicity. Other unitaries in Eq.~(\ref{Ufirst}) are defined similarly. One can check that $U$ is trivial in the bulk, while the left virtual Hilbert spaces $\mathbb{C}^l$ are left(right)-translated to the nearest one (separated by two physical sites) at the lower (upper) boundary (see the arrows at the edges of Fig.~\ref{figS8}(a) Left). In the second half of a Floquet period, we sequentially apply direct products of SWAP gates on the physical level as follows (see Fig.~\ref{figS8}(a) Right):
\begin{equation}
U'=(U'_{\begin{tikzpicture}[scale=0.125] \draw[thick,orange!75!black] (0,1) -- (1,0); \draw[fill=white] (0,1) circle (0.3); \draw[fill=gray!50!white] (1,0) circle (0.3); \end{tikzpicture}}\otimes U'_{\begin{tikzpicture}[scale=0.125] \draw[thick,orange!75!black] (0,1) -- (1,0); \draw[fill=black] (0,1) circle (0.3); \draw[fill=gray!50!white] (1,0) circle (0.3); \end{tikzpicture}}) (U'_{\begin{tikzpicture}[scale=0.125] \draw[thick,green!50!black] (1,1) -- (0,0); \draw[fill=white] (1,1) circle (0.3); \draw[fill=gray!50!white] (0,0) circle (0.3); \end{tikzpicture}}\otimes U'_{\begin{tikzpicture}[scale=0.125] \draw[thick,green!50!black] (1,1) -- (0,0); \draw[fill=black] (1,1) circle (0.3); \draw[fill=gray!50!white] (0,0) circle (0.3); \end{tikzpicture}}) (U'_{\begin{tikzpicture}[scale=0.125] \draw[thick,blue!50!black] (0,1) -- (1,0); \draw[fill=gray!50!white] (0,1) circle (0.3); \draw[fill=white] (1,0) circle (0.3); \end{tikzpicture}}\otimes U'_{\begin{tikzpicture}[scale=0.125] \draw[thick,blue!50!black] (0,1) -- (1,0); \draw[fill=gray!50!white] (0,1) circle (0.3); \draw[fill=black] (1,0) circle (0.3); \end{tikzpicture}}) (U'_{\begin{tikzpicture}[scale=0.125] \draw[thick,red!50!black] (1,1) -- (0,0); \draw[fill=gray!50!white] (1,1) circle (0.3); \draw[fill=white] (0,0) circle (0.3); \end{tikzpicture}}\otimes U'_{\begin{tikzpicture}[scale=0.125] \draw[thick,red!50!black] (1,1) -- (0,0); \draw[fill=gray!50!white] (1,1) circle (0.3); \draw[fill=black] (0,0) circle (0.3); \end{tikzpicture}}),
\end{equation}
where
\begin{equation}
U'_{\begin{tikzpicture}[scale=0.125] \draw[thick,red!50!black] (1,1) -- (0,0); \draw[fill=gray!50!white] (1,1) circle (0.3); \draw[fill=white] (0,0) circle (0.3); \end{tikzpicture}}\otimes U'_{\begin{tikzpicture}[scale=0.125] \draw[thick,red!50!black] (1,1) -- (0,0); \draw[fill=gray!50!white] (1,1) circle (0.3); \draw[fill=black] (0,0) circle (0.3); \end{tikzpicture}} =  \bigotimes_{\begin{tikzpicture}[scale=0.125] \draw[thick,red!50!black] (1,1) -- (0,0); \draw[fill=gray!50!white] (1,1) circle (0.3); \draw[fill=white] (0,0) circle (0.3); \end{tikzpicture}}\mathbb{S}_{\begin{tikzpicture}[scale=0.125] \draw[thick,red!50!black] (1,1) -- (0,0); \draw[fill=gray!50!white] (1,1) circle (0.3); \draw[fill=white] (0,0) circle (0.3); \end{tikzpicture}} \otimes \bigotimes_{\begin{tikzpicture}[scale=0.125] \draw[thick,red!50!black] (1,1) -- (0,0); \draw[fill=gray!50!white] (1,1) circle (0.3); \draw[fill=black] (0,0) circle (0.3); \end{tikzpicture}}\mathbb{S}_{\begin{tikzpicture}[scale=0.125] \draw[thick,red!50!black] (1,1) -- (0,0); \draw[fill=gray!50!white] (1,1) circle (0.3); \draw[fill=black] (0,0) circle (0.3); \end{tikzpicture}},
\end{equation}
with $\mathbb{S}_{\begin{tikzpicture}[scale=0.125] \draw[thick,red!50!black] (1,1) -- (0,0); \draw[fill=gray!50!white] (1,1) circle (0.3); \draw[fill=white] (0,0) circle (0.3); \end{tikzpicture}}$ and $\mathbb{S}_{\begin{tikzpicture}[scale=0.125] \draw[thick,red!50!black] (1,1) -- (0,0); \draw[fill=gray!50!white] (1,1) circle (0.3); \draw[fill=black] (0,0) circle (0.3); \end{tikzpicture}}$ acting on two copies of $\mathbb{C}^d$. This is exactly the model studied in Refs.~\cite{Po2016} and \cite{Harper2017b}, which is the bosonic counterpart of the anomalous Floquet insulator of free fermions \cite{Rudner2013}. Such a unitary is also trivial in the bulk, while the dynamics at the lower (upper) edge is the one-site right (left) translation (see the arrows at the edges of Fig.~\ref{figS8}(a) Right). The entire Floquet operator is given by $U_{\rm F}=U'U$, which again has a trivial bulk, but exhibits nontrivial edge dynamics described by the MPU given in Fig.~\ref{figS8}(b).

\end{document}